# SSVEP-BASED BCI WHEELCHAIR CONTROL SYSTEM

**ZHOU CE**

**A GRADUATION EXERCISE SUBMITTED IN PARTIAL FULFILLMENT OF THE REQUIREMENTS FOR THE DEGREE OF BACHELOR OF ENGINEERING (ELECTRICAL)**

**DEPARTMENT OF ELECTRICAL ENGINEERING**

**FACULTY OF ENGINEERING**

**UNIVERSITY OF MALAYA**

**SEMESTER 2, 2015/2016**

# DECLARATION BY THE CANDIDATE

I, Zhou Ce, hereby declare that except where due acknowledgement has been made, the work presented in this thesis is by my own, and has not been submitted previously in whole or in part, to qualify for any other academic award.

The contents of this graduation exercise are the result of the work I have been carrying out since the official commencement date of the approved thesis project.

Date:

Signature:

Full Name:

Passport No.:

Matric No.:

# Abstract

A brain computer interface (BCI) is a system which allows a person to communicate or control the surroundings without depending on the brain’s normal output pathways of peripheral nerves and muscles. A lot of successful applications have arisen utilizing the advantages of BCI to assist disabled people so-called assistive technology. Considering using BCI has less limitation and huge potential, this project has been proposed to control the movement of an electronic wheelchair via brain signals. The goal of doing this project is to help the disabled people, especially paralyzed people suffering from motor disabilities, to improve their life qualities.

In order to realize the project stated above, Steady-State Visual Evoked Potential (SSVEP) is involved. It can be easily elicited in the visual cortical with the same frequency as the one is being focused by subject.

There are two important parts in this project. One is to process the EEG signals and another one is to make a visual stimulator using hardware. The EEG signals are processed in Matlab using the algorithm of Butterworth Infinite Impulse Response (IIR) bandpass filter (for preprocessing) and Fast Fourier Transform (FFT) (for feature extraction). Besides, a harmonics-based classification method is proposed and applied in the classification part. Moreover, the design of the visual stimulator combines LEDs as flickers and LCDs as information displayers on one panel. Microcontrollers are employed to control the SSVEP visual stimuli panel.

This project is evaluated by subjects with different races and ages. Experimental results show the system is easy to be operated and it can achieve approximately minimum 1 second time delay. So it demonstrates that this SSVEP-based BCI controlled wheelchair has a huge potential to be applied for disabled people in the future.

# Acknowledgement

First, I would like to thank my supervisor in University of Malaya (UM), Prof. Mahmoud Moghavvemi, for giving me this opportunity to get to know so many innovative and interesting projects. His support, kind advice, patience and guidance are most appreciated.

Prof. Xueyu Zou, my supervisor in Yangtze University, I would like to thank you for giving me this opportunity to do my Final Year Project in UM.

My friend and collaborator, Alireza Safdari Ghandehari who contributed to the project all the time, I would like to thank you for helping me in so many occasions.

I would like to thank the Center of Research in Electronics (CRAE) for providing me with the best equipment and environment for doing this project.

Kuek JinHao who is the previous researcher on this project, I would like to thank him for kindly helping me in signal processing in Matlab.

I would like to thank Clement Kwan, who is also my collaborator in this project, helping me in doing experiments of operating wheelchair in real time.

I would like to thank Chen Jun Hui, Haroon Wardak, Hidayat Hambali as well as Alireza Safdari Ghandehari and Clement Kwan, for becoming the subjects during the experiments.

Finally, I would like to thanks all my friends for their moral and spiritual support. I also wish to express my deepest gratitude to my family for their constant support, belief and encouragement. Without the help of the people mentioned above, this work would never have come into existence. Thank all of you again.

# Table of Contents

# List of Figures

# List of Tables

# 1. Introduction

## 1.1. Overview and Motivation

At present, the main impetus and motivation of the BCI research is that this communication technology can benefit those who have severe or complete motor paralysis, such as amyotrophic lateral sclerosis (ALS), Guillain–Barre´ syndrome, brain stem stroke and severe cerebral palsy [1][3]. During the late-stage of ALS, some patients are not able to communicate any more, even eye movements become unreliable and impossible [3]. These patients are referred to as totally “locked-in” people [4]. Meanwhile, motor recovery is also impossible for patients with progressive neurodegenerative diseases which prevent nerve cells from the respond of controlling voluntary movement [5]. Nevertheless, in most cases, patients still remain cognitive and sensory functions work, which support them to be aware of their environment [6]. BCI provides an alternative way for people to communicate and exchange information with the surroundings. That is why applying BCI in real life becomes a meaningful and promising way to help people in need.

## 1.2. Objective

The aim of this project is to design and developed an SSVEP-based wheelchair controlled by BCI system. The scope and objectives of this project includes:

- Investigation of the factors affecting performance of SSVEP
- Development of LED-based visual stimulator
- Development of communication between laptop and smart wheelchair
- Improving the long response time
- Performing real-time evaluation of the system

# 2. Literature Review

## 2.1.Brain Computer Interface (BCI)

### 2.1.1. Definition of BCI

A brain computer interface (BCI) is a system which allows a person to communicate or control the surroundings without depending on the brain's normal output pathways of peripheral nerves and muscles [1][2].

An online typical and completed BCI consists of six stages, which are signal acquisition, signal preprocessing, feature extraction, classification, translation into a command, and feedback [7] [8].

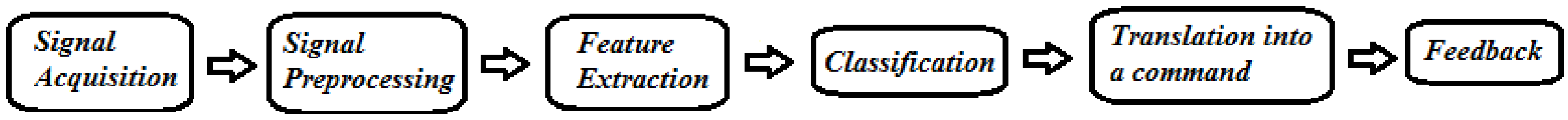

**Figure 1: Overview of a Typical and Complete BCI**

a. **Signal Acquisition**: This stage aims at acquiring the brain signal which reflects brain activities using specific sensor.

b. **Signal Preprocessing**: This stage consists reducing the noise of the input data and enhancing the useful information in the signals for further processing purposes.

c. **Feature Extraction**: The discriminative information in the input signal is extracted into relevant features in this stage.

d. **Classification**: This stage defines different features, obtained from the last stage, into different classes; which is called "classifiers". Each class corresponds to the specific mental state from the observed signals.

e. **Translation into a Command**: Once the classification part is done, a command associated to this kind of mental state is sent to the application such as a wheelchair, a robot or a speller.

f. **Feedback**: The final stage gives the subject a signal that the command has sent successfully which can help the subject control his action and brain activity.

This is the whole process of the online BCI interface. However, what must be mentioned is that calibration is an essential work to be done before operating every BCI system [9]. The reason is that calibration results in selecting the best features, which will contribute for the system operating accuracy.

### 2.1.2. Different Types of BCI

Nowadays, different BCI systems can be classified into three different categories: dependent BCI versus independent BCI, invasive BCI versus non-invasive BCI as well as synchronous BCI versus asynchronous BCI.

The first distinction which is generally highlighted is between dependent BCI and independent BCI [10]. A dependent BCI needs a certain level of control by patient, like controlling the direction of gazing [11], whereas an independent BCI does not need any motor control from patient who lost his/her motor ability totally.

The second distinction is invasive versus non-invasive according to the way that the brain signals are being measured in the BCI system [10]. If the electrodes used for getting brain signals are implanted over the surface of the cerebral cortex, the signal recorded from these electrodes is called electrocardiogram (ECoC) which does not damage any neurons, because no electrodes penetrate the brain. Besides, if the electrodes are placed deeply into the cerebral cortex, the signal recorded from electrodes penetrating brain tissue is called intracranial recordings. These two kinds of BCI system are assigned as invasive BCI. On the contrary, the sensor placed outside the cerebral cortex without performing surgery or breaking the skin is called non-invasive BCI. There are several non-invasive methods successfully applied in BCI systems, such as Electroencephalography (EEG), MagnetoEncephalography (MEG), Functional Magnetic Resonance Imaging (fMRI), and Functional Near-Infrared Imaging (fNIR) etc.

The third distinction often concerned is between synchronous BCI and asynchronous BCI [12]. In the synchronous BCI, the subject can only control the application during the certain time point, which means the whole procedure is imposed by the BCI system [10]. On the other hand, the asynchronous BCI seems like more flexible and user-friendly to use in the daily life without any external intervention. Patients are able to interact with the output devices at any time. However, it should be noticed that it is a lot harder to design an asynchronous BCI than a synchronous BCI because of the “self-paced” feature [12]. Generally, an asynchronous BCI should continuously do the signal processing to determine whether the patient is going to control the output device or not. Currently, designing an efficient asynchronous BCI becomes the biggest challenge which is urgently to be addressed [10].

## 2.2. Signal Acquisition

### 2.2.1. Neural Principles

There are approximately $10^{11}$ neurons or nerve cells making up the fundamental processing unit and forming the complex interconnected networks in the human brain in order to produce human behavior [13]. The four main parts of a typical neuron are: the cell body or soma, dendrites, axon and presynaptic terminals [14]. The cell body containing the nucleus which is the center of the neuron and the cell body is responsible for protein synthesis. The short branches extending from the cell body are called dendrites. Their function is to receive incoming signal sent by other neurons and as an input of the neuron. Every neuron has only one tubular output called axon. The end of axon are presynaptic terminal and they transmit electrical signals to other neurons. The junction where two neurons communicate with each other with the help of neurotransmitters is defined as synapse. Both of input and output parts include plenty of important information transferring among different neurons with the distances from 0.1mm to 2m [13].

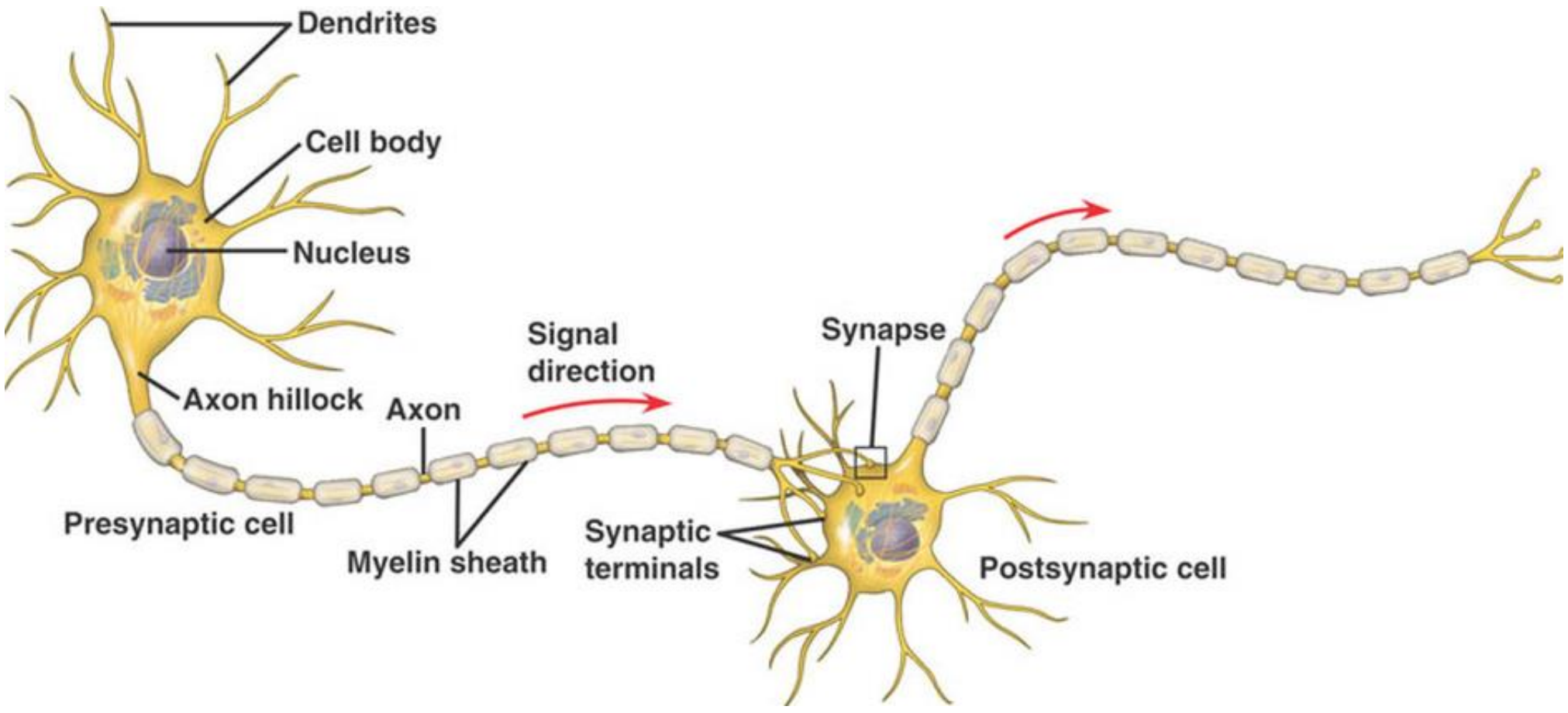

**Figure 2: The Structure of a Neuron [43]**

The central nervous system (CNS) together with the peripheral nervous system (PNS) plays a primary role in dealing with all mental activities so that it can control human behavior such as thinking, writing, speaking and moving. Human brain can be divided into left and right cerebral hemispheres, each of which has function of sensory and motor processing on the opposite side of the limbs and organs. In addition, it is impossible that similar cerebral hemispheres can have functionally equivalent or exactly symmetrical [15]. It should also be noticed that no conscious behavior can act alone which means that every function involves the entire cortex in some way [13].

From the anatomical view point, there are four lobes in CNS: Frontal Lobe, Parietal Lobe, Occipital Lobe and Temporal Lobe [15]. Figure 3 shows the functional areas of human brain.

a. Frontal Lobe: Executive functions, movement control.

b. Parietal Lobe: Multimodal sensory information integration.

c. Occipital Lobe: Visual processing center containing the visual cortex.

d. Temporal Lobe: Hearing and auditory signal processing, memory, emotion.

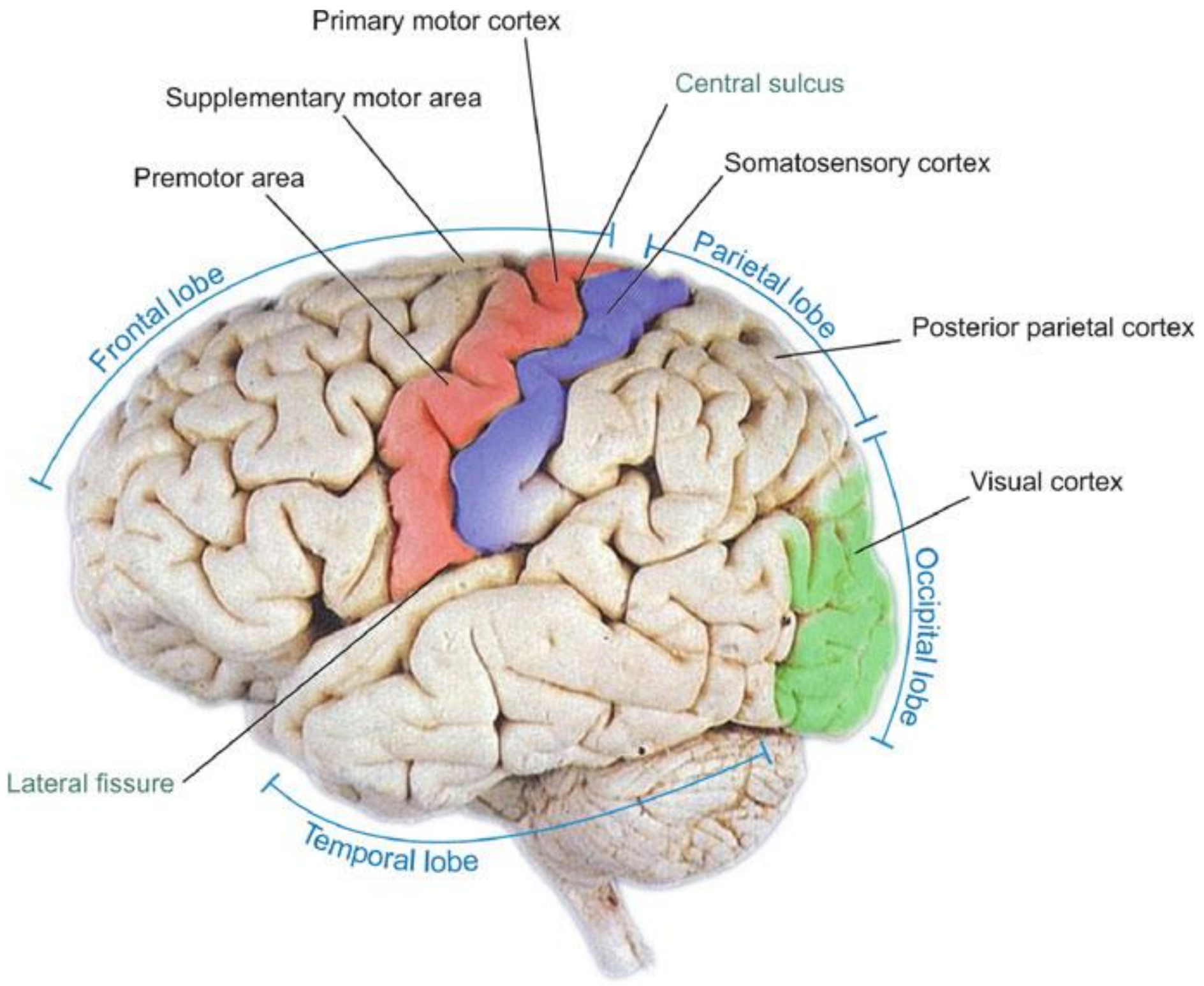

**Figure 3: Functional Areas of Human Brain [25]**

### 2.2.2. Different Types of Neuroimaging Techniques

There are different types of neuroimaging techniques being used in BCI, the functions of which are translating the meaning of specific brain signal into electrical signal.

The brain is bloody and electric. The materials changing between blood flow and electrical impulses make up the main activity of the brain. When the rate of activities within the neuron increases, it will cause the rate of metabolic demand for glucose and oxygen increase. Thus, it will also increase the rate of cerebral blood flow (CBF) to the active region. Therefore, the whole brain will start working by sending out tiny electrical impulses.

There are two major types of neuroimaging methods: haemodynamic and electrophysiological [17]. Haemodynamic is based on blood flowing and electrophysiological is relevant to electromagnetic [21].

On the one hand, haemodynamic response is a process which dependent on the oxygenation level in blood. Blood is more oxygenated in an activated region of the brain than in a non-activated region, which will cause two kinds of hemoglobin, deoxyhemoglobin (HbR) and oxyhemoglobin (HbO2). HbR and HbO2 differ in their magnetic susceptibility. HbR has a higher magnetization decay rate than HbO2 [22]. The typical Haemodynamic methods include Functional Magnetic Resonance Imaging (fMRI) and Functional Near-Infrared Imaging (fNIR).

- Functional Magnetic Resonance Imaging (fMRI): It provides quantitative hemodynamic information for both HbO2 and HbR, which can help to distinguish the activated region in the human brain when cognitive performance occurs [23].

- Functional Near-Infrared Imaging (fNIR): It qualitatively measures the information of functional activity in the human brain using optical techniques, such as scattered photons, which depend on the changes of the blood flow and tissue oxidation [23].

On the other hand, all the neurons connect into neural networks. The way of communicating among different neurons is by sending each other tiny electrical impulses thousands of times per second. When networks fire is synchronizing, the dynamics of the electric activity can be detected and recorded outside the skull [22]. The typical electrophysiological methods include Magnetoencephalography (MEG) and Electroencephalography (EEG).

- Magnetoencephalography (MEG): It uses sensors, in a tank containing liquid helium to enhance superconductivity, to measure the brain signal in magnetic fields. MEG signal is dominated by currents oriented tangential to the skull.

- Electroencephalography (EEG): It is the way to record electrical potentials over the scalp produced by plenty of neurons in the human brain.

Applying the fMRI and MEG in real life needs expensive and huge devices, which is not suitable most of the time. ECoG and intracranial recordings can acquire high quality signals including good topographical resolution and wide frequency ranges for BCI but they are not easily applicable due to their invasive nature. Long-term safety, stability and duration of the signal should be the main concerns before clinical use. Once implanted, their signal quality also can gradually become weaker because of tissue reaction issues [15]. Near-infrared spectroscopy (NIRS) is the theory of the fNIR and it is also a very new measurement modality noticed by research groups because of its portable, sensitive and non-invasive features [16]. As time passing by, the ease of installation and simple electronic design may make fNIR become a more and more popular acquisition method in BCI system.

### 2.2.3. Electroencephalography (EEG)

Electroencephalography (EEG) measures the electrical activity which is the sum of postsynaptic potentials generated by thousands of neurons having the same radial orientation with respect to the scalp. And the electrical activity can be obtained and recorded by using electrodes placed on the scalp [9] [18]. Figure 4 shows three different ways to measure brain signal, which are EEG, ECoG and Intracortical Recordings.

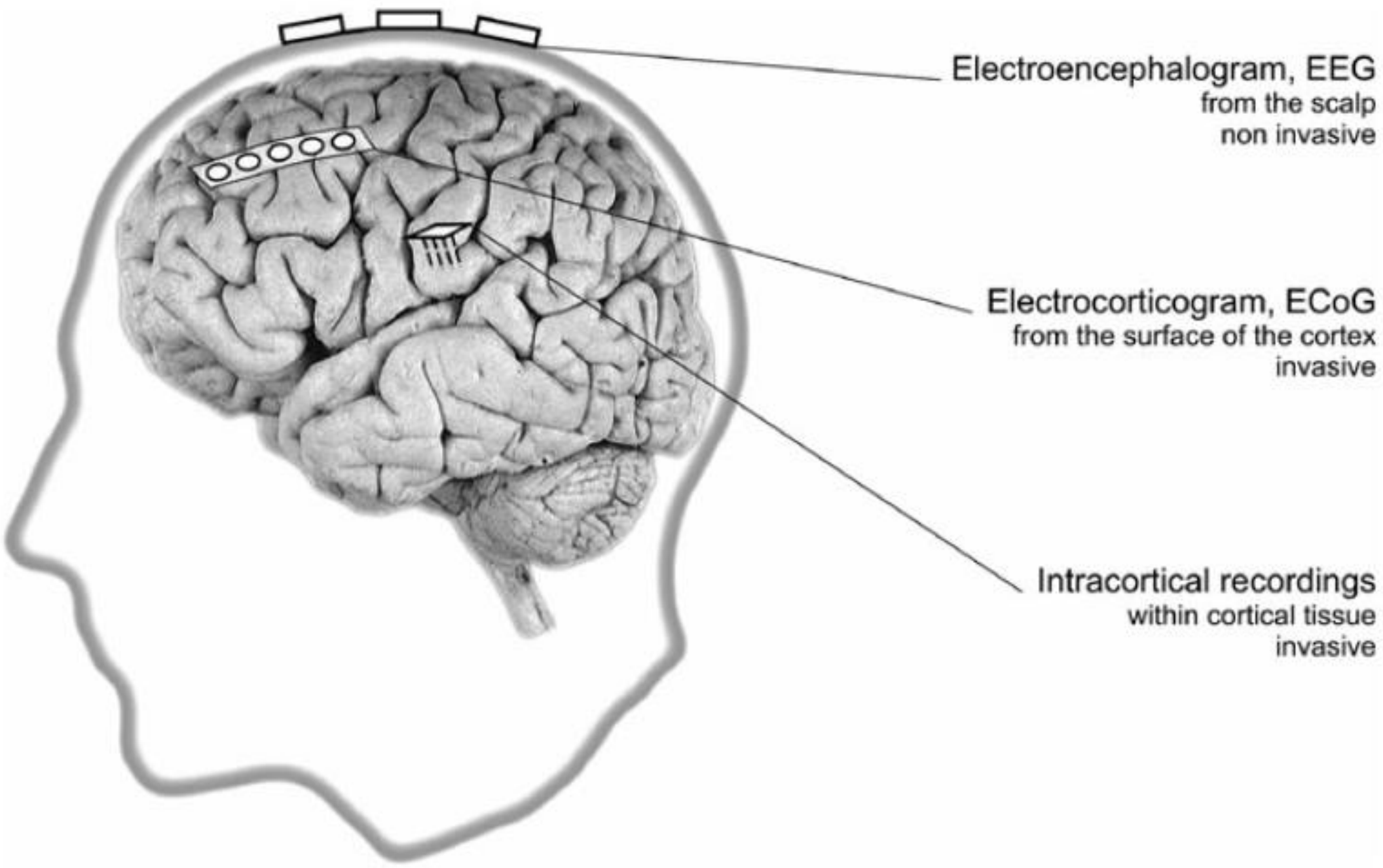

**Figure 4: EEG, ECoG and Intracortical Recordings [25].**

EEG also has both advantages and disadvantages. As it is shown in Table 1, in comparison to MEG or fMRI, EEG is more portable, inexpensive and determines a reasonable trade-off between temporal and spatial resolution. On the other hand, ECoG or Intracortical recordings have higher signal quality, frequency range and spatial resolution [18]. Furthermore, artifacts are less problematic. However, these advantages come with the serious drawback of requiring surgery, which is relevant to ethical, financial, and other considerations. It is also not clear whether invasive methods can provide safe and stable recording over years. Non-invasive BCI is easier and more convenient to implement EEG to real life because of no surgery requirement and lower associated risk. However, signals are attenuated by crossing the scalp, skull and many other layers before reaching the electrodes. Consequently, signals recorded by EEG have very weak amplitudes, in the range of 2 to 100 μV [19]. Meanwhile, the noise is generated both within the brain and over the scalp. And it is evaluated by signal-to-noise ratio (SNR). A low SNR means the quality of the signal is not good. Unfortunately, the main drawback of EEG is its low SNR.

**Table 1: Comparison of Neuroimaging Methods** [17]

| Method | Temporal Resolution | Spatial Resolution | Invasiveness | Activity | Portability |
|---|---|---|---|---|---|
| EEG | 0.05s | 10mm | Non-invasive | Electrical | Portable |
| MEG | 0.05s | 5mm | Non-invasive | Magnetic | Non-portable |
| ECoG | 0.003s | 1mm | Invasive | Electrical | Portable |
| Intracortical | 0.003s | 0.05-0.5mm | Invasive | Electrical | Portable |
| fMRI | 1s | 1mm | Non-invasive | Hemodynamic | Non-portable |
| fNIRS | 1s | 5mm | Non-invasive | Hemodynamic | Portable |

EEG signals are consist of various oscillations which are distinguished by different frequency ranges so-called “rhythms” [18]. There are six main brain rhythms with specific properties [9] [18] [20].

a. **Delta Rhythm (δ):** This is a slow rhythm lies within the range from 1 to 4 Hz, which is mainly found in healthy adults during deep sleep or while walking.

b. **Theta Rhythm (θ):** This is a slightly faster rhythm lies within the range from 4 to 7 Hz, observed mainly during consciousness slips towards drowsiness and it also appears in children. This rhythm is associated with creative inspiration and deep meditation as well.

c. **Alpha Rhythm (α):** This kind of rhythm is located in the range of 8 to 12 Hz frequency band, which are mainly found in the posterior regions of the head, known as occipital lobe. It appears when the subject has closed eyes or is in a relaxation state without any concentration. It is also observed that the amplitude is reduced or eliminated after opening eyes or starting some mental concentration. Alpha rhythm is the most prominent rhythm in brain activities as known so far.

d. **Mu Rhythm ($\mu$):** These are oscillations in the range of 8 to 13 Hz frequency band, mainly observed in the motor and sensorimotor cortex. The amplitude of this rhythm varies when subjects perform movements. Consequently, this rhythm is also known as the “sensorimotor rhythm”.

e. **Beta Rhythm (β):** This is a relatively fast rhythm, belonging approximately to the range of 13 to 30 Hz, which can be observed in awaken and conscious persons. This rhythm is also affected by movements or active thinking.

f. **Gamma Rhythm (γ):** This rhythm concerns mainly frequency range from 30 to 100Hz. It is associated with various cognitive and motor functions. Although the occurrence is very rare and the amplitudes are extremely low, it can help to determine some brain diseases.

#### 2.2.4. Electrodes placement

As discussed in the previous paragraphs, EEG is recorded by using electrodes. The number of electrodes used for EEG depends on the application, for instance in high-resolution EEG, 256 electrodes are used and in low-resolution EEG, few electrodes are used. Generally, the EEG used in BCIs is low-resolution EEG, where the placement of electrodes is commonly based on the International 10-20 system [24]. This system is standardized by the American Electroencephalographic Society and measures four points on the scalp, which are nasion, inion, left and right preauricular points [17]; moreover, it is suitable and can be achieved on different subjects. Figure 5 shows the International 10-20 electrode placement system.

Figure 5 illustrates the international 10-20 positioning system [24] [25]. The letters F, C, T, P and O stand for Frontal, Central, Temporal, Parietal and Occipital respectively. The distances between electrodes are typically 10 or 20 % of a half head circumference apart. For example, Fp is 10% from the Nasion, which is the intersection of the frontal and nasal bones at the bridge of the nose, and 20% from Fz3. Similarly O1 and O2 are 10% from Inion which is a small bulge on the back of the skull exactly about the neck.

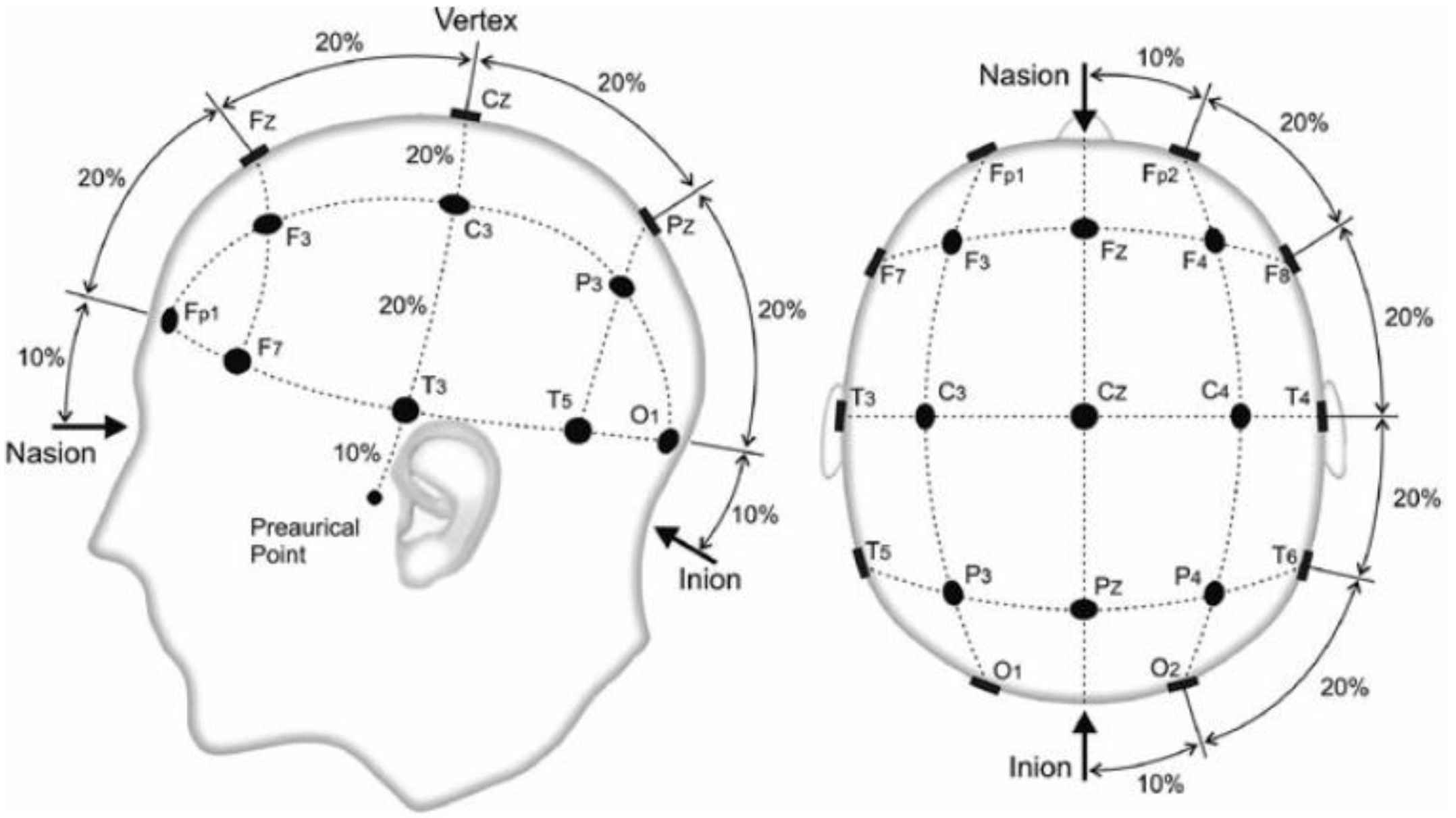

**Figure 5: The International 10-20 Electrode Placement System [25].**

All of the EEG signals are measured as the different potentials over time between active electrodes and constant reference electrode. The reference electrode is chosen by the willing of researchers, but in most cases, an inactive position which can provide constant electrical potential like midline electrodes, Cz and Fpz, are initially considered. A qualified electrode should always have constant impedance below 5 kΩ, so that it can record signals more accurately [19].

## 2.3. Steady-State Visual Evoked Potential (SSVEP)

A steady-state visual evoked potential (SSVEP) is a continuous visual cortical response with the pattern of stable voltage oscillation elicited by repetitive visual stimulus (RVS) with a constant frequency above 6Hz on the central retina [29]. When the subject focuses his attention on the RVS, a SSVEP matching the frequency or harmonics of this RVS is elicited and showed in the subject EEG signals. Generally, SSVEP has two different definitions. On the one hand, Ragan [30] suggests that SSVEP is a direct response in the primary visual cortex. On the other hand, Silberstein et al. [31] proposes that SSVEP is an indirect cortical response to the visual stimulus from the peripheral retina when the human brain performs cognitive task depending on cortico-cortico loops, and it generates a complex amplitude and phase topography at visual cortex simultaneously varying from different individual. In addition, information transfer rate (ITR) depends on the type of brain signals being used. SSVEP-based BCI has a lot of advantages, such as high ITR, high SNR and minimal user training comparing to the other kind of BCIs.

### 2.3.1. Neurophysiological and Electrophysiological Activities in BCIs

There are various signals with their own properties having been studied and researched so far. And it is found that brain signals can be categorized into two different patterns, evoked signals and spontaneous signals, which are based on the cognitive and behavioral mechanism [2] [15] [26].

a) *Evoked Signals*: are automatically generated when the subject is given the specific stimulus without any special training.

There are mainly two types of evoked signals:

- Steady-State Visual Evoked Potentials (SSVEPs): SSVEPs are evoked in the visual cortex, especially occipital part, in response to the visual stimulus. Focusing on one flickering stimuli can elicit the same frequency in the human brain corresponding its associated command [25].

- P300: It relies on continuously flickering stimuli which are normally flashing symbols or letters. Selective attention to a specific flashing symbols or letters elicits a peak of amplitude in time domain, which is defined as P300. And it can be generated at around 300ms in the parietal cortex in response to special frequent stimuli, such as auditory stimulus or somatosensory.

b) *Spontaneous Signals*: are voluntarily generated by the user, without external stimulation, following an internal cognitive process.

There are several different spontaneous signals:

- Slow cortical potentials (SCPs): SCPs occur in the frequency range from 0.1 to 2Hz as slow voltage shifts in the EEG activities. SCPs have the properties that positive SCPs are associated with mental inhibition, while negative SCPs relate to mental preparation [41].

- Cortical neuron action potential: The action potential is increased by the expected movements which are enhanced in the preferred direction of neurons [28].

- Oscillations (ERD/ERS of μ and β rhythms): Oscillations are yielded by the feedback loops in the brain neural networks. Both mu and beta rhythms originate from sensorimotor cortex without involving inputs and outputs. The decrease of oscillatory in μ and β band is called event-related desynchronization (ERD) while the power of rhythms increasing after a voluntary movement is called event-related synchronization (ERS). ERD/ERS patterns produced by motor imagery, which is the imagination of movement without real movement, are very similar to the patterns elicited by real movements [25].

- Movement-related potentials (MRPs): MRPs occur around 1s earlier than a movement's starts; they have the feature of bilateral distribution and best response in the parietal cortex. The closer to the movement, the better response they are.

- Readiness or Bereitschaft potential (RP or BP): This potential is generated by the brain neuron network consisting of movement-related potentials peaking in the motor potential [27].

### 2.3.2. Neurophysiology of the Human Visual System

A SSVEP-based BCI has a close relationship with the human visual system in visual processing. The human visual system consists of the eye and the brain. The eye acts like an external camera to detect and collect information while the brain is the main site to process the information sent by eyes. So it is very important and necessary to understand the detail of both eye and brain.

Human vision is one of the most complex visual systems among all the animals in the world. Eye, the main functional part of the visual system, consists of three main layers including the sclera, the choroid and the retina. Each layer also has its own structure. The cornea belongs to the layer of sclera. The pupil, iris, and lens are included in the layer of choroid. The layer of retina is the thin layer of cells at the back of the eyeball. It includes receptor cells where physical stimuli of light rays is taken in and converted into neural signals sent to brain. When brain receives the electrical and chemical signals from eye, brain translates them to construct physical image.

Based on the structure of eye, external electromagnetic rays with the right range of wavelength (from 300 to 700 nm) first hit the cornea. Then, they pass through the pupil, which controls the amount of light (the black aperture at the front of the eye). Iris is the pigmented area around the pupil and it can automatically adjust the size of the pupil. Two layers of iris muscles control the pupil to contract or dilate, so that it can control the amount of light entering eye. The lens with the cornea adjusts the focal length. There are two kinds of photoreceptors, which are cones and rods. Photoreceptors generate the color and shadow based on the data information received on retina. Then the information is transduced into neural impulses and they are sent to brain by the optic nerve for next processing step, visual signals processing which starts in brain.

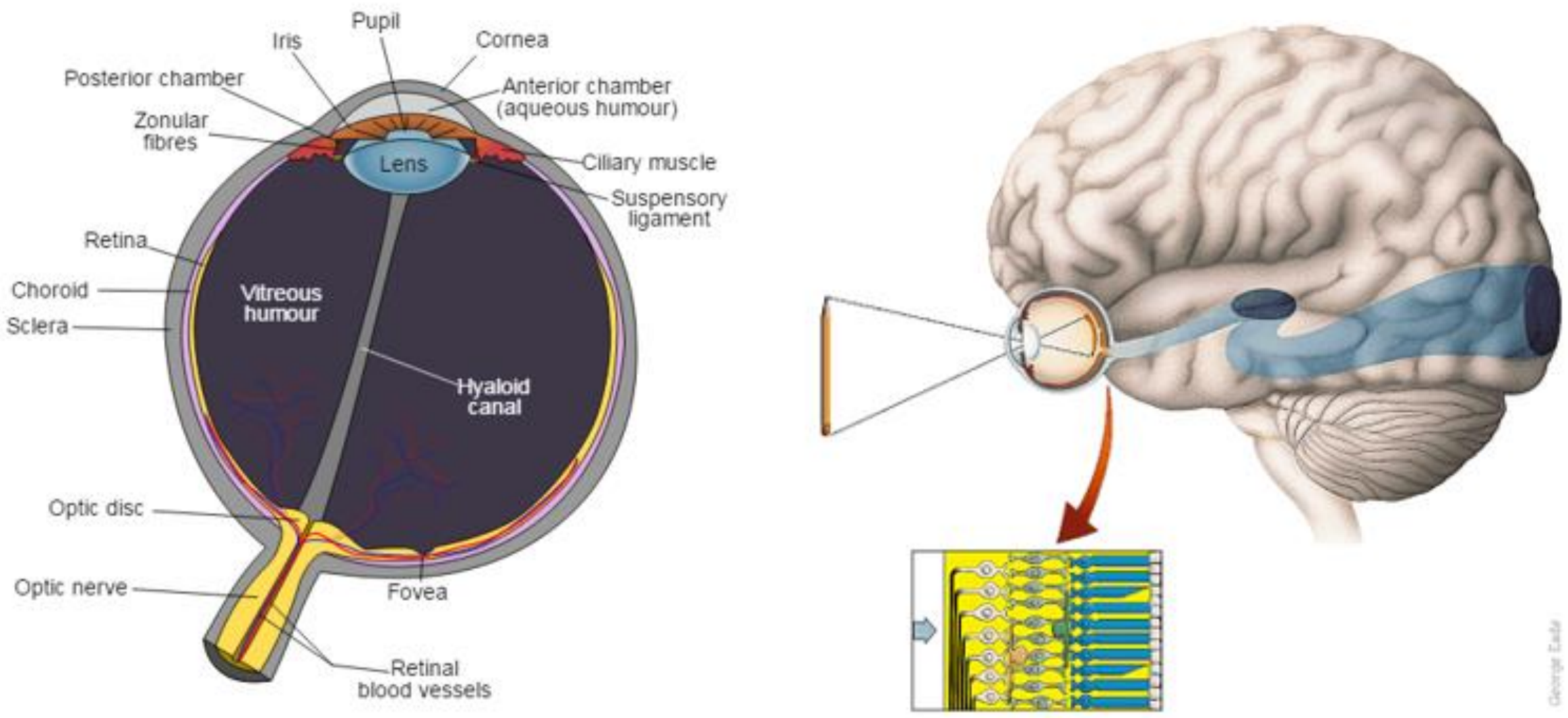

**Figure 6: Anatomy of the Human Eye (Left) and Computational Modeling of the Eye (Right) [48]**

Figure 7 shows the overview of the human visual system. Just like each hemisphere controls the half of opposite body, the visual system has the same mechanism. The left hemisphere of the brain maps to the right visual field, and vice versa. The function of the optic chiasm is to connect each eye to the associated hemisphere. This guarantees that visual cortex can receive the information from both eyes. The lateral geniculate nucleus is the main part used to process the visual information.

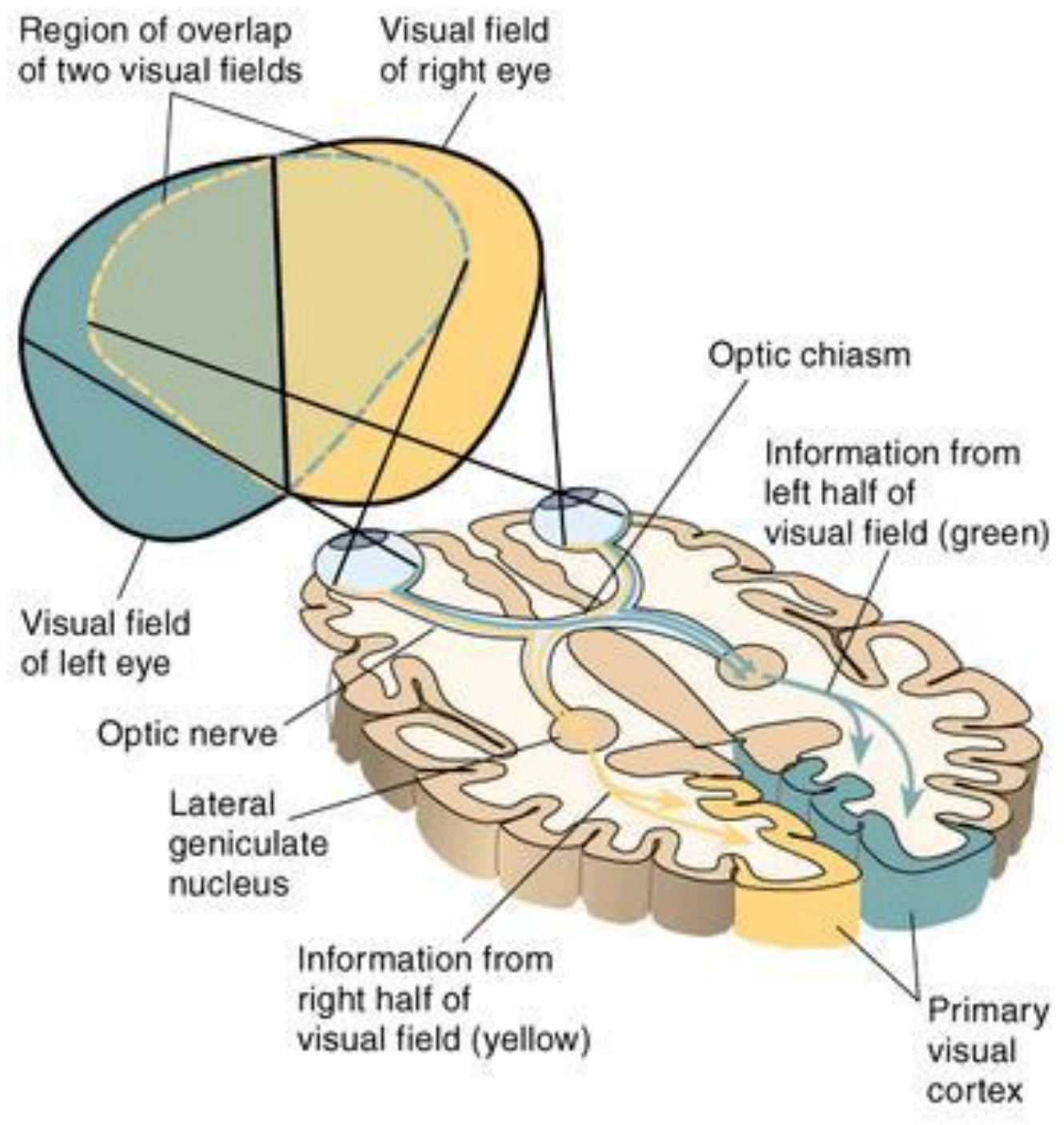

**Figure 7: The Overview of the Human Visual System [34]**

As referred above, there are two types of photoreceptors in the retina named after their shape as “rods” and “cones”. Their function is detecting and converting external electromagnetic rays to electrical signals. The rods are more sensitive and responsive to light. Figure 8 shows the range of light received by human eyes.

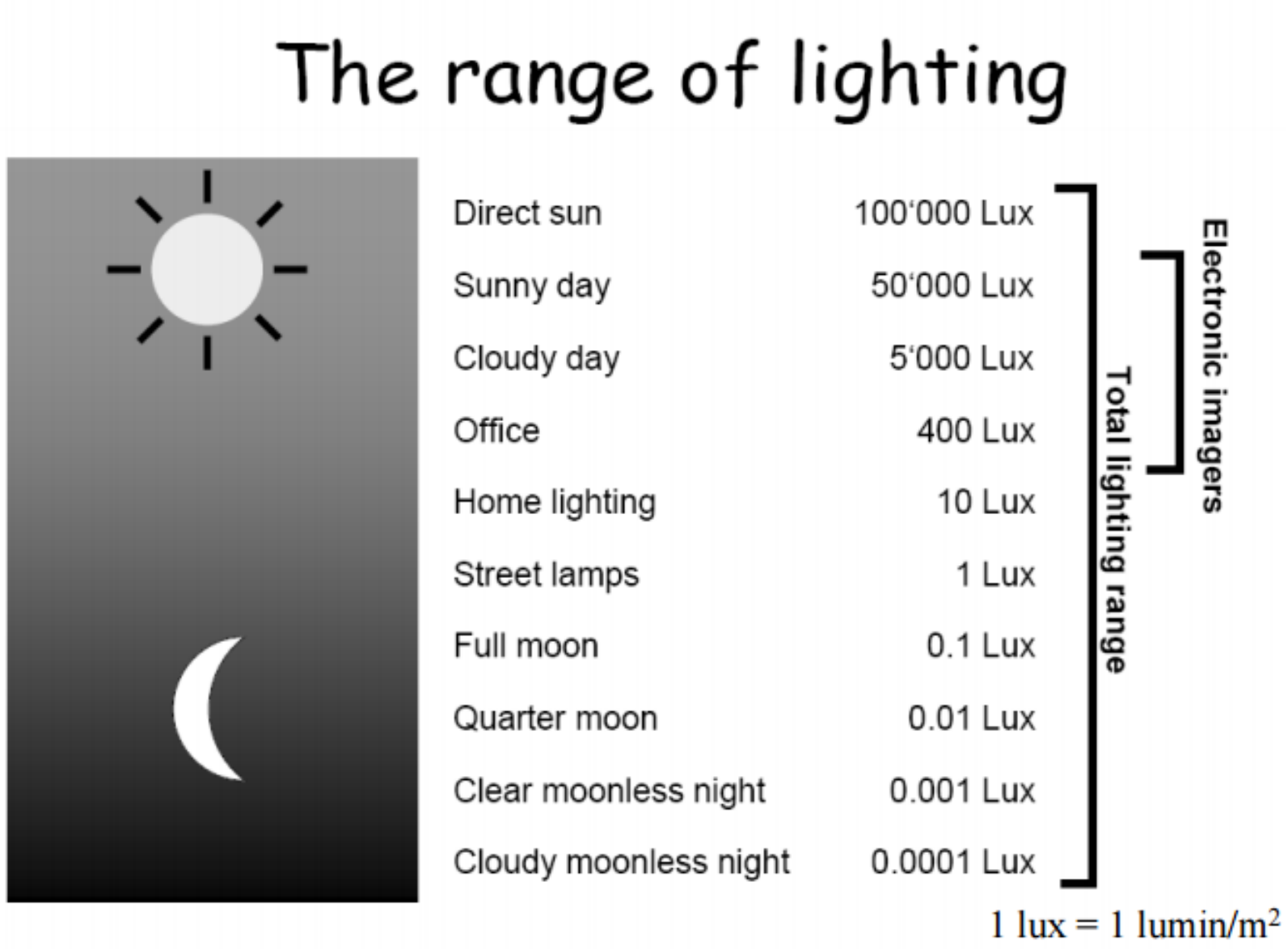

**Figure 8: The Range of Light [50]**

In the dark, rods cannot detect color and cones are inactive. So eyes “lose” the ability to distinguish colors and only shades of grey are perceived, this vision is called “scotopic” or “night vision”. In the daylight, the cones are most active; this vision is assigned as photopic or day vision. In dimly light, both rods and cones are active and it is defined as mesopic vision.

There are also three types of cones based on the sensitivity to different bands of the electromagnetic spectrum. Figure 9 shows the sensitivity of the three types of cones. In the range of 450 to 500 nm (short wave lengths), the blue cones are active in perceiving violet-blue, while the other two kinds of cones are active in higher wavelength of light. The green cones perceive the green in the region of 530 to 570 nm (medium wavelengths). And the red cones response to red in the region of 630 to 670 nm wavelengths (long wavelengths). The combination of three cones can perceive and describe any external color, which is called “trichromacy”.

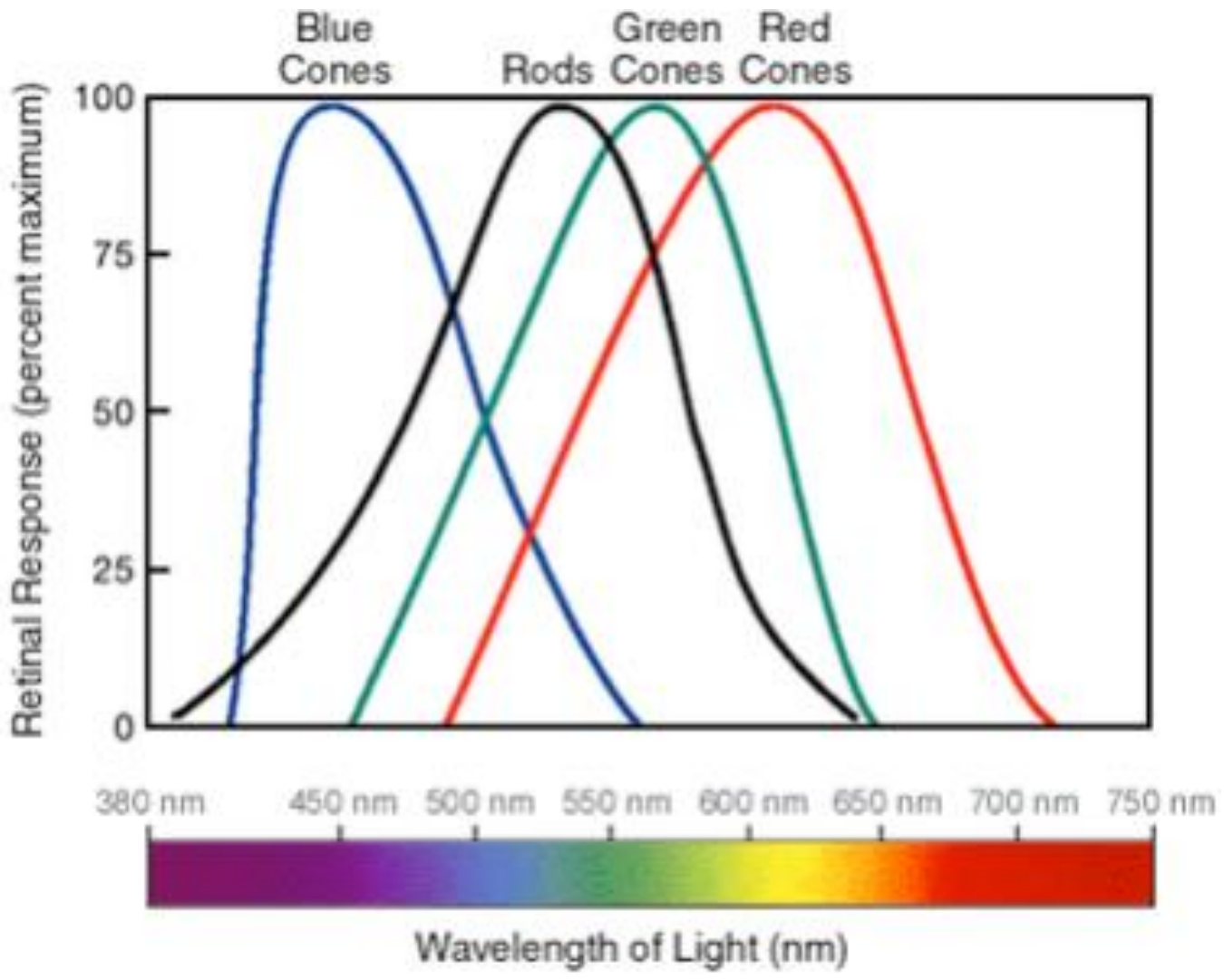

**Figure 9: The Sensitivity of the Three Types of Cones [50]**

### 2.3.3. Stimulators

The stimulator is used to display the RVS with distinctive properties. There are three main stimulators being applied in the SSVEP-based BCI system, which are light-emitting diode (LED), cathode ray tube (CRT) monitor and liquid crystal display (LCD) screen. All the three stimulators have their own features. It is known that improving the accuracy of the BCI is the most important parameter among all of the targets. Therefore, the selection of a suitable stimulator is very crucial in this regard.

In SSVEP studies, the accuracy of the BCI is influenced by a lot of factors, such as bit rate which refers to classification speed, the number of commands etc. But the primary factor is the strength of the SSVEP response in terms of the SNR and the properties of the stimuli [35].

Because of the different properties of RVS, they are categorized into three main groups, namely light stimuli, single graphics stimuli and pattern reversal stimuli [35].

a. *Light Stimuli*: It is proposed to utilize light sources such as LED controlled by dedicated electronic circuit to display sequences or waveforms in different frequencies. The factors of luminance and color are considered in this kind of RVS.

b. *Single Graphics Stimuli*: It is proposed to use computer screen also called LCD or CRT to generate the flickers with specific shape (e.g., circle, triangle, square etc.). The flicker frequency is the integrate number of the screen refresh cycles.

c. *Pattern Reversal Stimuli*: It is proposed to use computer screen to generate the flickers with at least two graphical patterns alternately appear and disappear. The typical pattern reversal stimuli is checkerboard with the color black and white.

In previous research, LEDs are the most widely used method to render light stimuli, while rectangles and checkerboards are the usual options utilized in the single graphic and pattern reversal stimuli [35]. From the viewpoint of most researches, it is shown that the number of researches using computer screens is a little bit more than using LEDs to generate light stimuli. According to Zhenghua Wu et al. [29], the SSVEP response evoked by LED is faster than by computer screen. It is also suggested that the bit rate is higher using LED stimulator than using computer screen based on Regan’s study [30]. However, the best advantage of using computer screen is no hardware involved, which means it fully depends on software so that it can reduce the complexity of the design and it also benefits the designers. Meanwhile, using computer screen makes it possible to fine-tune and adjust frequency and pattern of the stimuli in the different sessions of BCI. However, the disadvantage is the refresh rate of the computer screen (LCD is usually 60Hz), which limits the number of frequencies to be used for stimulation. Furthermore, only the frequencies below half of the refresh rate and integer times of refresh rate can be considered as implementation choices [36]. Otherwise, errors will appear. In most of the situations, the frequency displaying on computer screen is very low and it should be avoid becoming other frequencies’ harmonics. So the accuracy of the response declines due to unpredictable delays and inaccurate stimulations. On the other hand, comparing to LCD or CRT, LED has larger potential to display plenty of different waveforms with different frequencies without any limitation. In addition, one of the ways to improve the number of frequencies in computer screen is to get higher refresh rate screen, such as 120Hz, which will partially ameliorate the problem.

Zhenghua Wu et al. have investigated the spectrum differences of three kinds of flickers and the differences in SSVEPs evoked by three different stimulators, which are LED, LCD and CRT. The result is shown below.

- High - complexity BCI: It is above 20 choices BCI system. As consideration above, if LCD or CRT is applied in this kind of BCI, there is a high chance that some components may have some overlaps with other fundamental frequencies or harmonics. It is very hard to avoid this phenomenon appearing because of the properties of computer screen. Instead of using LCD or CRT, it is better to use LED without the number of frequency limitation.

- Low - complexity BCI: It is referred to 10 choices below BCI system. Prior to use LED and CRT, LCD is the better option as stimulator which can combine itself with the brain signal processor in one computer. Arranging the frequencies used in LCD, the problem of overlap can be avoided. Meanwhile, comparing the LCD with CRT, LCD is more user-friendly which diminishes eye tiredness. It is also widely believed using LCD as stimulator in SSVEP-based BCI will become a promising way in the future.

- Medium - complexity BCI: The range of choices in this kind of BCI is from 10 to 20 choices. LED has hardware part which will contribute the complexity of the BCI comparing to CRT’s convenience. Considering the accuracy of the system and the overlap problem in LCD, CRT is better to employ in terms of a lot higher accuracy.

However, there is no distinct boundary among each kind of BCI. It is more dependent on the practical applications. Meanwhile, the algorithms used in other stage also influence the selection of stimulator significantly.

#### 2.3.4. Stimulus Frequency

Although it still needs a lot of exploration to find the real mechanism of SSVEP, it is certain that SSVEP can be elicited by different frequencies ranging from 1-100Hz [32]. Frequencies are roughly divided into three main regions, considering the maximum amplitude of SSVEP. Regan [30] has obtained that low frequency region is from 5 to 12Hz, medium frequency region is from 12 to 25Hz and high frequency region is from 25 to 50Hz [30].

WANG Yijun et al. [33] have done some research on the classification of different stimulation frequencies. But they have gotten the different result from Regan. Figure 10 is the result from WANG Yijun et al.

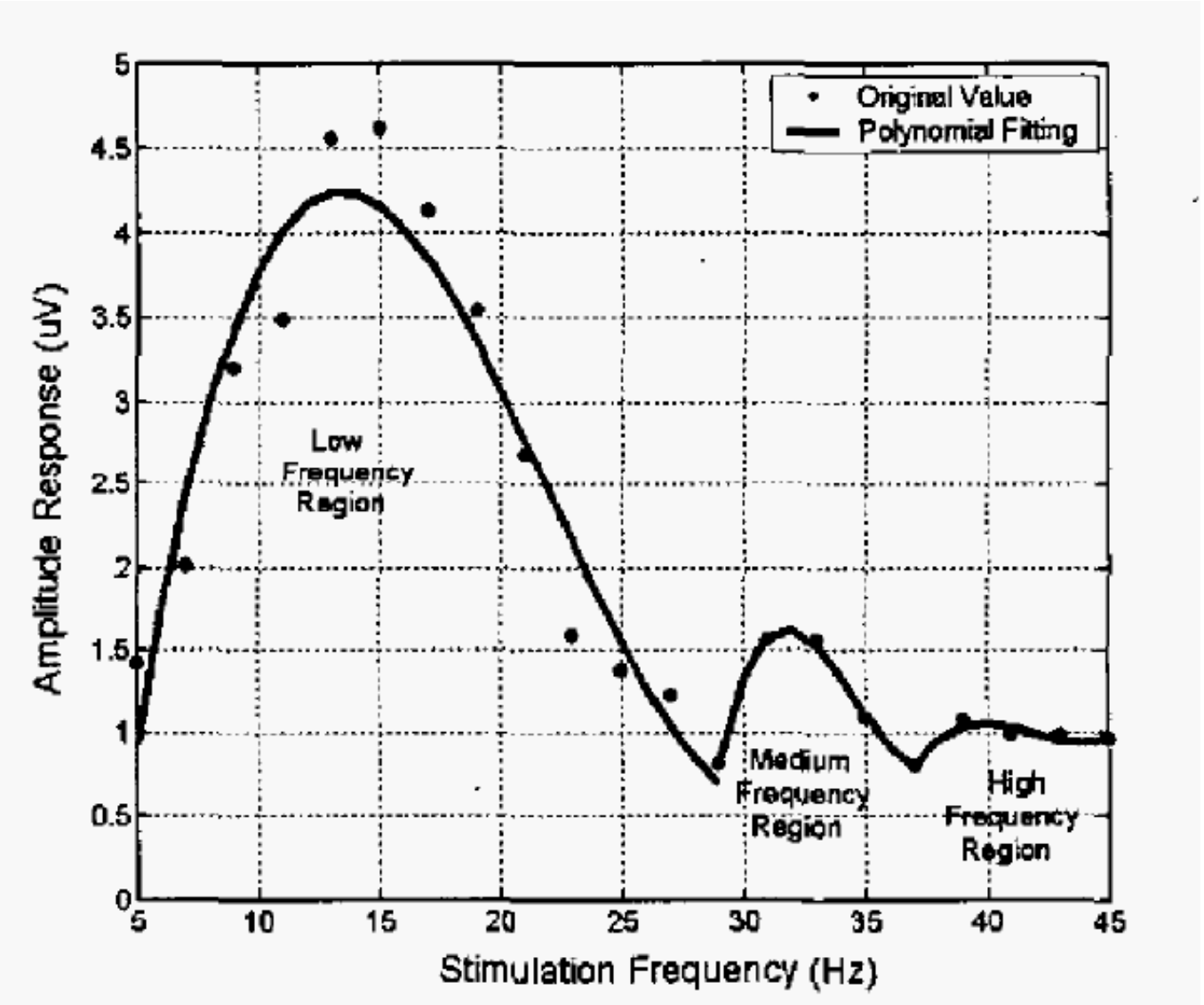

**Figure 10: The Amplitude Response of Different Frequency Regions [33]**

The figure reveals the result of the amplitude response for different frequency regions. It shows that the low stimulation frequency region gives the highest amplitude response followed by medium frequency region and high frequency region. In addition, there are several factors which should be considered, such as color, luminance and electrode position. Currently, low frequency is the most popular range used in the SSVEP-based BCI, especially alpha band, due to its high ITR. However, it has some limitations such as visual fatigue, interference from alpha band, and some possibility of leading to photosensitive epileptic seizure. WANG Yijun et al. believe that the stimulations with the frequencies above 20Hz can overcome all of these problems. In their work, a SNR curve is drawn to evaluate whether high frequency SSVEP can be used in BCI systems. Figure 11 shows the relationship between SNR and stimulation frequency.

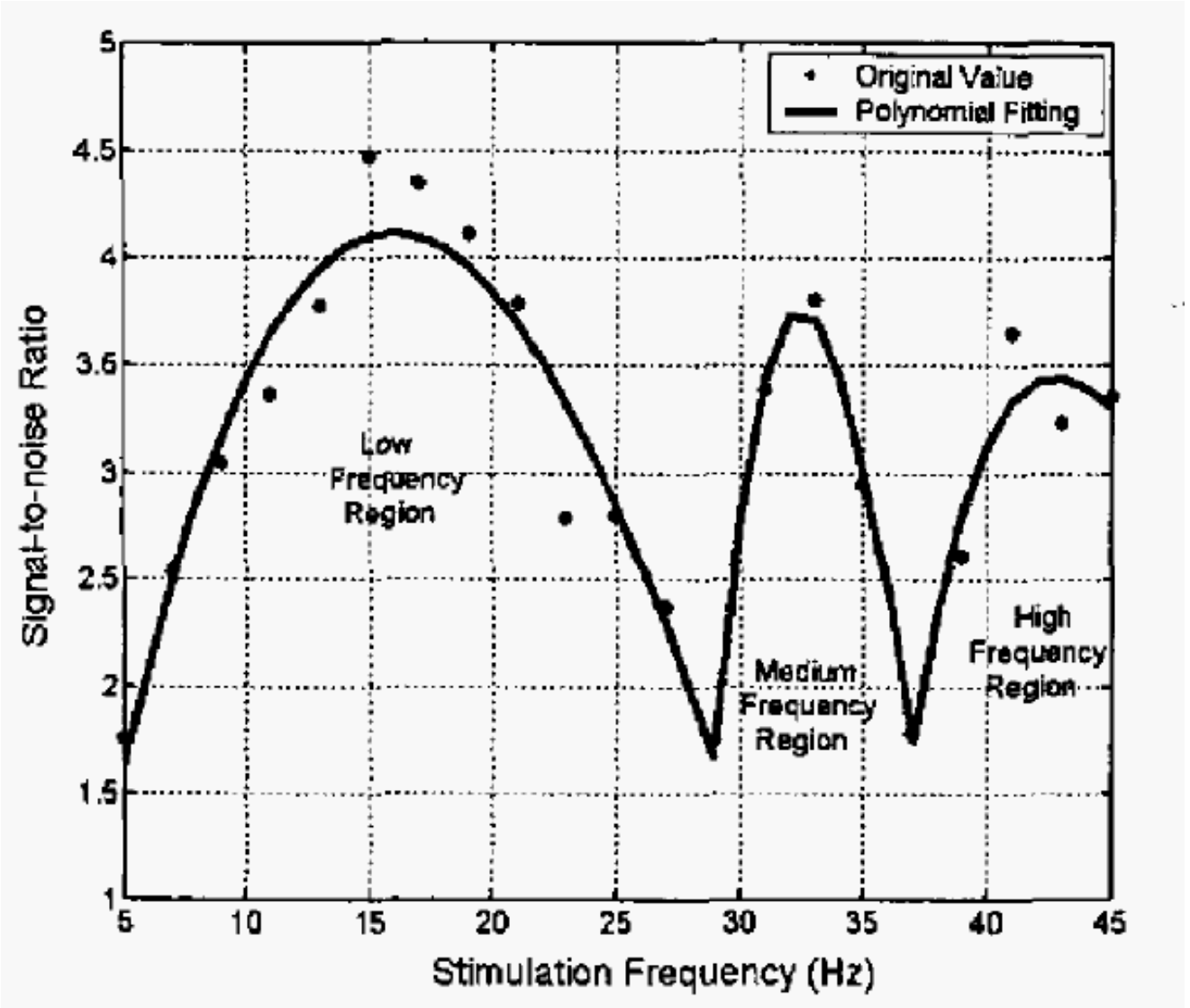

**Figure 11: The Relationship between SNR and Stimulation Frequency [33]**

The result suggests that high frequency SSVEP has lower background noise than the low frequency SSVEP. WANG Yijun et al. used FFT with 1024 points and the frequency resolution 0.25Hz to analyze the EEG data offline. Stimulation frequency should correspond to the maximum amplitude spectra. The detection accuracy was above 95% for all stimulation frequencies, which means high frequency has a large potential to be used in the future research neglecting the low amplitude response.

Ivan Volosyak et al. [37] have done a demographic analysis of stimulation frequency. It shows that most of people even without any previous BCI experience can use the BCI in a very noisy surrounding. Considering all of the factors such as tiredness, alcohol and caffeine consumption, the performance of the subjects is not influenced significantly. Only few subjects indicate fatigue. Meanwhile, the result also shows the medium frequency SSVEP has accuracy of 92.26 ± 7.8% with the mean ITR of 17.24 ± 6.99 bit/minute for 84 out of 86 subjects. Whereas high frequency SSVEP has accuracy of 89.16 ± 9.29% with the mean ITR of 12.10 ± 7.31 bit/minute for 56 out of 86 subjects.

### 2.3.5. Stimulus Color

Stimulus color is also an important and confusing factor in SSVEP. The reason of "importance" is that the mechanism of the human visual system is very complex, and the cones are able to sense color in red, green and blue. So there should be some differences in different color SSVEP. Thus, there are many researchers exploring the best and most user-friendly color in SSVEP. Meanwhile, the reason of confusion is that there are various conclusions being drawn by different research group, which means it still needs a long way to get the unanimous result.

Surej Mouli et al. [38] have done an investigation on the effect of different colors and different types of LEDs. There are two kinds of LEDs, which are clear and frosted. The target of their research is to find the best color in the stimulation and then compare the different types of LED in the chosen color. Red, green and blue are used as three colors and 7Hz, 8Hz, 9Hz and 10Hz are used as four frequencies without phase consideration. Statistical algorithm, t-test, is implemented in the signal processing. The results show that 7Hz green clear LED gives the highest response and all the subjects agreed that the frost LED feel more comfortabl.

On the contrary, the other research groups working on the stimulus color have published different result. Teng Cao et al. [39] use five different colors, which are white, green, red, gray and black, encoded with different phases and frequencies as stimuli. They implement canonical correlation analysis (CCA), which is a multivariable statistical method, in the feature extraction for multichannel analysis instead of the one-channel based FFT. CCA coefficient means the correlation with the stimuli signals. Higher the coefficient means better correlation. The result shows that white color has the highest CCA coefficient and it is also the best one encoded with different phases. Highest accuracy and ITR also come to white color. Overall, it means the white color is the best choice to elicit SSVEP based on their research.

The problem in stimulus color still cannot reach a consensus because the performance of stimuli colors is based on environment and individuals. Interestingly, one of the suggestions put forward by [35] is that changing the stimuli color dynamically depends on different situations could resolve many issues. In conclusion, more research is required to find the most suitable color.

### 2.3.6. Stimulus Waveform and Harmonics

Waveform is another factor in stimuli, which also affects the performance of the SSVEP. In most researches, a periodic square wave with 50% duty cycle is used.

Fei Teng et al. [40] have done a test to compare the successful rate of different waveforms in SSVEP performance. The waveforms are square wave with different duty cycles, triangular wave and sinusoidal wave. Meanwhile, they also analyze the harmonics and propose that harmonics can be utilized in SSVEP-based BCI system. The result shows that square wave with 50% duty cycle has the highest accuracy and best response at the fundamental frequency. And it is also found that harmonics are generated by both fundamental frequency and artifacts. The first harmonic is mainly generated by fundamental frequency while the second one is mainly produced by the artifacts.

It is found that only few research group works on this area, therefore more reseach is required to for more reliable results.

## 2.4. Signal Processing

### 2.4.1. Signal Preprocessing

The EEG signals through electrodes are recorded from the scalp, but the signals have a lot of noise and very low amplitudes. Therefore, there is a need to reduce the noise and enhance the signal before signal processing stage. There are three steps in signal preprocessing, which are referencing, temporal filtering and signal enhancement.

First of all, referencing means the EEG signals measured by different electrodes should be compared to another reference electrode which has constant voltage. Most of the time, the brain activity voltage at the selected referencing electrode is almost zero [26]. After solving the reference issue, internal and external noises mixing in the EEG signal must be considered. Therefore, a filter is needed to remove these noises. There are several filters always being used in the temporal filtering step, such as Infinite Impulse Response (IIR) filter and Finite

Impulse Response (FIR). FIR filter has very good quality in the frequency domain while IIR filter is not as good. However, IIR filter is a recursive which means output value are included in the filter processing [42]. Therefore, the number of coefficients is lower in IIR in comparison to FIR. Low pass, high pass and band pass can be employed in IIR filter. After temporal filtering, the EEG signals become clear from noise, but their amplitudes are still very low because the scalp weakens the EEG signals. Signal enhancement step can improve the amplitudes of the signals. Among the numerous enhancement methods, some methods are more commonly used in BCI system, such as Common average referencing (CAR), Surface Laplacian (SL), principal component analysis (PCA), independent component analysis (ICA), common spatial patterns (CSP), common spatial subspace decomposition (CSSD), frequency normalization (Freq-Norm) and so on [28]. Every method has different performance, but all of them can be used to enhance the EEG signal depending on the associated BCI application.

### 2.4.2. Feature Extraction

There are numerous values, which are also called features, describing the properties of EEG signals [9]. In order to get the best features, extracting the features from all of these values is necessary. And one or several extracted features can be combined into a vector called feature vector. Generally, feature vector before being concatenated into a single feature vector has high dimensionality; because several features are generally extracted from several channels and from different time segments [44]. It is also known that feature extraction is a core step in the design of BCI system. So it is important to reduce the dimensionality and select the best feature vector to avoid wrong classification result in the following stages of BCI system.

Various feature extraction algorithms are available in BCI research. Some methods focus on temporal variations of the EEG signal, some focus on frequency information and some even focus on both criteria. The properties of temporal features include signal amplitude, autoregressive parameters and Hjorth parameters. On the other hand, the frequency information can be shown in the methods of Band power and Power spectral density. In order to deal with the problem in both time and frequency domains, it is better to use Fast Fourier Transform (FFT) or Wavelets Transform (WT). The function of FFT and WT are very similar. The main difference between FFT and WT is that FFT computes the FFT result in a fixed size sliding window while WT is in a flexibly dynamic window size. In addition, the frequency resolution of WT is also dynamically changed at the same time with the window size. However, FFT is more utilized because of its lower computational tasks. Indeed, it still depends on the practical BCI to select the feature extraction method.

### 2.4.3. Classification

Another crucial step is classification. The algorithms of classification define classifiers to classify the feature vectors into different classes which correspond to different commands. However, the classification also needs a training step to train the classifiers so that they can learn how to classify the different feature vectors. After training step, the classifiers automatically match the different feature vectors to the specific classes in the real test.

The classification algorithms are categorized into several main groups based on the classifiers, namely linear classifiers, artificial neural networks, nonlinear bayesian classifiers, nearest neighbor classifiers and combinations of classifier [44]. Considering the most used classifiers, two of the classifiers are introduced here.

Among all of these classifiers, linear classifiers are used in most of the application, which employ linear algorithm to classify the feature vectors. The simplicity of linear algorithm reduces the rate of overlap. The standard classifiers are Linear Discriminant Analysis (LDA) and Support Vector Machines (SVM). Especially, LDA is very suitable for online BCI because of its low computational requirement and simplicity. But its main shortcoming is the low accuracy, which leads to bad classification result. SVM and LDA are fundamentally similar, but SVM outperform LDA in many ways. SVM can provide good generalization, which is not sensitive to overtraining and curse-of-dimensionality; as result SVM has higher accuracy and achievement.

Neural networks are also common in BCI systems. Generally, MultiLayer Perceptron (MLP) is one of the algorithms mostly used. MLP architecture consists of an input layer, number of hidden layers and an output layer in the neuron. The output decides the class of the input feature vectors. MLP can adapt to large variety of problems. However, MLP may provide low accuracy because of its sensitivity to overtraining. Rajesh Singla et al. have done an investigation to compare the accuracy of SVM with MLP, which proves that SVM outperforms MLP [45].

## 2.5. BCI Application

Once the classification is done, the BCI system will translate the different classes to different commands according to the classification result. BCIs are applied in so many different applications, so the output device varies, such as wheelchair, speller etc. Most of the applications are designed for disabled people. Thus, feedback is needed for the patients using BCI. Furthermore, the purpose of feedback is to make sure that the subject knows whether the command is successfully sent or not. It also reduces the training time if a good feedback is provided. In addition, the feedbacks have different manifestations, some provide the visual feedbacks, and some provide the auditory feedbacks or haptic feedbacks. However, the feedback design does not exist in most of the projects currently. The most probable reason is that it is difficult to select and design a proper feedback.

### 2.5.1. A Self-Paced and Calibration-Less SSVEP-Based Brain-Computer Interface Speller

A self-paced and calibration-less SSVEP-based BCI speller has been proposed [46]. It fuses the visual stimuli and the application in one single and homogeneous way. The system consists of two main parts, which are graphical user interface (GUI) and signal processing part. The visual stimuli are displayed on the LCD screen. Figure 12shows the speller GUI, and the example of selecting the letters “A”, “B” or “C”.

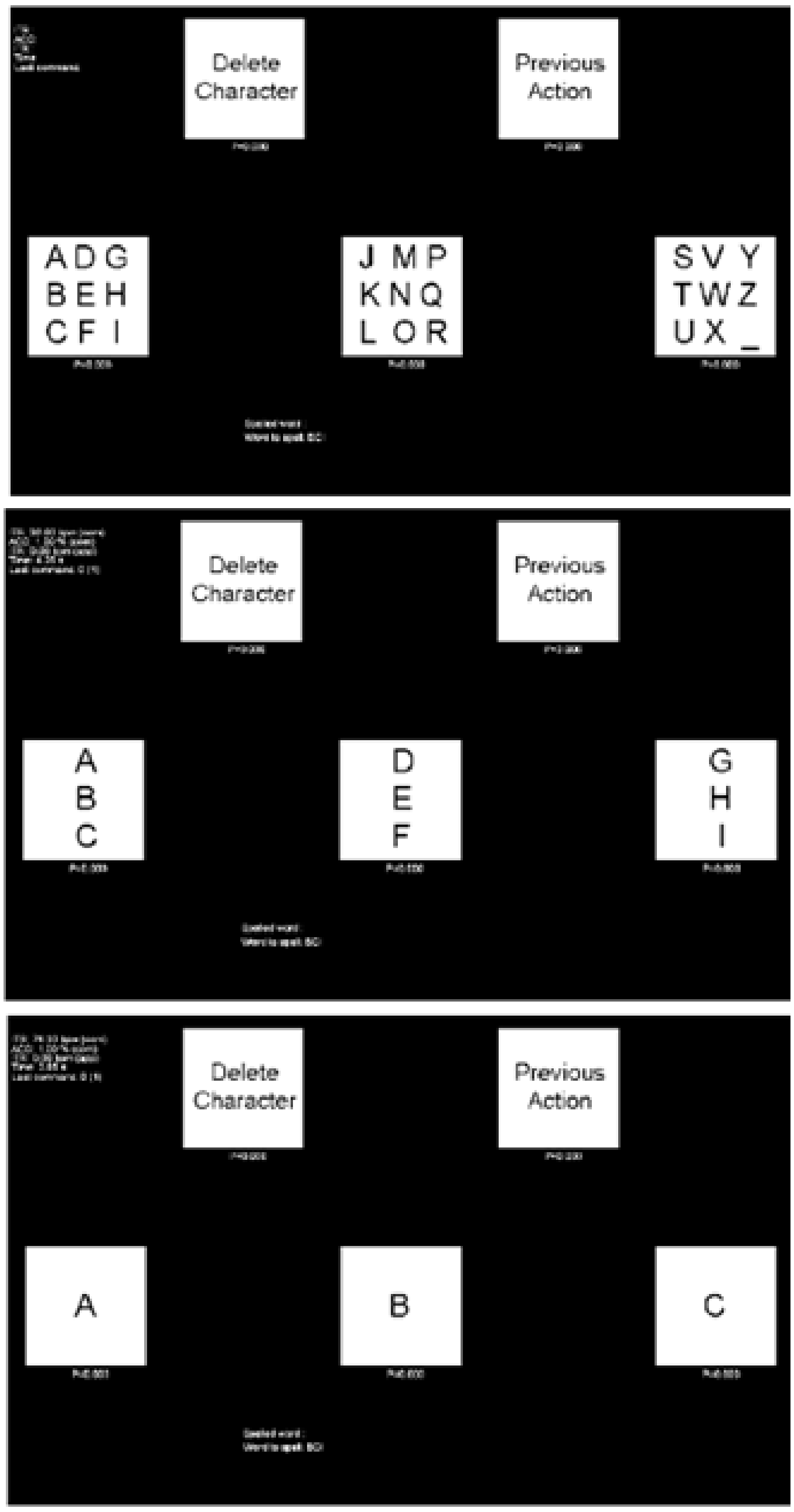

**Figure 12: the speller GUI and the example of selecting the letters “A”, “B” or “C” [46].**

This BCI allows writing 27 characters including the 26 Latin characters from A to Z and "_"for separating the words. The selection of a letter is based on five main BCI commands. Three commands, $COM_1$, $COM_2$ and $COM_3$, correspond to the three boxes containing all the letters. $COM_4$ is defined for canceling the previous action or going back one step, while $COM_5$ is used for deleting the last written characters. The five boxes have different flickering frequencies based on the number of frames (one frame means one image displayed on the screen). After each command, all the five boxes stop flickering for 1 second. Not only can it give the subject visual feedback, but also give some time for subject to relax which makes it more user-friendly. For commands $COM_1$, $COM_2$ and $COM_3$, the animations are different. Once the command sent, the other two boxes disappear and the target box split in three new boxes. The advantage of this is to smooth the gaze of user. In addition, the tests show that it the mean speed of selecting letters is 5.51 letters per minute. Meanwhile, the mean ITR is 37.62 bpm with the mean accuracy of 92.25%.

#### 2.5.2. Control of an Electrical Prosthesis with an SSVEP-Based BCI

A two-axis electrical prosthetic hand using SSVEP-based BCI is proposed in [47]. It has four classes and it is self-paced controlled. The visual stimulation is composed of four red LED bars with different flickering frequencies which can be controlled and changed independently by microcontroller. The prosthetic hand is capable of grasping and wrist rotation. Figure 13 shows the structure of the prosthetic hand.

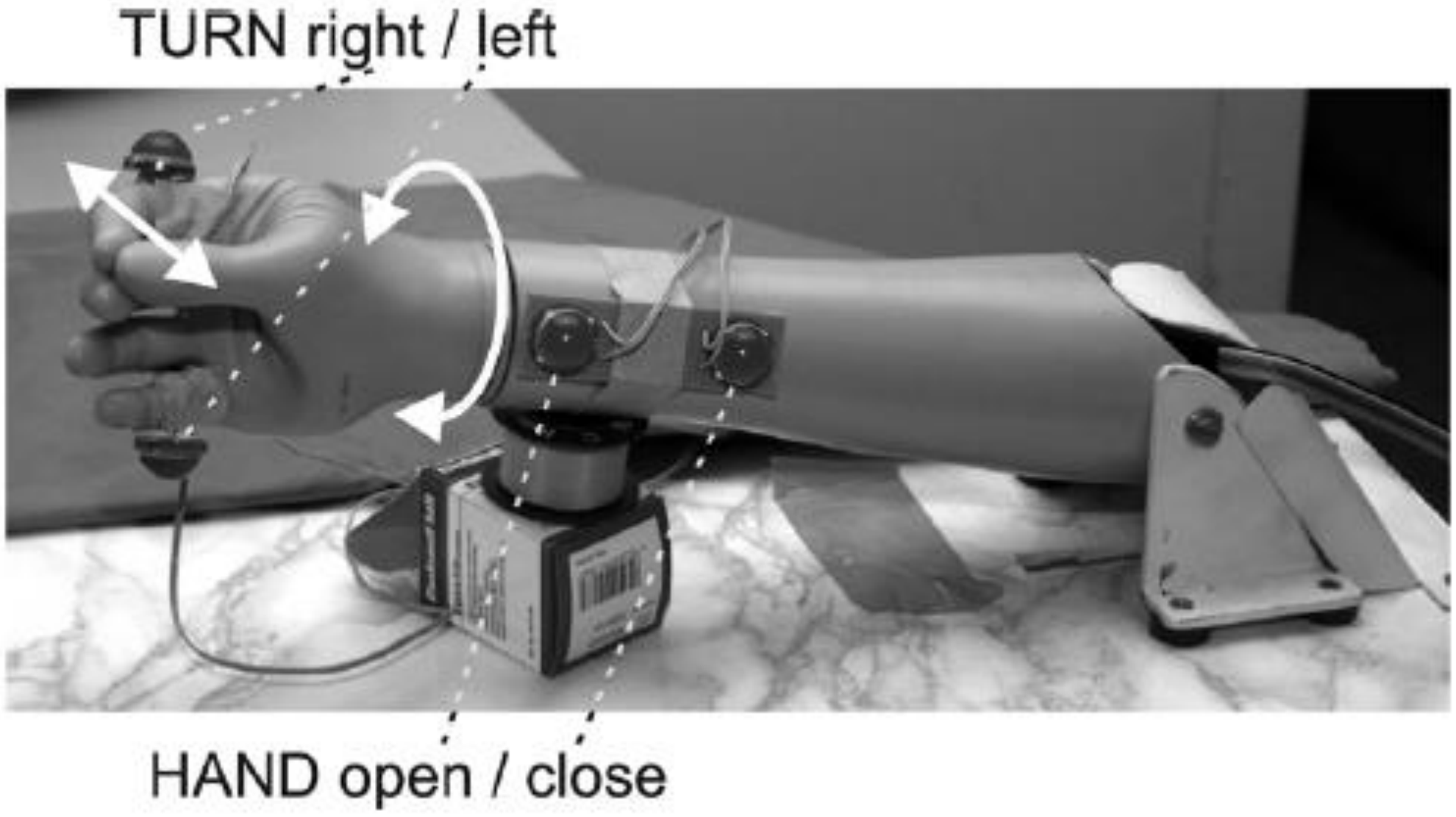

**Figure 13: The Structure of the Prosthetic Hand Device [47]**

The EEG recording equipment uses four electrodes to place 2.5cm anterior and posterior to O1 and O2. At the signal processing stage, the feature extraction uses discrete Fourier transform (DFT) and an algorithm named harmonic sum decision (HSD) is applied for the classification part. Additionally, HSD means that the total number of harmonics at each stimulation frequency varies depending on their frequency range in the synchronous mode. However, one second as dwell time is needed if one class was detected in asynchronous control and after one second, no further decision is considered from the BCI for the next eight seconds. Therefore, the system can prevent the new selections while the device is performing the movement. One of the configurations for LED frequencies used is shown in Table 2.

**Table 2: Each Flickering Light Corresponds to One Movement [47]**

| LED Frequency (Hz) | Position | Function |
|---|---|---|
| 6 | lateral at fifth finger | turn left (TL) |
| 7 | lateral at index finger | turn right (TR) |
| 8 | wrist distal | hand open (HO) |
| 13 | Wrist Proximal | hand close (HC) |

The result shows that the classification accuracy is from 74% to 88% for most of the subjects. During hand prosthetic control test, it was demonstrated that subjects can control it very well based on the ratio of the false actions and right actions.

## 2.6. Summary

This chapter is aimed at reviewing the current theories on the SSVEP-based BCI, including five sections.

First of all, a brief introduction to the BCI structure is presented. Six stage of BCI, signal acquisition, signal preprocessing, feature extraction, classification, translation into a command and feedback, are mentioned. Meanwhile, three different types of BCI are also referred and compared, which are the dependent BCI versus independent BCI, invasive BCI versus non-invasive BCI as well as synchronous BCI versus asynchronous BCI.

In the second section, the items relating to the signal acquisition are covered. The neural principles are discussed briefly from macroscopic and microscopic point of view. They show the relationship between the functions of human brain and neurons. Besides, different types of neuroimaging techniques are analyzed. Moreover, various methods of using EEG are discussed and compared; also advantages and disadvantages are mentioned. A table of comparison, covering most of the methods, is placed in this chapter in for convenient and effective data extraction. Several different rhythms in EEG signal corresponding to different mental activities are explained. Besides, as different electrodes map to different cortexes, the principle of electrodes placement is introduced.

In addition, SSVEP is mainly introduced in the third section. Several neurophysiological and electrophysiological activities are discussed, including evoked signals and spontaneous signals. Furthermore, a human visual system is illustrated in detail, which consists of eye and brain. It describes briefly how human visual system works and the relationship between eye and brain. Eyes act like an external camera to detect and collect information while brain is the main site to process the information from eyes. Several factors influencing the quality of SSVEP are referred. Stimulators, stimulus frequency, stimulus color, stimulus waveform and harmonics are presented and explained one by one. Many different research groups work on these different factors. Among all of the factors, it is found that LEDs are widely used and have a large potential to display more frequencies. Besides that, the frequency in the low frequency region has the highest amplitude response. Meanwhile, it is also proved that medium frequency region and high frequency region can be used in the future due to its ability to overcome the problem with the current BCI systems. For the stimulus color, the investigation results vary because BCIs are individual and environment dependent. There is lack of research on how the waveforms affect the SSVEP performance.

The fourth section is used to discuss about signal processing, which refers to signal preprocessing, feature extraction and classification. Various algorithms are included in signal processing section. IIR and FIR filters are the most popular methods used in signal preprocessing step. Considering that IIR filter has the lower number of coefficients, so it is simpler to use comparing to FIR filter. In the step of feature extraction, FFT and WT are compared and it is known that WT can analyze the EEG signal dynamically in the time domain while FFT is easier to use because of its lower computational level. Besides, two kinds of classifiers are mainly explained in the signal classification stage. One is linear classifier, and another one is neural networks classifier. It is shown that SVM is a better way in comparison with LDA and MLP, because it can provide good generalization and high accuracy.

In the last section, two types of BCI applications are introduced, which are a Self-Paced and Calibration-Less SSVEP-Based Brain-Computer Interface Speller and Control of an Electrical Prosthesis with an SSVEP-Based BCI. Both of these two projects show that subjects can control the output devices very well and the BCI systems’ performance is promising for future work. So it implies that SSVEP-based BCI can be applied in real life.

# 3. Methodology

## 3.1. SSVEP-Based BCI System

There are six stages in this SSVEP-based BCI control system. As mentioned before, these stages are signal acquisition, signal preprocessing, feature extraction, classification, translation into a command, and feedback. An overview of the project is shown in Figure 14.

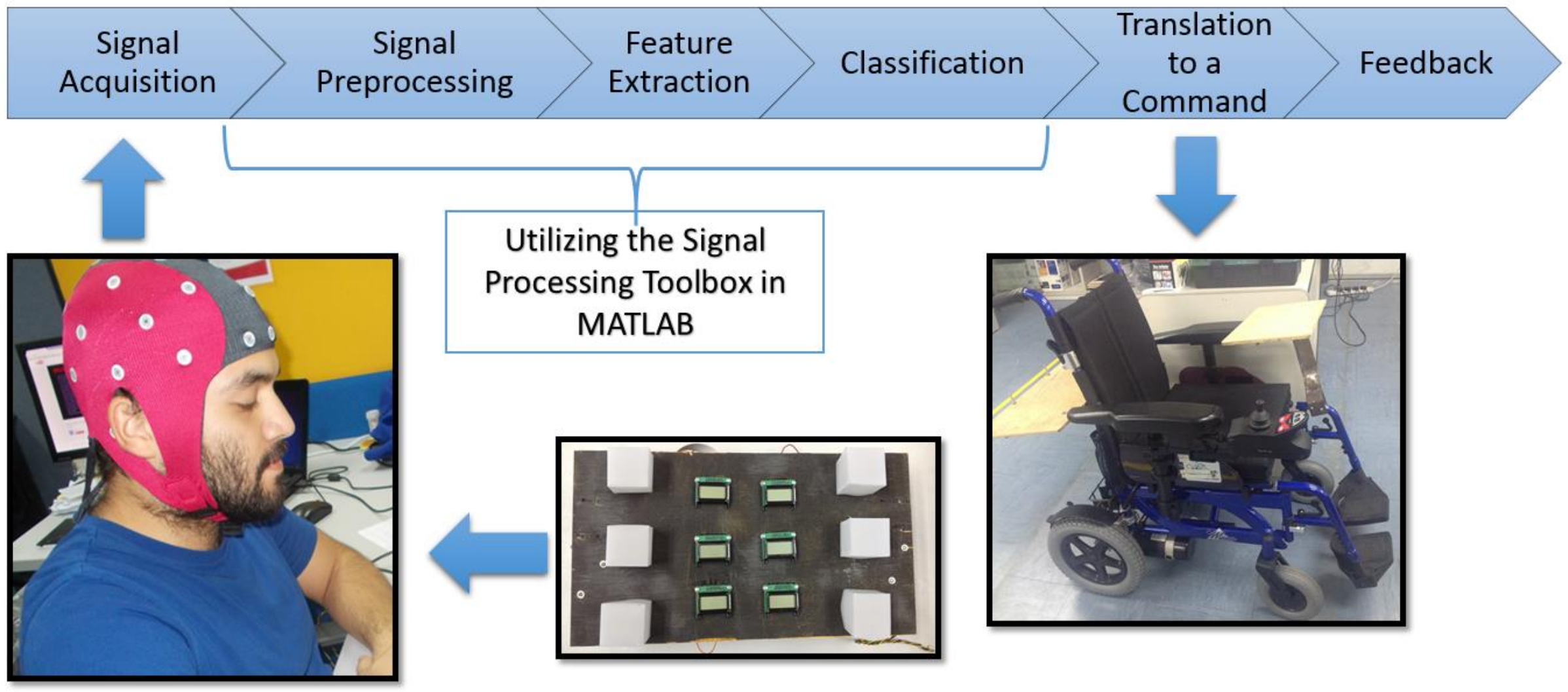

**Figure 14: Overview of the Project**

In order to implement each stage, different sets of software and hardware are involved. SSVEP RVS are displayed by the LED panel. The acquisition of EEG signals is performed by using a 32-channel ASA-LAB$^{TM}$ EEG recording system. MATLAB is employed in the stage of signal processing. The output device is a smart wheelchair. The feedback is given by the text information displayed on LCDs on the SSVEP visual stimuli panel.

## 3.2. EEG Recording System

The EEG signal is recorded by Waveguard$^{TM}$ electrode cap with Ag / AgCl sintered electrodes and shielded co-axial cables that keep external noise away. The EEG recording system is equipped with a 32-channels amplifier to amplify the EEG signal. In this project, 3 channels related to visual cortex are used which are $O_1$, $O_Z$ and $O_2$ with the recording sample rate of 256 Hz. The reference electrode is the default ground between the position of Fpz and Fz. Figure 15, shows the Waveguard$^{TM}$ electrode cap layout and the electrodes inside the red ellipse are used in this project. Figure 16 illustrates the potential difference between active and reference electrodes, these signals are raw EEG signal acquired in real time with ASA-LAB.

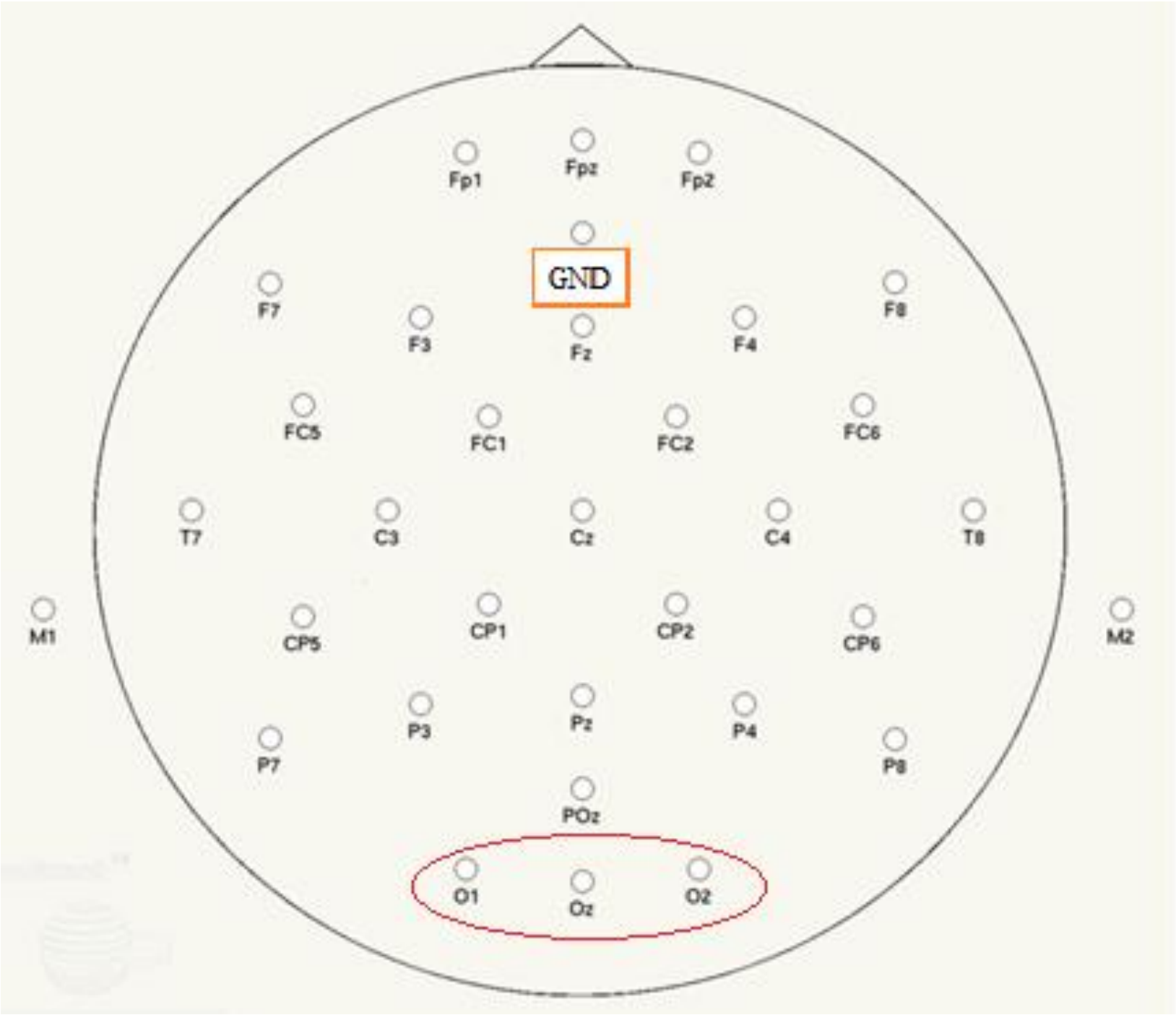

**Figure 15: Waveguard$^{TM}$ Electrode Cap Layout**

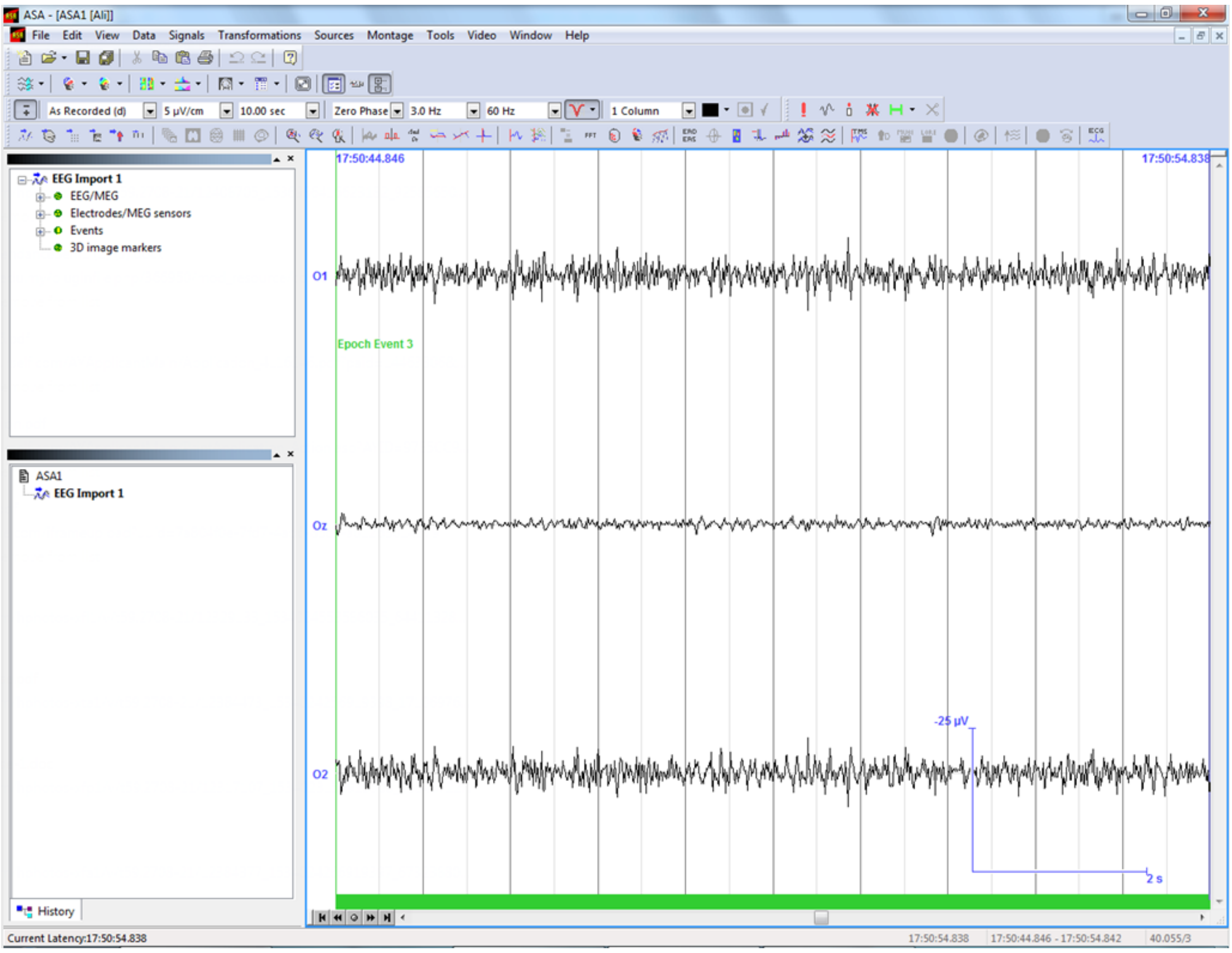

**Figure 16: EEG Signals in Real Time**

## 3.3. Algorithms for Signal Processing

Signal Processing Toolbox in MATLAB is used for signal processing. Raw EEG signals should go through several steps in the stage of signal processing so that the useful information embedded inside the raw signals can be utilized for next stage. Different algorithms are employed in the steps of signal preprocessing, feature extraction and classification.

### 3.3.1. IIR Bandpass Filter Implemented for Signal Preprocessing

Butterworth Infinite-Impulse Response (IIR) bandpass filter is applied for signal preprocessing. IIR filters are digital filter with infinite impulse response. In this project, the main function of the IIR bandpass filter is to eliminate the useless components including DC level and extra unused frequencies. The system function of IIR is

$$H(z) = \frac{\sum_{k=0}^{M} b_k z^{-k}}{1 - \sum_{k=1}^{N} a_k z^{-k}}$$

The difference equation is

$$y[n] - \sum_{k=1}^{N} a_k y[n-k] = \sum_{k=0}^{M} b_k x[n-k]$$

It has the feedback which is known as a recursive part of the filter. So the output is weighted sum of current and previous M inputs and previous N outputs. This feature guarantees the IIR filters have better frequency response than FIR filter with the same order. However, the phase characteristic of IIR filter is non-liner. Therefore, it causes the needs of linearity for the system.

Discrete-time IIR filters are designed using continuous-time prototype filters, including Butterworth filter, Chebyshev filter and so on. The advantage of Butterworth filter is that it has shaper cutoff comparing to other types of filters so that it makes the filter has a frequency response as flat as possible in the passband. The function *[b,a] = butter(n,Wn,ftype)* is for designing an Nth order Butterworth filter and it returns the filter coefficients in length N+1 vectors *b*(numerator) and *a*(denominator). Additionally, the coefficients are listed in descending powers of z. The cutoff frequency, *Wn*, must be in the range of 0.0 to 1.0 corresponding to half of the sample rate. In this work, a bandpass Butterworth filter is required which means the cutoff frequency is a two-element vector (Wn=[w1 w2]). *butter* returns an order 2N bandpass filter with the low cutoff frequency w1 and high cutoff frequency w2.

After establishing the parameters of Butterworth IIR bandpass filter, the returned coefficients are given to the function *y = filtfilt(b,a,x)*. *filtfilt* creates a non-causal, zero-phase filtering approach which can eliminate the nonlinear phase distortion of an IIR filter so that the problem of the non-linear phase existing after Butterworth IIR bandpass filter can be solved. *filtfilt* filters the data in vector x with the filter described by vectors *a* and *b*, to create the filtered data *y*. Variable *y* contains the useful data for feature extraction.

A 6th-order Butterworth IIR bandpass filter is used in the project, with the low and high cutoff frequency of 3Hz and 50Hz respectively. The implementation of filter on Oz is shown below.

```
Fp1 = 3;                    % Bandpass filter passband 1
Fp2 = 50;                   % Bandpass filter passband 2
order = 3;                  % Bandpass filter order, n = 2^order
% Creating a butterworth bandpass filter (recursive - IIR filter)
[b,a] = butter(order,[Fp1 Fp2]/(Fs/2),'bandpass');
filt_eeg_Oz = filtfilt(b,a,mEEG(:,31));
```

### 3.3.2. FFT Implemented for Feature Extraction

In order to get the discriminative information in the input signal, using a high-efficient way to extract the huge input data is important. When the subject is looking at one of the flickering LEDs with the specific frequency, the SSVEP response in his/her brain has the same exact fundamental frequency as the flickering frequency. Meanwhile, the harmonics, which are integer multiple of fundamental frequency, are also elicited from the SSVEP response. Therefore, the method of feature extraction selected in this project should determine the power spectrum in frequency domain. Considering FFT can achieve high computation with short time delay, so it is selected to apply in the step of feature extraction. Additionally, the main delay in this project is due to the window size which is the time waiting for the new EEG coming inside the window. So comparing to the WT, fixed window size results in the fixed delay time and this is beneficial for the subjects to adapt themselves to the time delay; especially it is more suitable for the subjects to control the BCI wheelchair.

Discrete Fourier Transform (DFT) is defined as follows:

$$X(k) = \sum_{n=1}^{N} x(n) e^{\frac{-j2\pi kn}{N}} \text{, } where\ k = 1,2,\ldots,N$$

FFT computes the DFT of a sequence. The transform function of FFT is defined as function shown:

$$Y(k) = \sum_{j=1}^{n} X(j)\, W_n^{(j-1)(k-1)}, where\ W_n = e^{(-2\pi i)/n}\ is\ one\ of\ n\ roots\ of\ unity$$

FFT can manage to reduce the complexity of computing the DFT from $n^2$ to $n\log n$ so that the computational speed can be highly improved. In Matlab, the function *fft(x,n)* is utilized, where vector *x* stands for the filtered data *y* and vector *n* stands for n-points in FFT.

For algorithm performance purposes, *fft* allows to pad the input with trailing zeros. In this case, pad each row of X with zeros so that the length of each row is the next higher power of 2 from the current length. In order to further improve the computational speed and make it flexible for changing the FFT windows in the real tests, the routine is implemented as follows:

$$p = 2\text{^}nextpow2(n).$$

The function *nextpow2(L)* can find the nearest power of two sequence length for FFT operations and return it back.

The FFT result dividing the total points inside single window gives the average value of amplitudes in the power spectrum. However, the useful part of FFT result is the absolute values in single-side amplitude spectrum multiplying by 2, which are the exact values of single-side amplitude spectrum. After getting the useful part of FFT result, they will be sent to the step of classification.

### 3.3.3. Classification Utilizing Harmonics

Considering the cost of the time in processing, most existing classification methods are better used in classifying more than one dimension’s data. Meanwhile, the harmonics show a huge potential to be used for classification based on the data recorded in this project. Figure 17 shows the SSVEP response for the 8Hz square wave visual stimulus whereas the Figure 18 shows the SSVEP response for the subject right before he look at visual stimulus. Both data have gone through the filter and FFT, but the analyses of the power spectral density are very different. The fundamental frequency and the harmonics are clearly seen in Figure 17, where the fundamental frequency is 8Hz, the first harmonic is 16Hz and the second harmonic is 24Hz. However, it is hard to see any relationship between frequencies corresponding to the peaks in Figure 18. So it can be concluded that the harmonics only exist when the subject is looking at the RVS. In other word, the harmonics will be elicited when the fundamental frequency of the SSVEP response can remain stable.

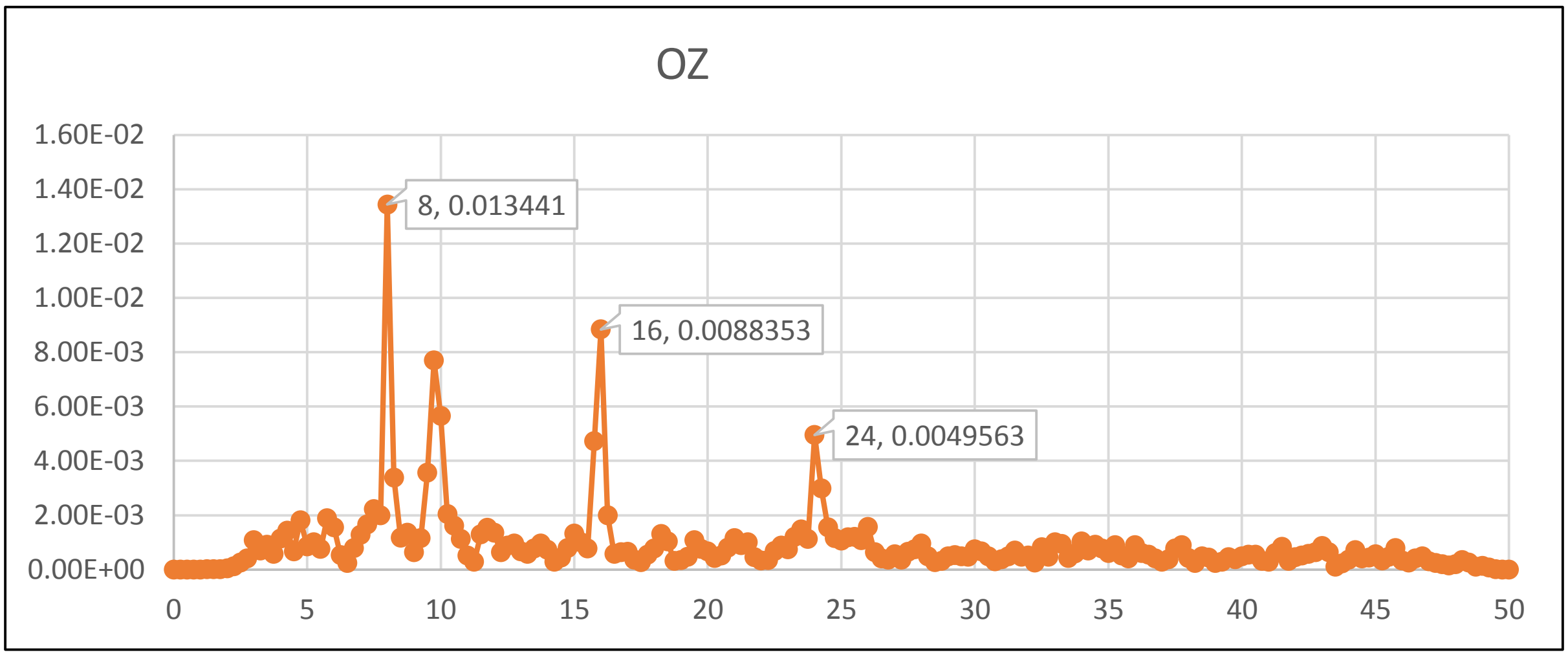

**Figure 17: 8Hz Square Wave with 256Hz Sampling Frequency and 4s Windows (from Oz electrode)**

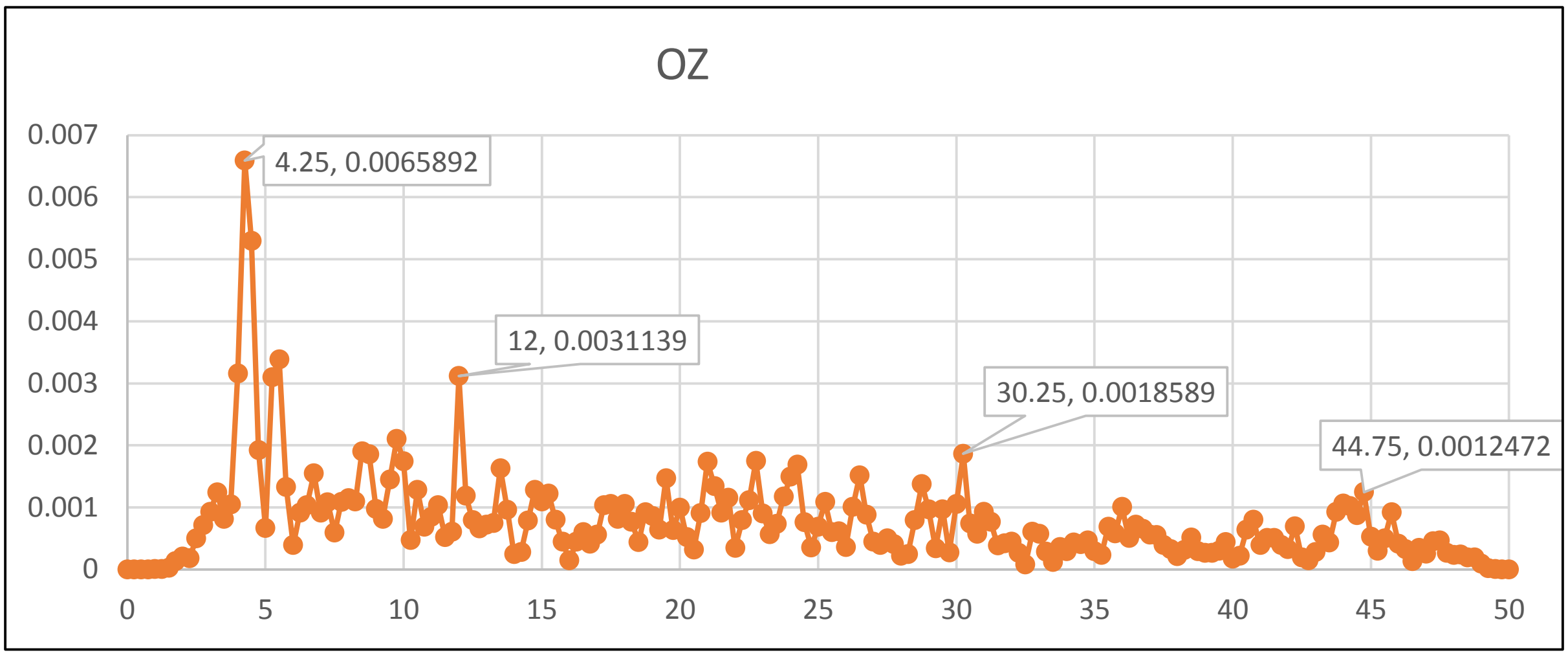

**Figure 18: The Noise with 256Hz Sampling Frequency and 4s Windows (From Oz electrode)**

Based on the feature of the harmonics' activity, a harmonic-based classification method has been proposed. The main principle of the method is to find out the frequency with the maximum harmonic points. The harmonic points are equal to the number of harmonics observed in the power spectrum density divided by the total possible harmonics (existing between 5Hz to 60Hz) for each frequency. Meanwhile, thresholds of harmonic points for each frequency need to be set based on different frequency values and the subject. When the maximum harmonic points exceed the threshold, the command of the corresponding frequency will be sent to the serial port.

In this project, the window size of FFT is 2 seconds. So, the frequency revolution is 0.5Hz. First of all, an array of flickering frequencies with six elements which stands for six different frequencies should be generated. The next task is to find frequency search spaces for each fundamental and harmonic frequency (each row represent one frequency and its harmonics). Also the value of fundamental and harmonics frequencies should be calculated for the future calculations. For search spaces, each row is used for specific frequency and the values in each row are the border values. In order to calculate the associated index in FFT result, a *search_space_point* is created based on *search_space*. Additionally, the function *round(X)* is to guarantee the elements of X is rounded to the nearest integers, this is to make sure that the index is not a non-integer value. The code for last three tasks is shown below.

```
for i = 1:6
    for j = 1:10
        search_space(i,j) = flickering_frequency(:,i)*(j - 0.5);
        fundamental_and_harmonics(i,j) = flickering_frequency(:,i)*j;
    end
end
search_space_point = round(search_space.* dWindow + 1);
```

After the matrix of the *search_space_point* has been created, the process goes into the main loop. The routine below is for finding the maximum value in Oz, and it is the same routine for O1 and O2. There are 9 intervals in each row. After making sure the border values in *search_space* does not exceed 50Hz, the function *[Y, I] = max(X)* is used in selecting the maximum values of power spectral density. *max* function returns the indices of the maximum values in vector I and the maximum values in vector Y. According to the relationship between maximum index and its corresponding frequency, the maximum frequency is obtained and saved into a matrix.

```
for i = 1:6
    for j = 1:9
         if search_space(i,j) <= 50
            [max_value, f_max_ind]=…
            max(Y_Oz(search_space_point(i,j):search_space_point(i,j+1)));
           f_max_Oz(i,j) = (f_max_ind + search_space_point(i,j)-2)/NFFT*Fs;
         end
     end
 end
```

The next step is comparing the frequency with maximum power with the fundamental and harmonics frequency. The comparison routine for Oz is shown below. If the conditions can be matched, the point of the corresponding flickering frequency will be added one. For O1 and O2 the same method is applied.

```
for i = 1:6
        for j = 1:10
            if (f_max_Oz(i,j) == fundamental_and_harmonics(i,1)||...
                f_max_Oz(i,j) == fundamental_and_harmonics(i,2)||...
```

```
        f_max_Oz(i,j) == fundamental_and_harmonics(i,3)||...
        f_max_Oz(i,j) == fundamental_and_harmonics(i,4)||...
        f_max_Oz(i,j) == fundamental_and_harmonics(i,5)||...
        f_max_Oz(i,j) == fundamental_and_harmonics(i,6)||...
        f_max_Oz(i,j) == fundamental_and_harmonics(i,7)||...
        f_max_Oz(i,j) == fundamental_and_harmonics(i,8)||...
        f_max_Oz(i,j) == fundamental_and_harmonics(i,9)||...
        f_max_Oz(i,j) == fundamental_and_harmonics(i,10))
        point(i) = point(i) + 1;
    end
```

Finally the normalized points must be calculated. Since the number of harmonics (between 5Hz and 50Hz) for each frequency is different, the total points divided by the total theoretical possible harmonics on the three electrodes becomes the normalized points as shown below. The function of *floor(X)* has been implemented to make sure the elements of X are rounded to the nearest integers towards minus infinity. This function is used to correctly calculate all the possible harmonics below 50Hz. It is important to mention that the main routine check for harmonics below 50Hz only.

```
for i = 1:6
       point(i) = point(i)/(3*floor(50/flickering_frequency(i)));
end
```

Finally, if the maximum value in the *point* is bigger than the corresponding threshold, then the relative command will be sent to the serial port.

The advantage of harmonic-based method is that it can reduce the possibility of the fasle detection compare to when fundamental frequency method is used. Meanwhile, it reduces the time delay because the FFT windows can be reduced. It is important to mention that if fundamental frequency method is used, the window time cannot be reduced, because the number of false detections increases significantly.

## 3.4. Design of SSVEP Visual Stimuli Panel

### 3.4.1. Overview of Panel Design

There are two ideas for panel design based on how the panel information is presented. The user experience and the future potentials for these designs are very different, although the only hardware difference is that one has 6 small LCDs. In order to satisfy wider range of applications, both of the designs are capable of controlling more than the wheelchair alone.

### *a. First Idea:*

The initial idea of panel design is using six LED units to flicker at six different frequencies and each frequency stands for a specific task. LCDs are not used in this idea. Each pair of frequency can control one device. Three devices are considered in this project, wheelchair, TV and air conditioner.

- **Main Menu**: The main panel interface is consisted of three pairs of LED units which control the operating status (on/off) of the corresponding devices. When the subject focuses on one of the LED units, 'f1', 'f3' or 'f5'; the wheelchair, TV or air conditioner is turned on respectively. Figure 19 represent the main menu on the panel.
- **Wheelchair Menu**: When the wheelchair is turned on by the subject, the panel menu is changed to the menu in Figure 20. Four LED units are assigned as the wheelchair control commands, which are 'f1', 'f2', 'f3' and 'f4'. Through focusing on LED units, the subject can control the wheelchair to go to the expected directions. 'f6' is arranged for going back to the main menu.
- **TV Menu**: When the TV is turned on by the subject, the menu on the panel change to the menu in Figure 21. Through looking at the LED units 'f1', 'f2', 'f3' and 'f4'; the TV channels and volumes can be changed. Without operating for 10 seconds, the interface will automatically go back to the main menu.
- **Air Conditioner Menu**: When the air conditioner is turned on by the subject, the menu changes to the menu in Figure 22. Through concentrating on the LED units 'f1', 'f2', 'f3' and 'f4'; the subject can control both temperature and fan speed of the air conditioner. Without operating for 10 seconds, the menu will automatically go back to the main menu.

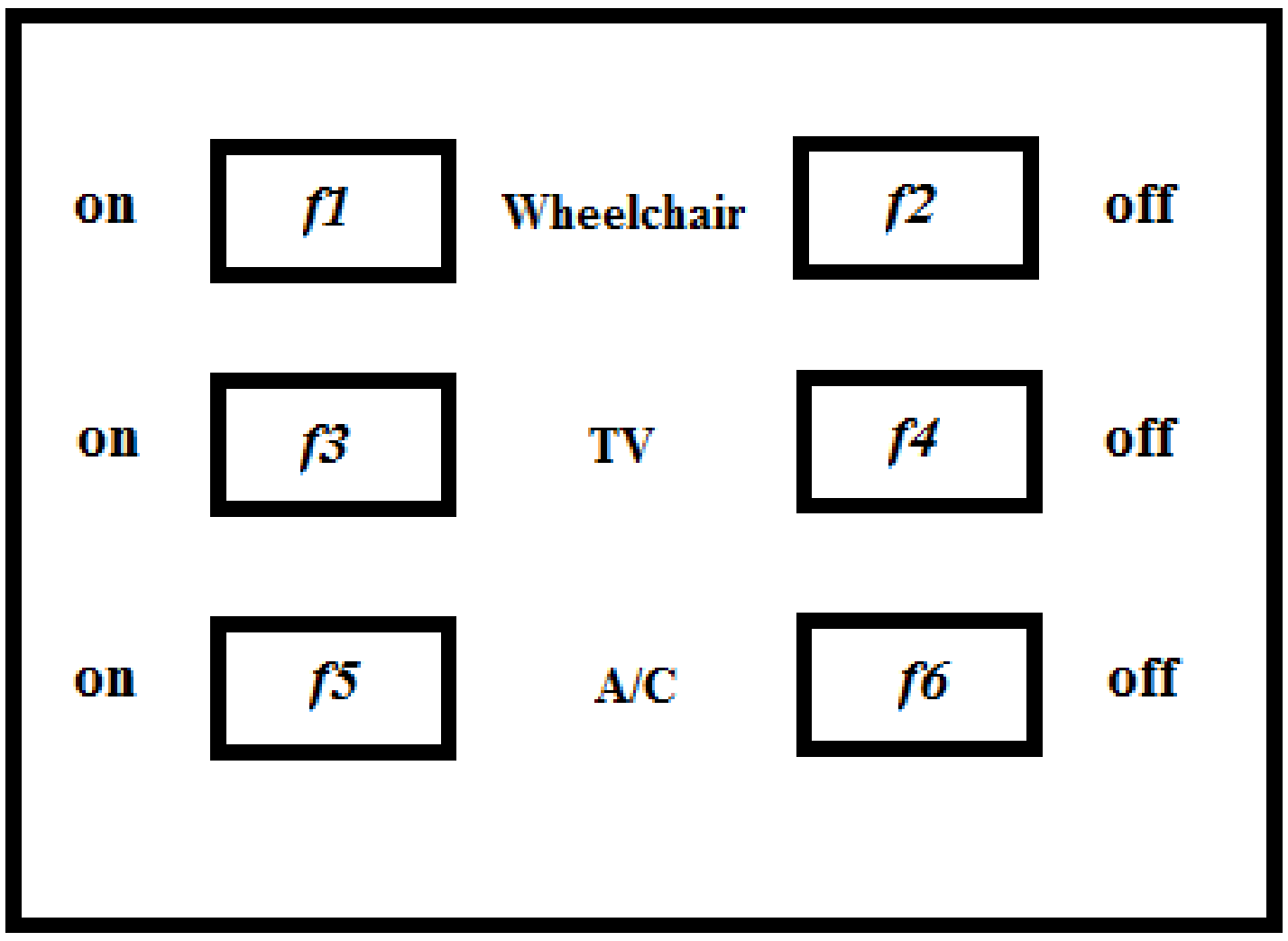

**Figure 19: Main Menu**

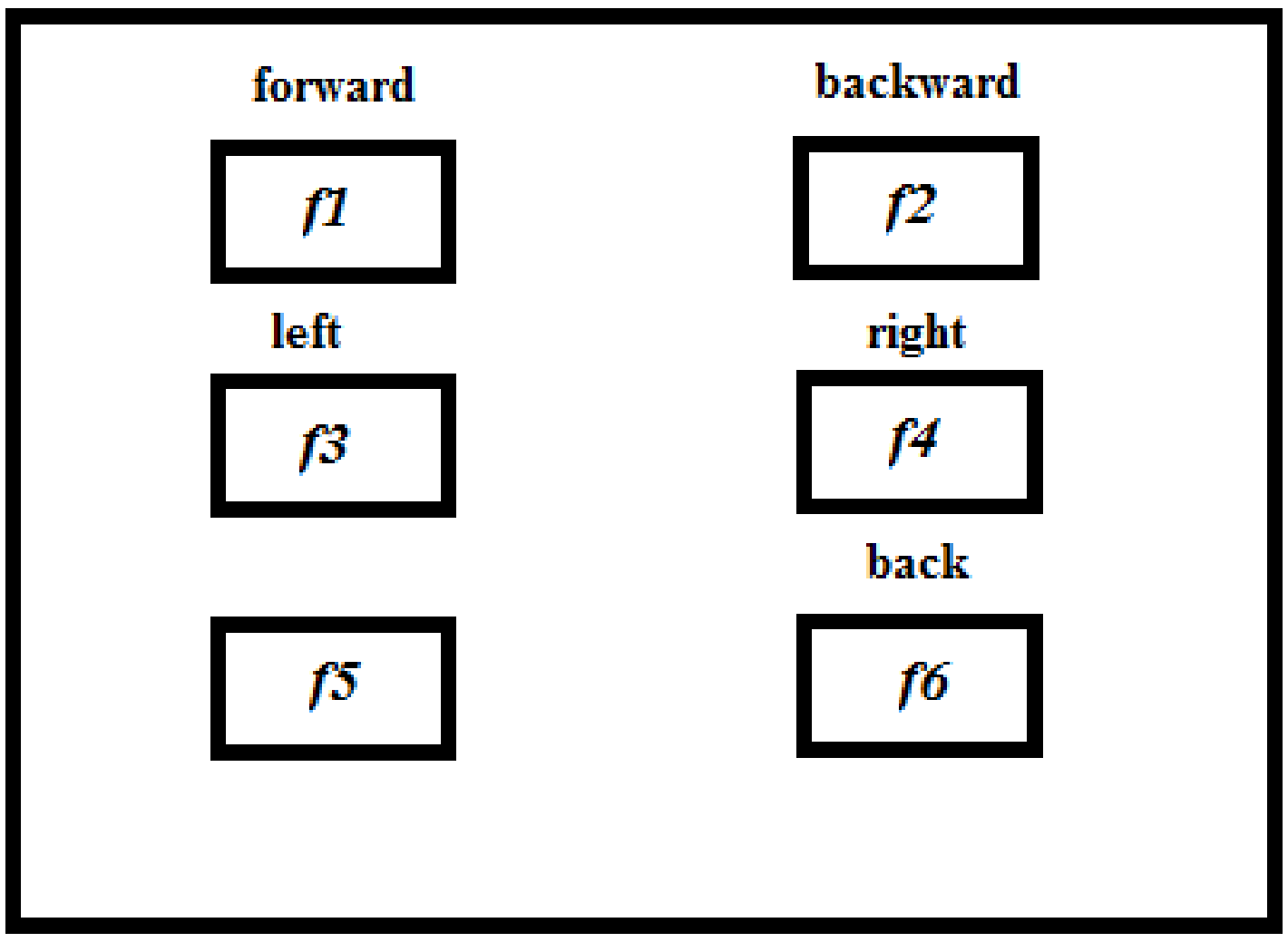

**Figure 20: Wheelchair Menu**

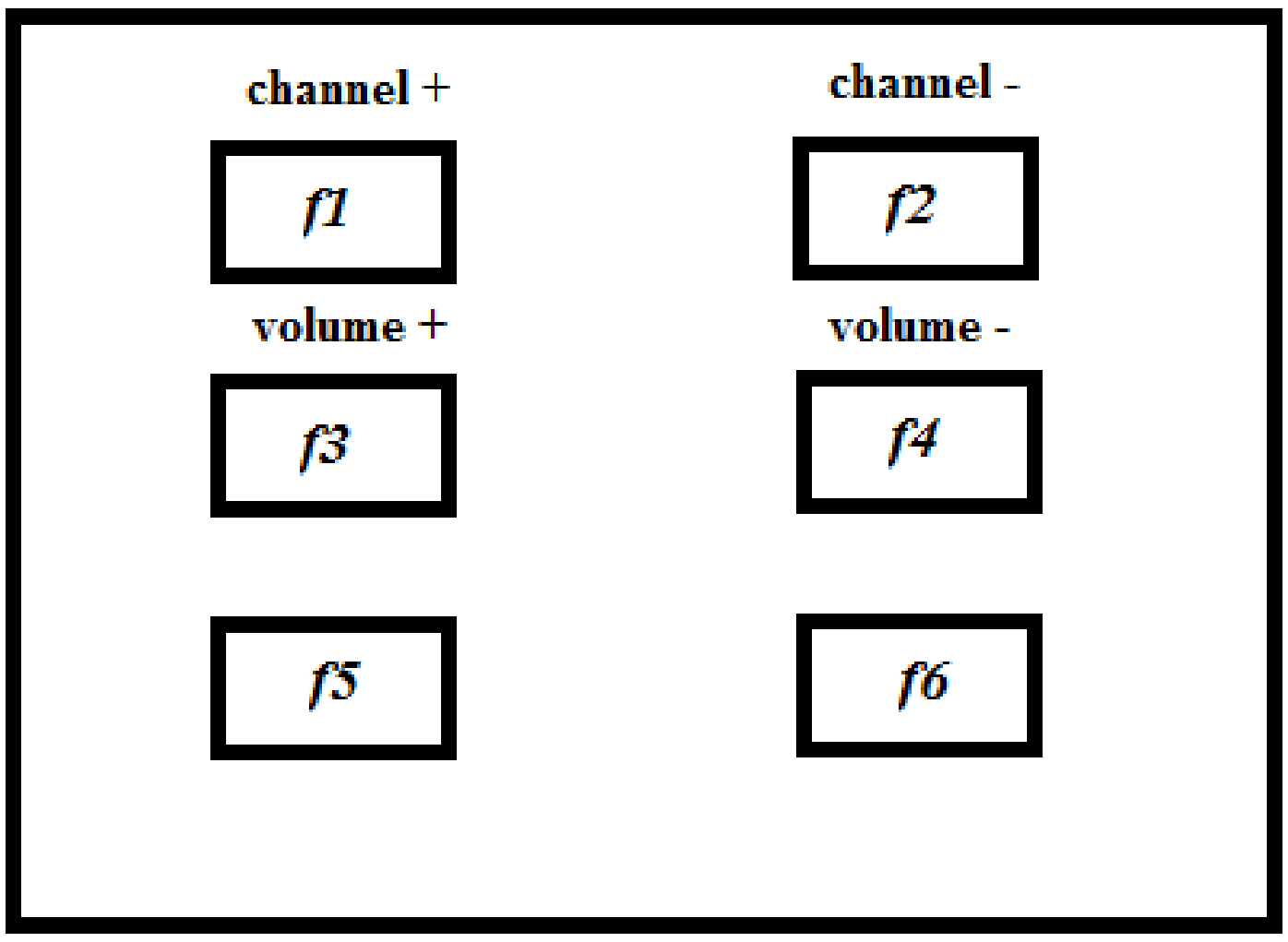

**Figure 21: TV Menu**

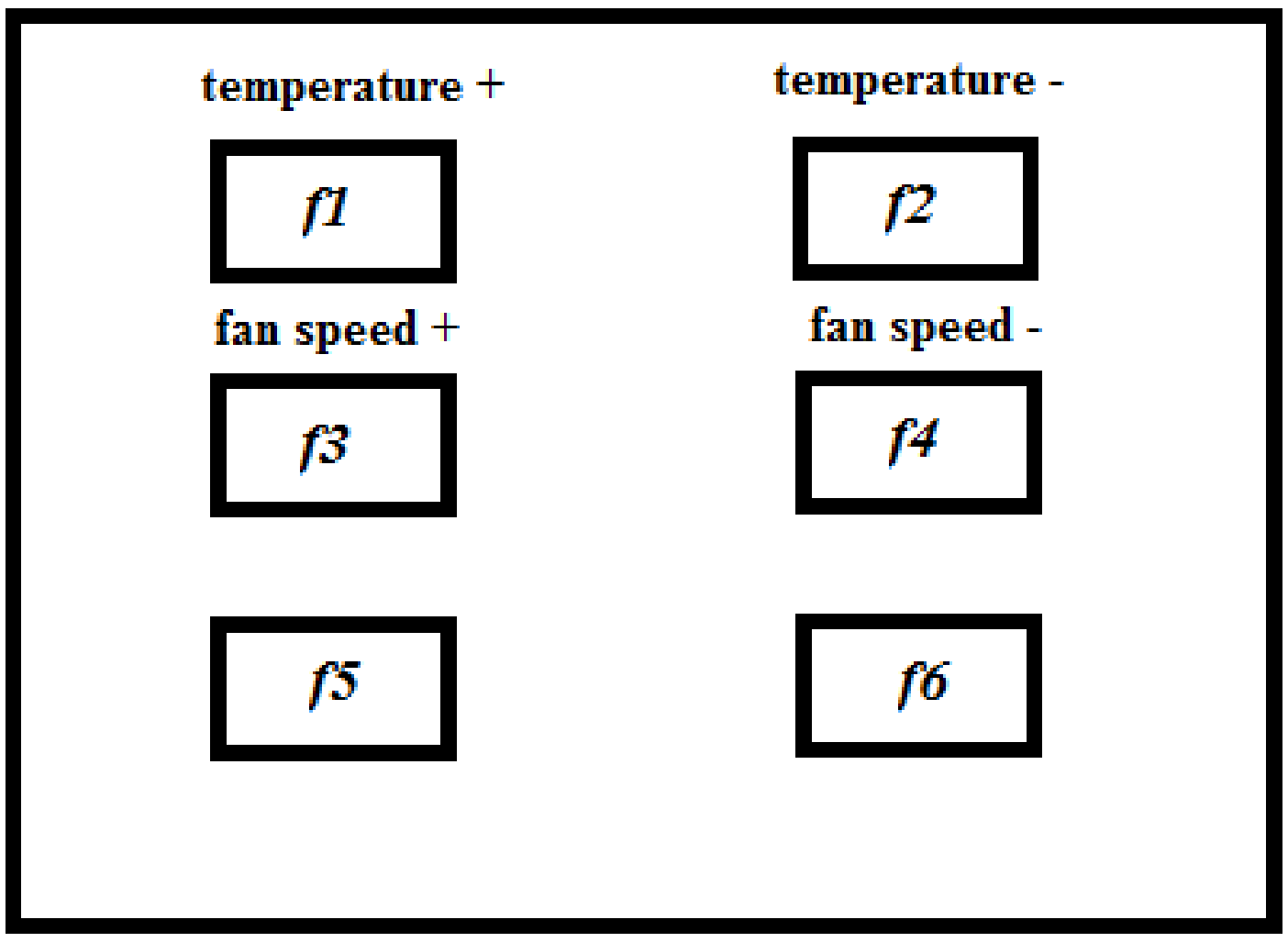

**Figure 22: Air Conditioner Menu**

***b. Second Idea:***

Six LED units with different flickering frequencies are still used in this version. The panel is improved by adding six LCDs to display the menu information for each LED unit. In order to control more devices, the panel menus are also changed. The maximum of 5 devices can be controlled in this version.

- **Turn On Menu**: As shown in Figure 23, the six rectangles in the middle are representing the LCDs and the other rectangles are LED units denoted by '*f*'. 'f6' is assigned as the master switch of the whole panel system. In this menu all the LED units are off except for f6. When the subject focuses on the 'f6' LED unit, the panel system is turned on and the panel menu is changed to the main menu.
- **Main Menu**: The main menu is designed for the selection of wheelchair, TV, air conditioner, and light as shown in Figure 24. Once one device is selected, the panel menu is changed to the corresponding device's menu. If the subject focuses on f6 LED unit, the panel goes to turn on menu.
- **Wheelchair Menu**: When the wheelchair is selected by the subject, the panel menu is change to the wheelchair menu shown in Figure 25. Wheelchair should be turned on by looking at 'f1'. Then, the subject can control the movement of the wheelchair. Additionally, when the wheelchair is turned off, the interface goes back to the main panel interface simultaneously. The off bottom in every sub-interface has the same function as the wheelchair one.
- **TV Menu**: After selecting the TV in the main menu, the panel information is transferred to TV menu as shown in Figure 26. Instead of using 10 seconds timer (as it was proposed in the first idea), the subject has the full control over all the menus. Subject can turn the TV on or off.
- **Air Conditioner Menu**: When the subject chooses air condition (by focusing on associated LED unit) in the main menu, the air condition menu appears. From this menu, the subject can adjust the operating status, temperature and the fan speed as shown in Figure 27.
- **Light Menu**: When the subject chooses light (by focusing on associated LED unit) in the main menu, the air condition menu appears. The light menu is the simplest menu because the light has two possible statuses, which are on and off. This menu is shown in Figure 28.

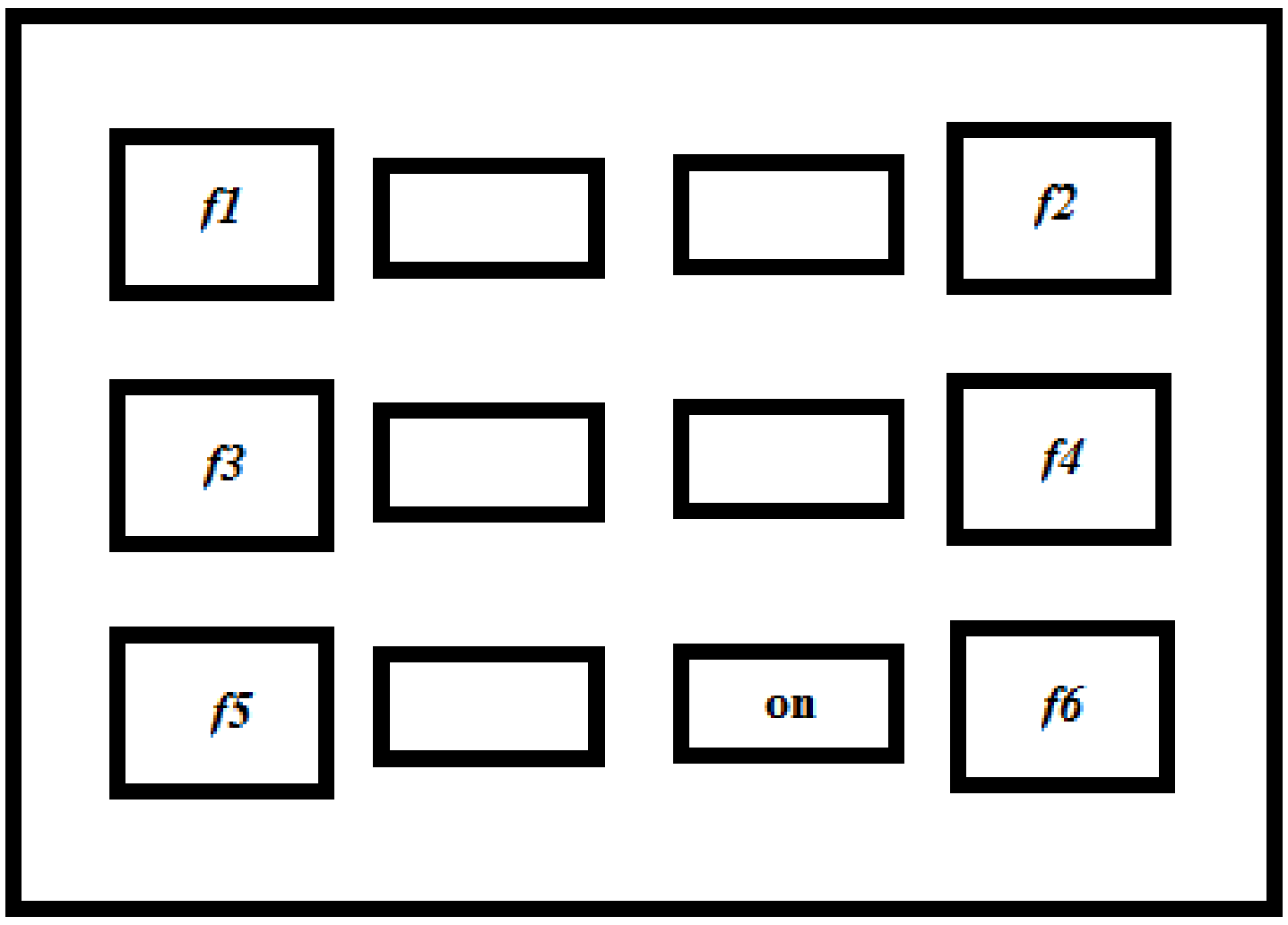

**Figure 23: Turn On Menu**

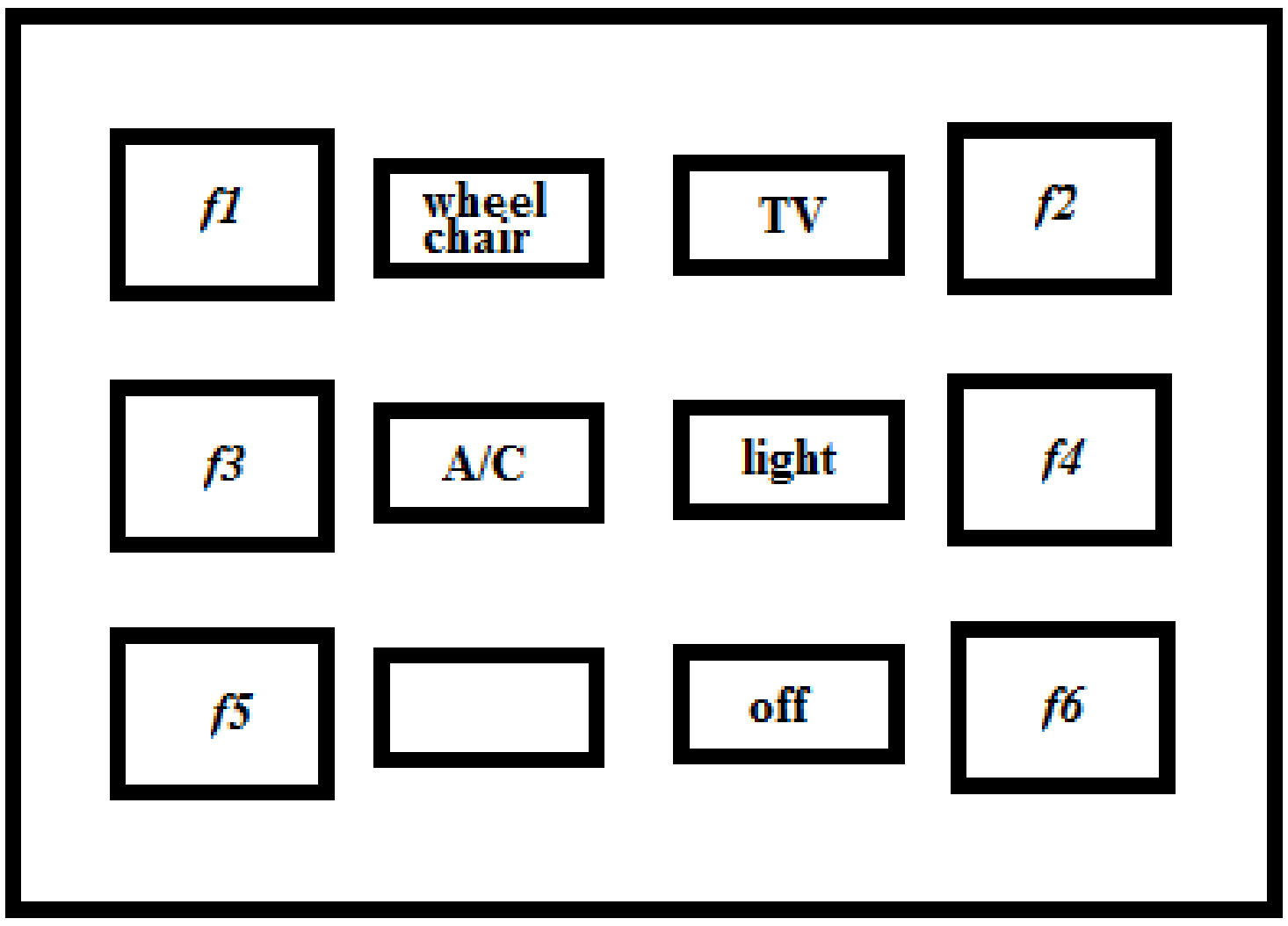

**Figure 24: Main Menu**

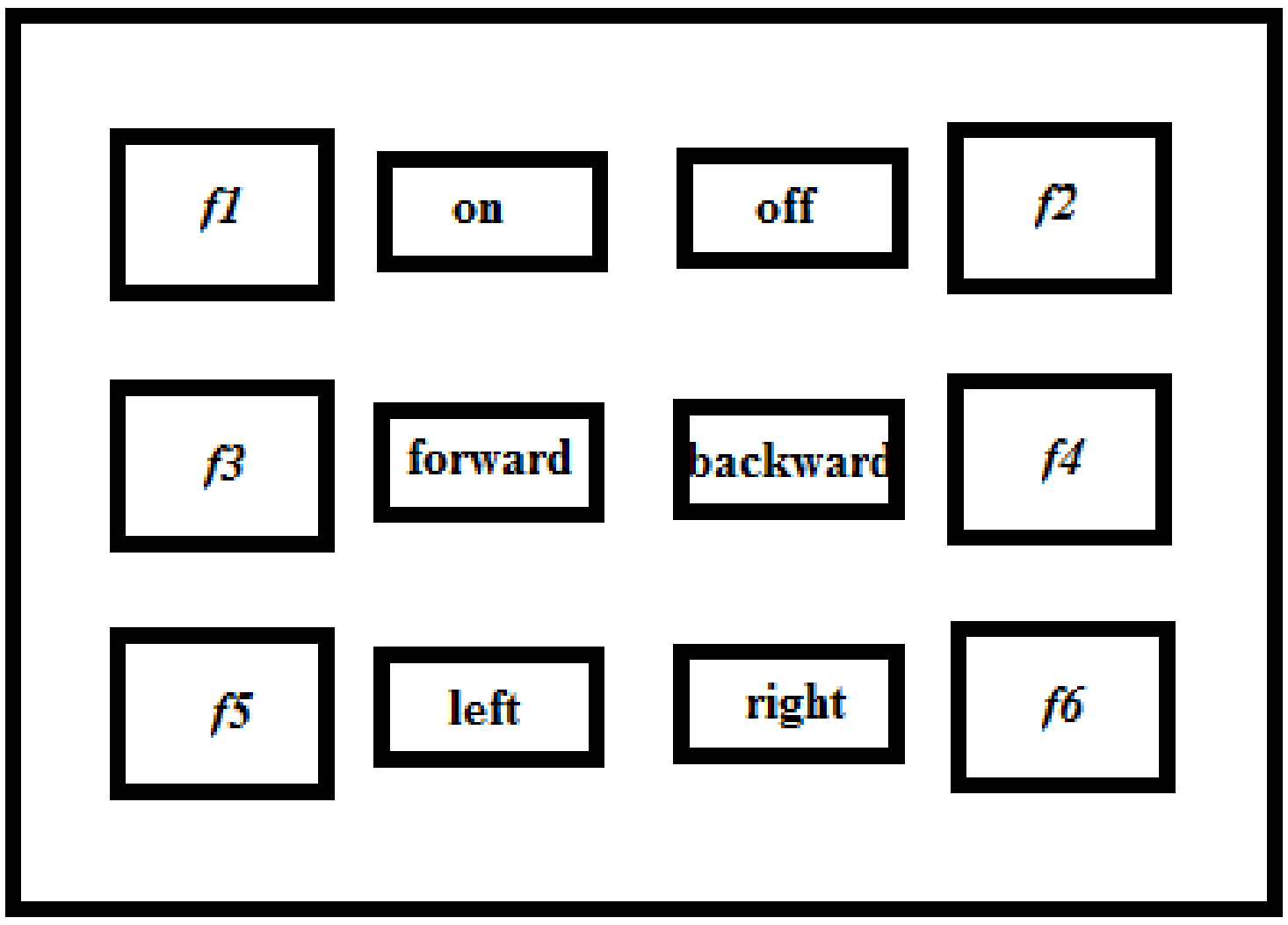

**Figure 25: Wheelchair Menu**

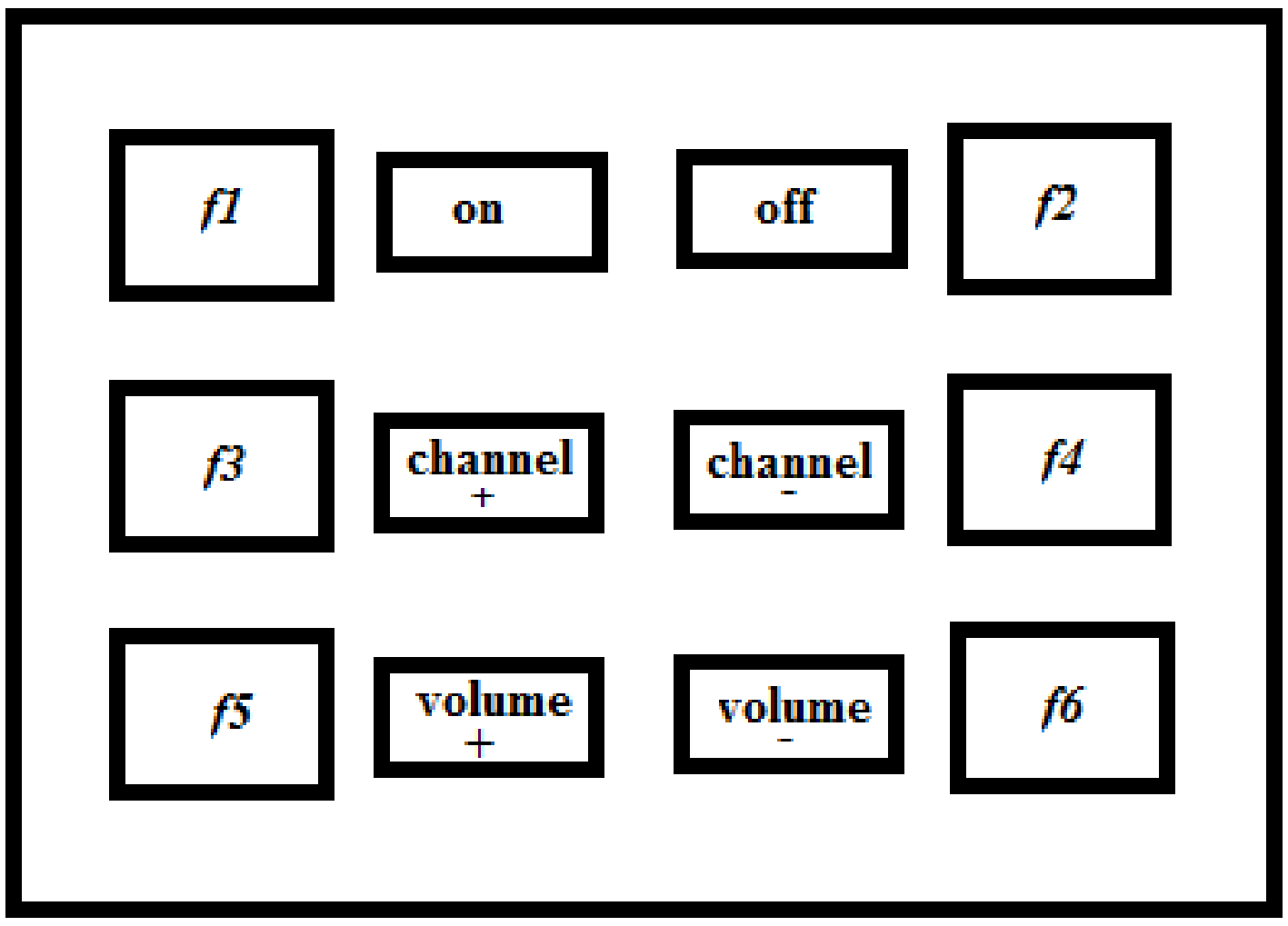

**Figure 26: TV Menu**

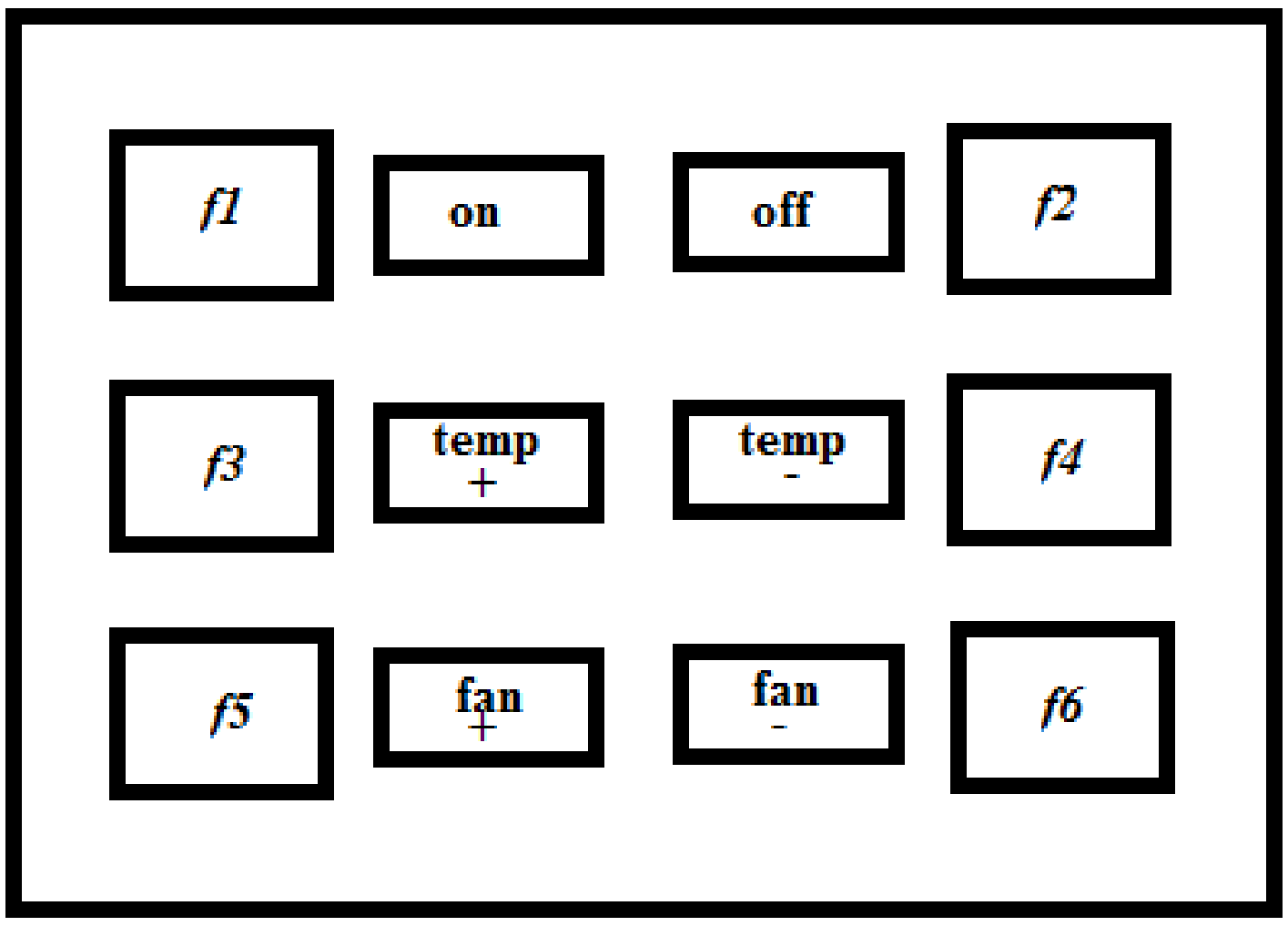

**Figure 27: Air Conditioner Menu**

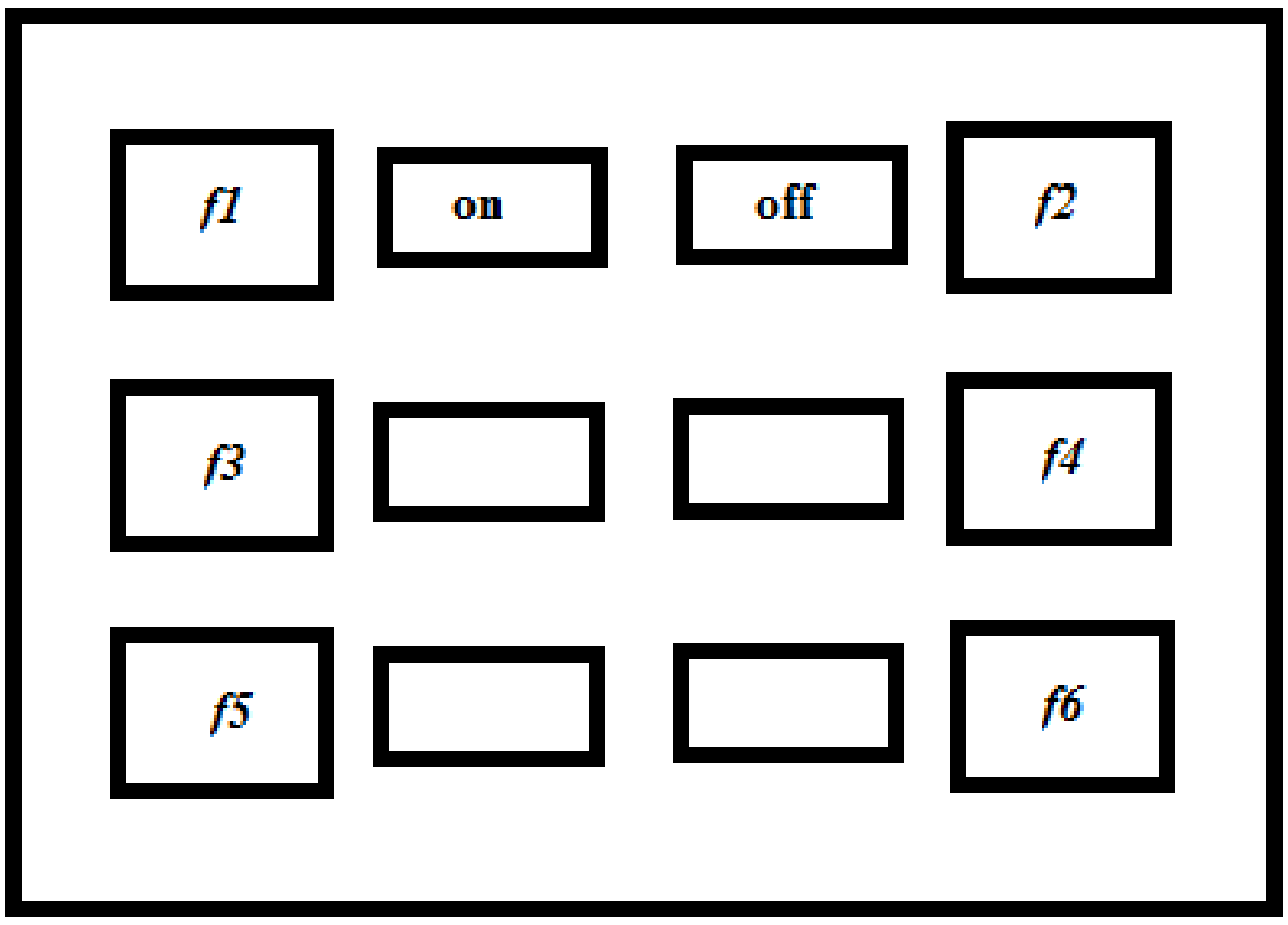

**Figure 28: Light Menu**

### 3.4.2. Overview of the Circuit Design

The goal of using hardware to make LED flickers is to generate the signal with high stability and low noise. The current design is using microcontroller to control the frequency generation. Meanwhile, Digital-to-analog converter (DAC) is controlled by microcontroller to generate analog signals required. The combination of operational amplifier (op amp) and LEDs make the current through the LED adjustable. Figure 29 shows the block diagram of signal generation process.

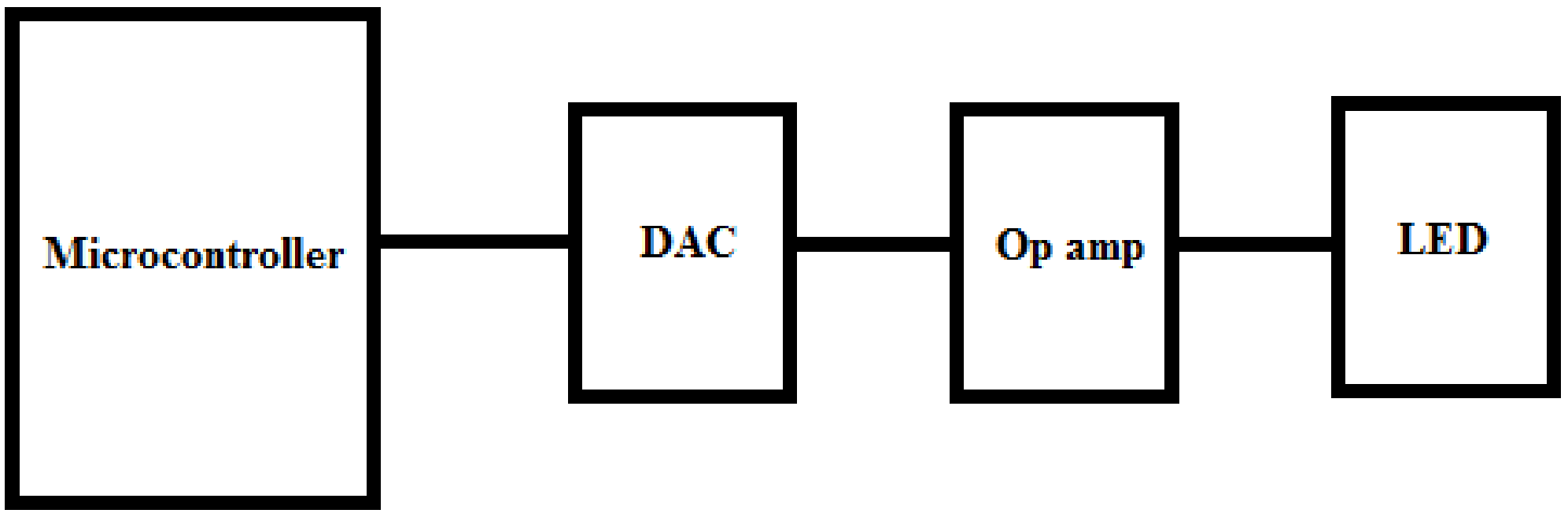

**Figure 29: Block Diagram of Signal Generation Process**

Another microcontroller is used to control the 6 LCDs. At the same time, if Matlab sends a command to the microcontroller, the information on the LCDs can be changed. Each LCD needs an unshared enable signal (colorful lines), while the data bus (black lines) can be shared between all LCDs. Figure 30 shows the block diagram of LCDs and microcontroller connection.

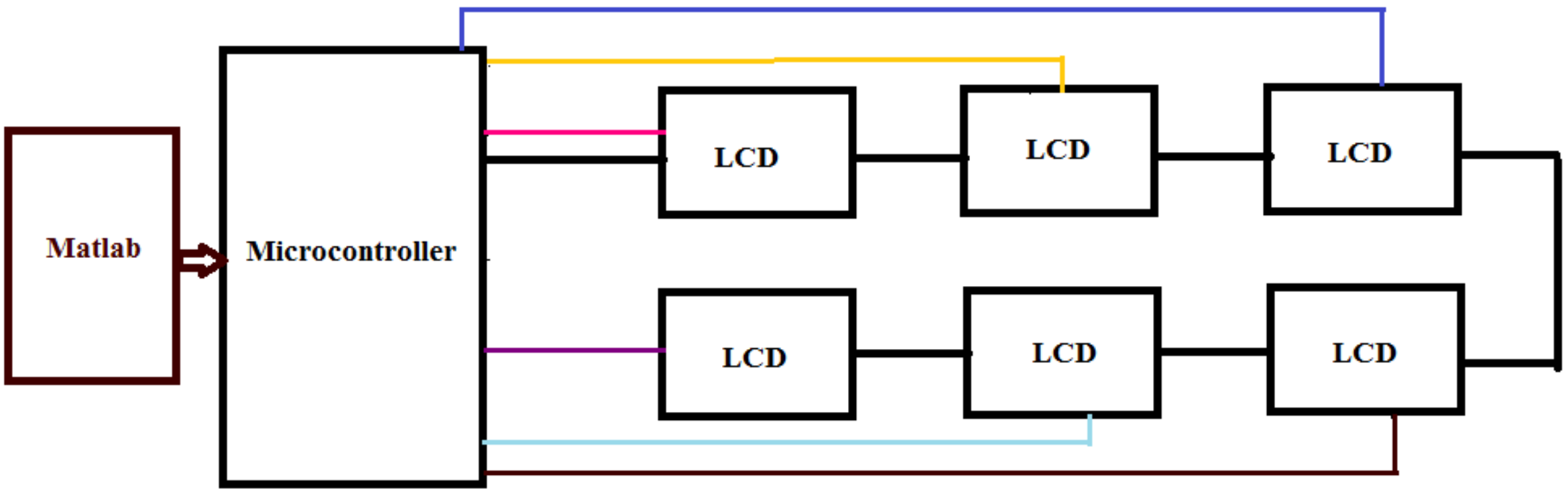

**Figure 30: The Block Diagram of LCDs Connection**

The combination of these two microcontrollers and associated circuits makes up the SSVEP visual stimuli panel. The function is that when the subject is looking at one of the flickering LEDs, the Matlab processes the EEG signal and send the corresponding commands through serial port to the microcontroller. The microcontroller takes action based on the command sent by Matlab and change the LCDs information.

### 3.4.3. Main Components in Panel Design

The components of SSVEP visual stimuli panel are sleeted while considering the price and the performance. The components in Table 3 are the main components purchased for the project. By using the main components and the microcontrollers, frequency generation and LCD management is achievable.

**Table 3: Components**

| | |
|---|---|
| LED | ASMT-YTD7-0AA02 |
| Op Amp | TCA0372BDP1G |
| Power Resistor | 352147RFT |
| Voltage Regulator | LD1085CDT-R |
| LCD | MC20805B6W-FPTLW |
| DAC | DAC088S085CIMT/NOPB |

### 3.4.4. Hardware for Frequency Generation

As referred previously, the frequency generations for the SSVEP visual stimuli panel needs an Arduino Uno (as microcontroller) with DACs, op amps and LEDs. Considering that the currents through LEDs are very high, power resistors are needed for dissipating voltage and adjusting current in the LEDs. The voltage regulator is also needed for regulating the input voltage, so that one battery can supply all the components with different operating voltages.

The purchased LED (ASMT-YTD7) is RGB (red, green and blue) type; therefore they can create wide range of colors. Each color has independent channel so that red, green and blue can be controlled independently. All three channels have the typical forward current of 20mA with the corresponding luminous intensity 650mcd, 1900mcd and 384mcd for red, green and blue respectively. Because of the high luminous, these LEDs are suitable for both indoor and outdoor; therefore the wheelchair can be used in outdoor. With the respect of typical forward current, the forward voltage is 2.1volts for red channel and 3.1 volts for the other two channels. In this project, 4 LEDs are combined together as a LED unit in one pack.

The selected op amp (TCA0372) can be operated below 40 volts. The output of the op amp is up to 1A and the requirement of one LED unit is less than 100mA; therefore, it can satisfy the current requirement. Moreover, the power dissipation of the op amp is within the limitation based on the op amp datasheet (the calculation is in 4.2.1).

The resolution of the DAC (DAC088S085) purchased is 8 bit and it can operate from 2.7 to 5.5 volts. This DAC has 8 channels and they are controlled through a digital interface consisting 3 lines. The interface is compatible with SPI and it can operate at clock rates up to 40MHz. $\overline{SYNC}$ is a Frame Synchronization Input pin to synchronize the DAC with other events. When this pin goes low, data is written into the DAC's input shift register on the falling edges of SCLK. After 16th falling edge of SCLK, a rising edge of $\overline{SYNC}$ causes the DAC to be updated. The most important feature of the DAC is the Daisy Chain operation. It allows communication with other DACs of same type, using a single serial interface. Figure 31 shows the Daisy Chain configuration.

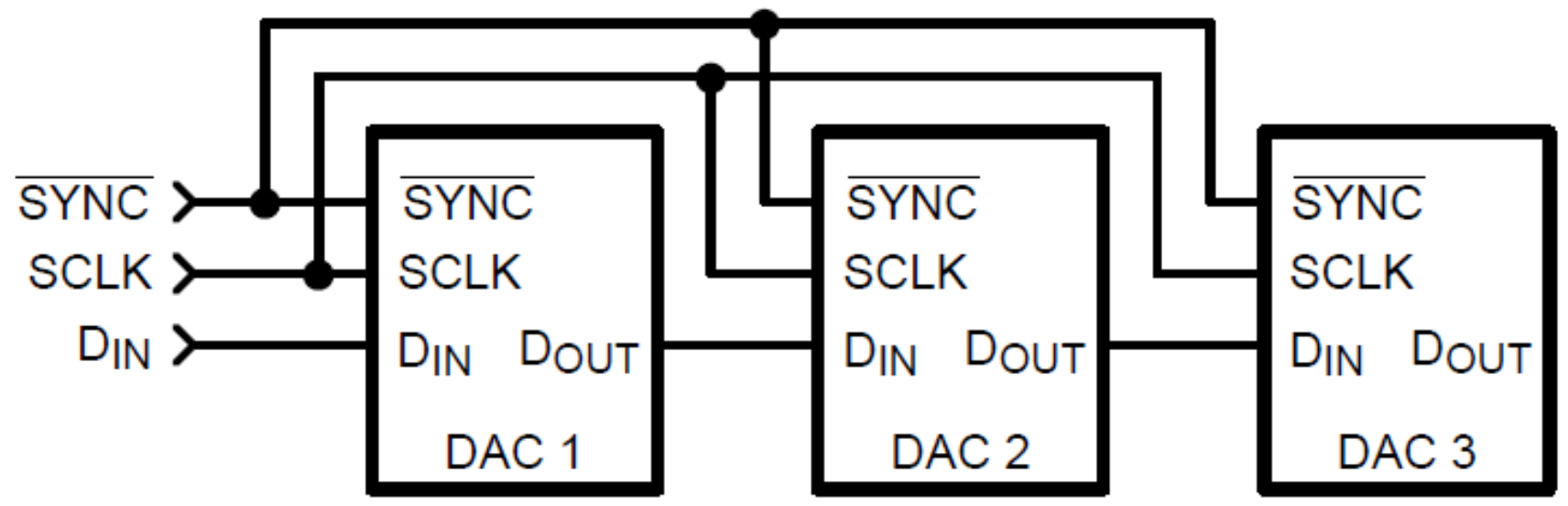

**Figure 31: Daisy Chain Configuration**

To support multiple devices in a daisy chain configuration, SCLK and $\overline{SYNC}$ are shared across all DACs and digital output of the first DAC in the chain is connected to digital input of the second one. The same configuration is applied between second and third DACs. Similar to a single channel write sequence, the conversion for a Daisy Chain operation begins on a falling edge of $\overline{SYNC}$ and ends on a rising edge of $\overline{SYNC}$. A valid write sequence for N devices in a chain requires N times 16 falling edges to shift the entire input data stream through the chain. Daisy Chain operation is specified for a maximum SCLK speed of 30MHz. In this project, 3 DACs are connected in Daisy Chain configuration and the Daisy Chain communication protocol is shown in Figure 32.

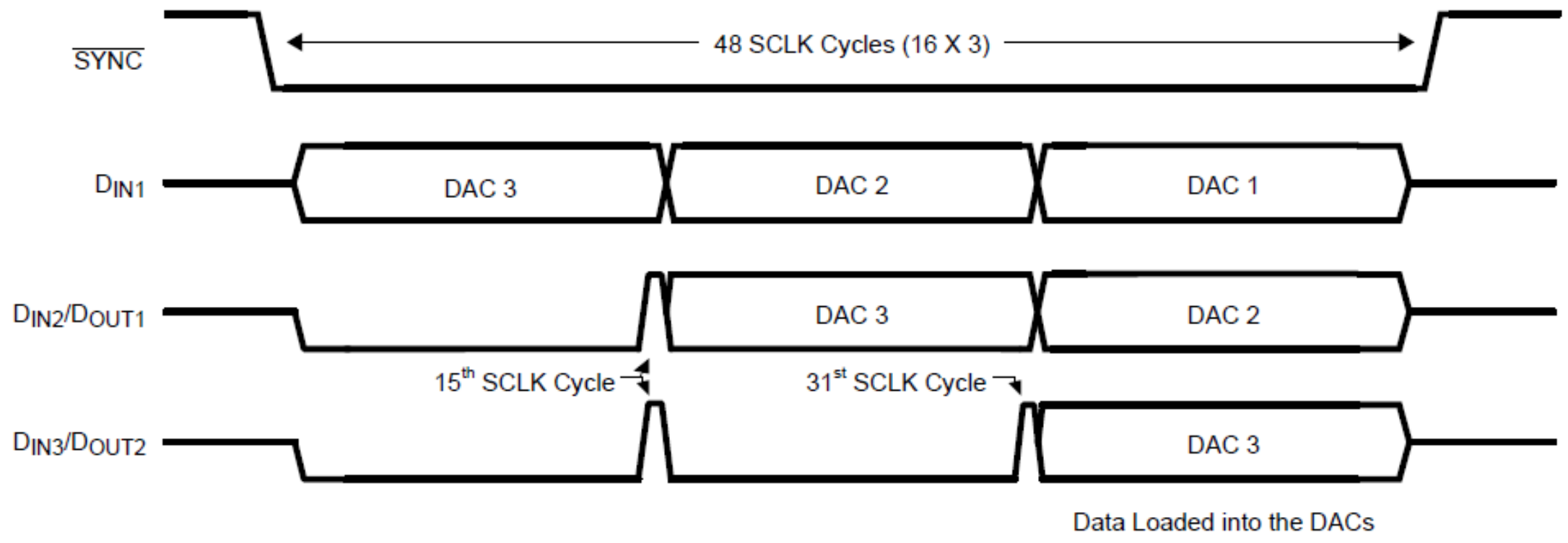

**Figure 32: Daisy Chain Communication Protocol**

In a write sequence, digital output remains low for the first fourteen falling edges of SCLK before going high on the fifteenth falling edge. Subsequently, the next sixteen falling edges of SCLK will output the first sixteen data bits entered into digital input. On the rising edge of $\overline{SYNC}$, the programmed function is executed in each DAC simultaneously. Besides, two modes of operation are available in DAC which are Write Register Mode (WRM) and Write Through Mode (WTM). WRM means that the registers of each DAC channel can be written to, without changing their outputs simultaneously while the meaning of WTM is the opposite.

### 3.4.5. Hardware for LCD Management

The main devices used in the information display part are Arduino Mega 2560 and LCDs. As mentioned before, the LCDs do not share their enable pins, but they share the same data bus. For this project, a special cable has been made as shown in Figure 33. In this special cabala seven connecters are connected to the ribbon wire. Six of them are connected to LCDs, and the last one is connected to microcontroller. The wires connected to enable pin of LCDs is cut out between each two connectors. Then, each of the cut out wires was separately connected to different output pin of Arduino Mega 2560 through the colorful wires shown in Figure 33. Thus, one Arduino Mega 2560 is able to control all the 6 LCDs at the same time.

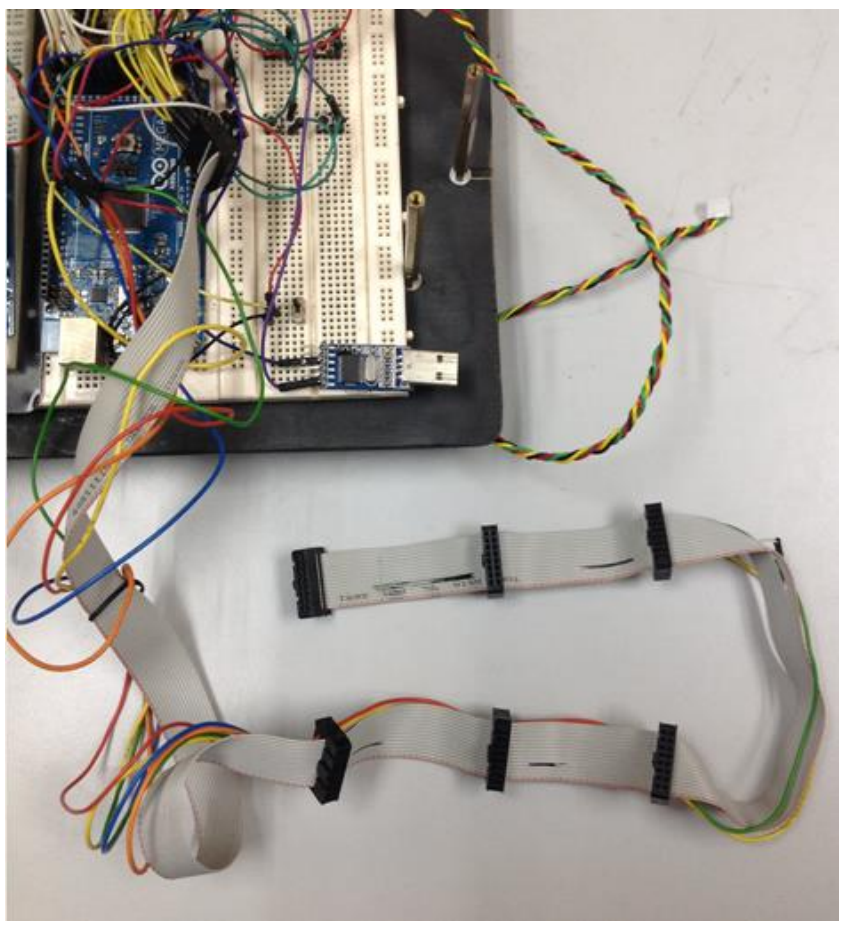

**Figure 33: The Cable Made for Interfacing 6 LCDs**

In order to communicate between Matlab and Arduino Mega 2560, an USB to UART converter is needed. The pin of transmitter (TXD) and ground (GND) are used for serial communication from Matlab to Arduino Mega 2560. The pin TXD is connected to one of the receiver pin (RX1) on Arduino Mega 2560. Since the data communication is only from Matlab to Arduino Mega 2560, the configuration explained above is sufficient. In the future two ways communication may be implanted.

### 3.4.6. Stimulation of Frequency Generation

The stimulation is based on the circuit design of frequency generation using Proteus 8.0.

Figure 34 shows the circuit schematic for the stimulation.

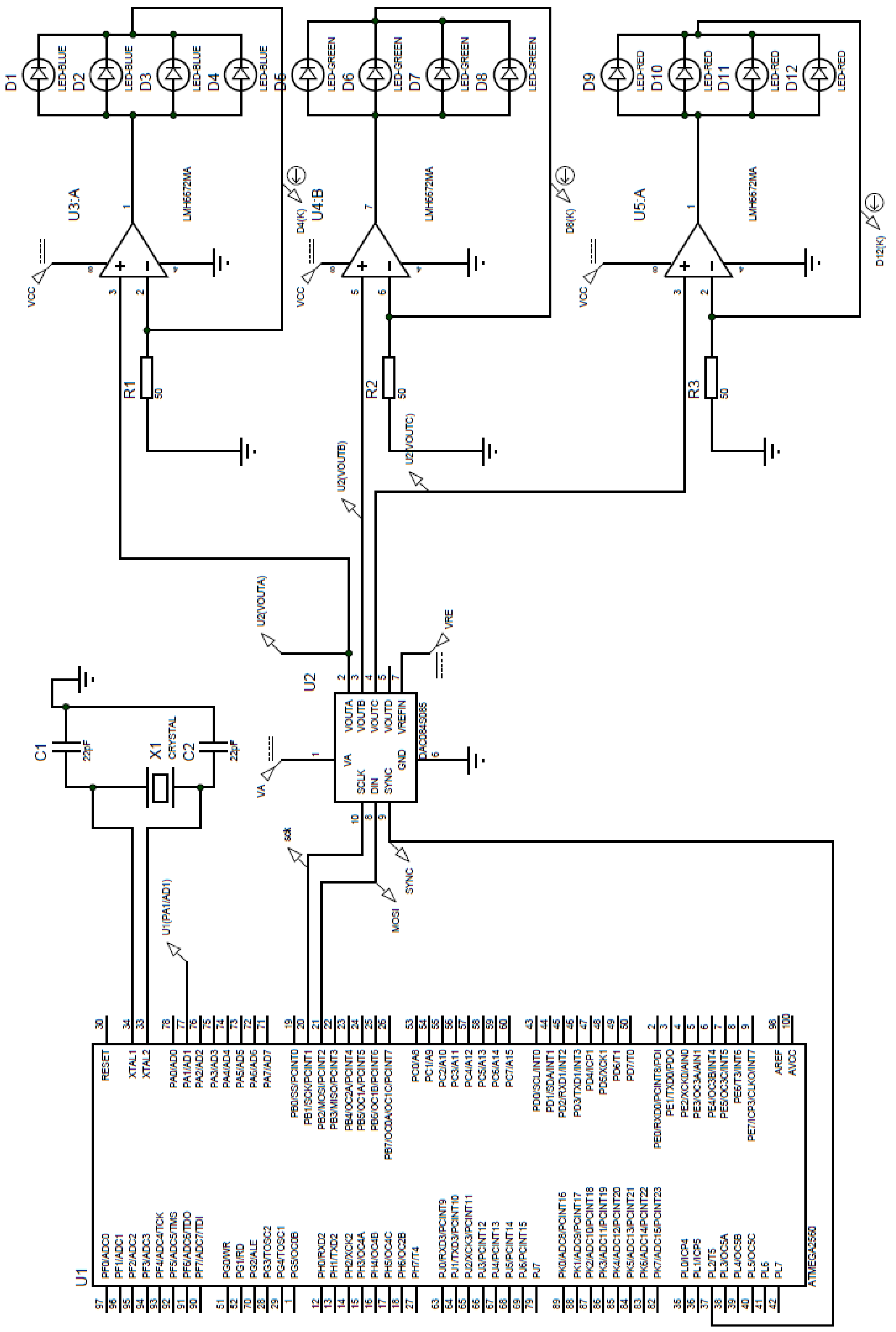

**Figure 34: Frequency Generation Circuit Schematic**

### 3.4.7. Algorithms for LED Flickering Frequencies

The LEDs should flicker in the different frequencies using only one microcontroller (Arduino Uno). In order to realize it, Daisy Chains and interrupts are applied in this part. The software used in this project for microcontroller programming is Arduino Software (IDE).

The interface of DAC is compatible with SPI, so in the process of setup, SPI transaction function should be initialized using the routine below:

SPI.beginTransaction( SPISettings( 30000000, MSBFIRST, SPI_MODE1 ) );

The SPISettings object is used to configure the SPI protocal. All 3 parameters are combined to a single SPISettings object, which is given to SPI.beginTransaction(). The first parameter defines the maximum speed of communication. Because DAC can operate at clock rates up to 30MHz, the maximum speed here should be 30000000. The second parameter is data order which means the order of the bits shifted into the SPI bus. Two possible values for the second parameter are Most Significant Bit First (MSBF) and Least Significant Bit First (LSBF). Most significant bit means the bit position has the greatest value in a binary number and least significant bit has the opposite meaning. In Daisy Chain operation, most significant bit is the first bit based on the datasheet, so MSBF should be chosen. The third parameter configure the clock polarity and phase, based on the DAC's datasheet, the interface used in the project must be configured with CPOL bit set to zero and its CPHA bit set to one. This configuration causes data on the MOSI output to be valid on the falling edge of SCLK. By checking the datasheet of Arduino Uno, SPI_MODE1 is selected.

The next step is to set a reference time interval to generate the flickering frequencies. After measuring the time taken for a complete write to all DACs and their 8 channels, it was concluded that 0.4 millisecond is long enough for both DAC write operation and the other tasks introduced in the future. Therefore, 0.4 millisecond interrupt for generating different waveform with high accuracy was implemented. The clock speed of Arduino Uno is 16MHz. So the top value of the interrupt is calculated as follows:

$$\frac{n+1}{16 \times 10^6} = 0.4 \times 10^{-3}$$

Thus, $n = 6399$. The n value must be converted to hexadecimal. The conversion result is 18FF which is the top value in Fast PWM mode. The interrupt is configured so that it generates interrupt every 0.4 millisecond.

After setting and enabling the interrupt, a Daisy Chain function of DAC is made for updating the DACs from the main loop of the code. WRM mode is the default mode in DAC and a special command called "Channel A Write" is available in this mode. Channel A Write allows the user to alter all DAC outputs when Channel A Write is performed, in this way all the outputs and channel A output get updated to the last value simultaneously. Therefore; in order to update several DAC outputs simultaneously (including Channel A), channel A should be written after setting other channels to their future value. This special command is exercised by setting data bits DB[15:12] to "1011" and data bits DB[11:0] to the desired control register value. The Daisy Chain function of DAC on channel A is shown as below:

```
void se
t_channel_A_update_all(uint8_t value_DAC1, uint8_t value_DAC2, uint8_t value_DAC3 )
{
   uint16_t command = 0xB000, command_DAC1, command_DAC2, command_DAC3;
   command_DAC1 = command | ( (uint16_t) value_DAC1 << 4);
   command_DAC2 = command | ( (uint16_t) value_DAC2 << 4);
   command_DAC3 = command | ( (uint16_t) value_DAC3 << 4);
   digitalWrite (SYNCpin, LOW);
   SPI.transfer16(command_DAC3);
   SPI.transfer16(command_DAC2);
   SPI.transfer16(command_DAC1);
   digitalWrite (SYNCpin, HIGH);
}
```

So at the beginning of the interrupt, the function set_channel_A_update_all should be called to update all channels. An interrupt counter to count the number of interrupts is necessary. Before the interrupt counter reaches the half value of the total number of interrupts for one specific frequency, the value given to DAC is high. When the interrupt counter reaches the half value, the value given to DAC will become low. Sometimes, the number of interrupts in one frequency is not an integer, so a function *round( )* is implemented. After running the codes, six LEDs can flicker in the given frequencies.

### 3.4.8. Algorithms for LCD Management

For the LCD management six LCDs and one Arduino Mega 2560 have been used. The LiquidCrystal library is implemented for controlling LCDs. The routine used in this project for initializing the connection between microcontroller and LCDs is as below:

```
LiquidCrystal lcd(rs, rw, enable, d0, d1, d2, d3, d4, d5, d6, d7)
```

Among all the parameters, rs and rw stand for Register Select Signal and Data Read/Write respectively. Enable pin is in the third parameter and d0 to d7 is the data bus line. After connecting LCDs with the corresponding pins on Arduino Mega 2560, the microcontroller can start controlling the information on LCDs. Routine *lcd.begin(cols, rows)* is used to initialize the interface between Arduino and the LCDs based on the LCD dimensions (maximum number of characters printed horizontally and vertically). Routine

*lcd.print(data)* is used to print text to LCDs. By connecting the enable pins with different pins on microcontroller and sharing the rest pins, the LCDs can display different text.

## 3.5.Serial Communication and Command Translation

### 3.5.1. Sending Commands from Matlab

After signal processing stage, the Matlab sends the associate commands through serial port and the commands are received by microcontroller. Then microcontroller can take actions and send commands to different devices.

First of all, the serial port needs to be established. The function *serial('PORT','P1',V1,'P2',V2,...)* is used for stablishing the port. It constructs a serial port object associated with the port and with the specifications. Then, connect the serial port object to the device using the *fopen* function. After the object is connected, the port is ready for receiving data from Matlab. So Matlab can use function *fwrite* to send the commands to serial port. The main loop for signal processing and command transmission is set to repeat every 0.1 second. In this 0.1 second the signal processing is done and large portion of the time is spent for transmitting the command, because Matlab does not handle the serial port efficiently. Although theoretically the command should be sent in less than 1 millisecond, Matlab needs up to 80 milliseconds (in rare occasions) to transmit the command. Fortunately, data can be written asynchronously with the *fwrite* function. The advantage of using asynchronous mode is that when the signal processing is in progress, the *fwrite* function can execute simultaneously. Since asynchronous mode is used, it is safe to conclude that 100 milliseconds interval is sufficient even for the rare occasions. Moreover considering that human reaction time is below 0.3 second, the speed of sending command is 3 times faster than human reaction time and this is sufficient for controlling wheelchair and other appliances. In order to stop the asynchronous write, Function *stopasync* is applied. It is used for stopping asynchronous write operation that is in progress with the device connected to serial port object. Meanwhile, function *fclose* and *delete* are implemented to close and delete the serial port.

### 3.5.2. Decrypting Communication

The movement of the wheelchair is controlled by the same microcontroller as the one controlling the LCDs. it also gets the commands from Matlab through serial port. Figure 35 shows the block diagram of this microcontroller (Arduino Mega 2560).

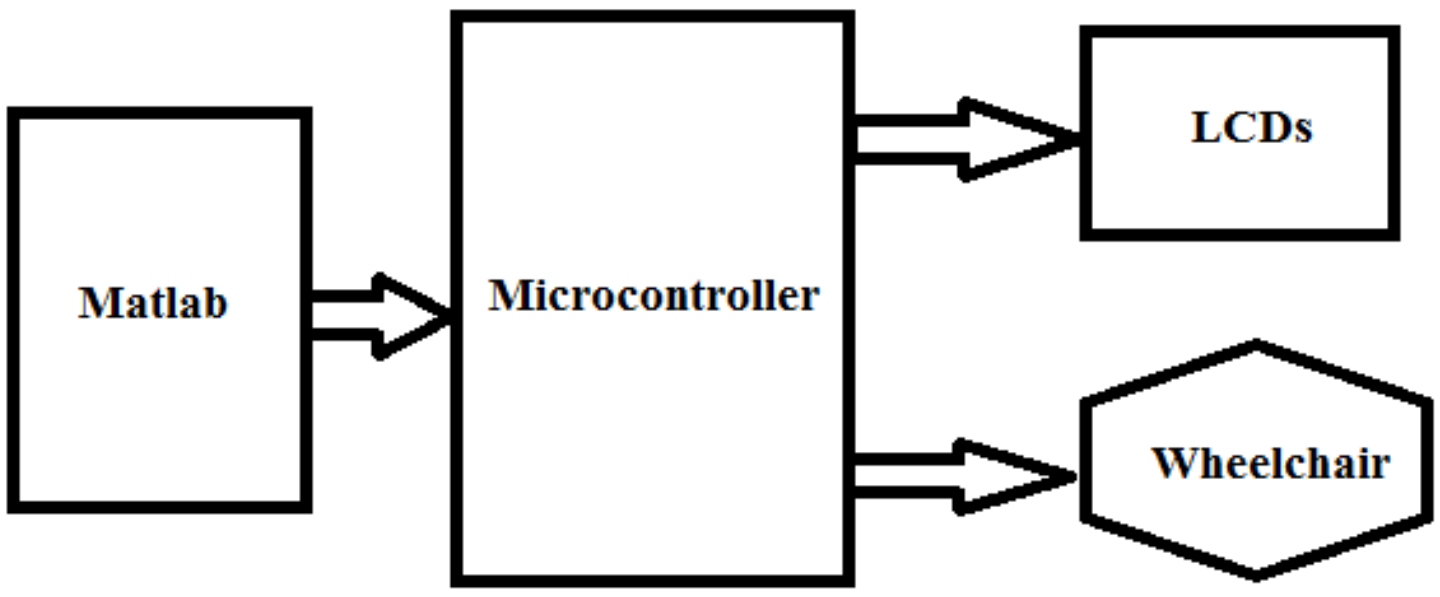

**Figure 35: The Complete Block Diagram of Arduino Mega 2560 Connected to the LCDs, Matlab and the Wheelchair**

Arduino Mega 2560 gets the command from serial port and start decrypting the receiving information. Function *available* gets the number of bytes available for reading from the serial port. Function *read* reads the first byte of incoming serial data. There are five data bytes in any command received from Matlab. Function *read* has been called five times in order to get the complete incoming serial data. Based on the *read* results, variable *EEG_command* will be given specific value through a [switch…case] structure. Both information on LCDs and the movement of wheelchair are decided by the value of *EEG_command*.

The electric wheelchair is originally controlled by a joystick. The joystick has four terminals which are connected to ground, +5 volts, forward/reverse control and left/right control. When the wheelchair is stable, the voltages of last two terminals are around 2.55 volts and 2.56 volts respectively. The voltage changes in these two terminals results in the wheelchair movements. For this project, a DAC, controlled by microcontroller is needed for generating the output voltage for these two joystick terminals, as a result of DAC outputs the wheelchair can be operated by EEG signal.

There are five different status associated with wheelchair, including staying stable, going forward, reversing, turning left and turning right. In order to control the wheelchair, the DAC need to be set to WTM mode using routine *SPI.transfer16(0x9000)*. In this mode the outputs of the DAC are updated independently. The forward/reverse movement is controlled by channel A and left/right movement is controlled by channel B. Channel A and channel B can be set to different values. For safety peruses a manual mode has been implemented where the wheelchair is controlled by push buttons. Both manual and brain signal mode use the same functions to set the channel value.

In order to easily select between manual and brain signal mode, a 3-pins radi switch is used in the circuit. The first pin and the third pin are connected to +5 volts and ground respectively. Depending on the position of the switch, two different types of controls can be selected. By changing the switch position the voltage on the second pin can change from high to low or vice versa depending on the initial position.

When the manually control mode is selected, the text on LCDs will be changed to "manually" simultaneously. There are four push buttons connected to four pins on microcontroller (Arduino Mega 2560). Each pin is assigned as an input with the internal pull-up resistor activated. When the button is pressed down, the corresponding pin on the microcontroller becomes low and the wheelchair moves towards the associated direction. The output of channel A and B have the specific values which result in moving towards associated direction. So the wheelchair can be manually operated by pressing and holding the desired button. When the brain signal mode is chosen, the wheelchair moves based on the *EEG_command* received from Matlab. The output of channel A and B changes, so that the wheelchair moves towards expected direction.

With the movement of wheelchair, the information on the LCDs is being changed simultaneously. When the specific direction is selected by the subject, the LCD corresponding to this direction shows a "selected" message on its second line. A new variable called *selection* is involved with the initial value zero. For example, if *EEG_command* is equal to 1, then the routine runs as below:

```
if (selection == 1)
{
}
else
{
lcd_1.setCursor(0,1);
lcd_1.print("selected");
lcd_2.clear();
lcd_3.clear();
lcd_4.clear();
lcd_4.print("right");
lcd_3.print("left");
lcd_2.print("backward");
selection = 1;
}
```

This means if the value of selection is equal to 1, then the previous *selection* and *EEG_command* values are equal to 1. So the texts on the LCDs are correct and do not need any changes. If the value of selection is not equal to 1, which means the previous values of *selection* and *EEG_command* value are not equal to each other. In this case, the texts on the LCDs are not correct and should be updated. Therefore, after clearing the LCDs' previous texts, function *setCursor* and *print* are used to set the position of the cursor and print the correct and new texts on the LCD. Finally, the value of *selection* is set to 1, so that the previous value of selection for next time becomes 1.

## 3.6.Summary

This chapter mainly introduces the algorithms and methods to realizing the SSVEP-based BCI control system. There are five sections in this chapter.

The first two sections show the overview of the achieved BCI system and the equipment used in this project. In this project three electrodes related to the area of visual cortical are used for acquiring the EEG signals.

Then the third sections, primarily focuses on explaining the algorithms employed to process the EEG signal in Matlab. Butterworth IIR bandpass filter with 5Hz low cutoff and 50Hz high cutoff frequency is implemented for preprocessing, so that the useless components, including DC level and extra unused frequencies, can be eliminated. In order to extract the useful information embedded in the filtered data, FFT is applied to get the power spectrum in frequency domain. Therefore, the fundamental frequency and harmonics can be used in the step of classification. Particularly, utilizing harmonics to classify the extracted vector can narrow the FFT window size. Besides, the delay of the system is mainly caused by the large window size. Therefore, the time delay can be improved by reducing the window size to 2 seconds.

The fourth section is about the methods of designing a SSVEP visual stimuli panel. First of all, two types of panel designs are proposed. In this project, the second design is utilized because of it is more convenient for controlling more applications using EEG signals. The panel consists of LEDs and LCDs. They are programmed by frequency generation circuit and LCD information circuit. Additionally, both two circuits need microcontroller. In the circuit for frequency generation, microcontroller controls the DAC using SPI. Meanwhile, Daisy Chain is used for the DAC serial interface so that three DACs can be controlled through one serial interface. As a result six LED units can flicker with different frequencies and colors. For the LCD management circuit, a special designed ribbon wire is used to manage the six LCDs and their information simultaneously.

Every part of the project must have a communication link to other parts, so the fifth section is focused on the serial communication and command translation. The communication between Matlab and microcontroller is achieved through serial port, so that Matlab can send commands to microcontroller. When microcontroller receives the commands, it starts decrypting the information inside the commands. Finally, the commands are translated to different actions. The wheelchair and LCD are both controlled through these commands, which are generated based on EEG signals. Therefore the wheelchair moves to the expected direction and LCDs display the corresponding information.

# 4. Result and Discussion

## 4.1.SSVEP Visual Stimuli Experiments

In order to find out the most efficient way to display SSVEP visual stimuli, different frequencies, colors, waveforms and harmonics are tested in this section.

### 4.1.1. Experiment Setup

In SSVEP visual stimuli experiments, four single color integrated LEDs in one package covered by a 5cm*5cm*5cm frosted paper box is used; this item is referred to as LED unit and shown Figure 36. Each LED unit is rated at 12V volt and the rated current for these units can be up to 60mA, however at that level of current the intensity is very strong. A switch is implemented for controlling the on/off of the LED unit as shown in Figure 37. A function generator GFG-8015G (shown in Figure 38) is connected with LED unit generating different waveforms with different frequencies. A four channel digital storage oscilloscope TDS 2024B (shown in Figure 39) is employed for measuring frequency, waveform shape and pk-pk voltage.

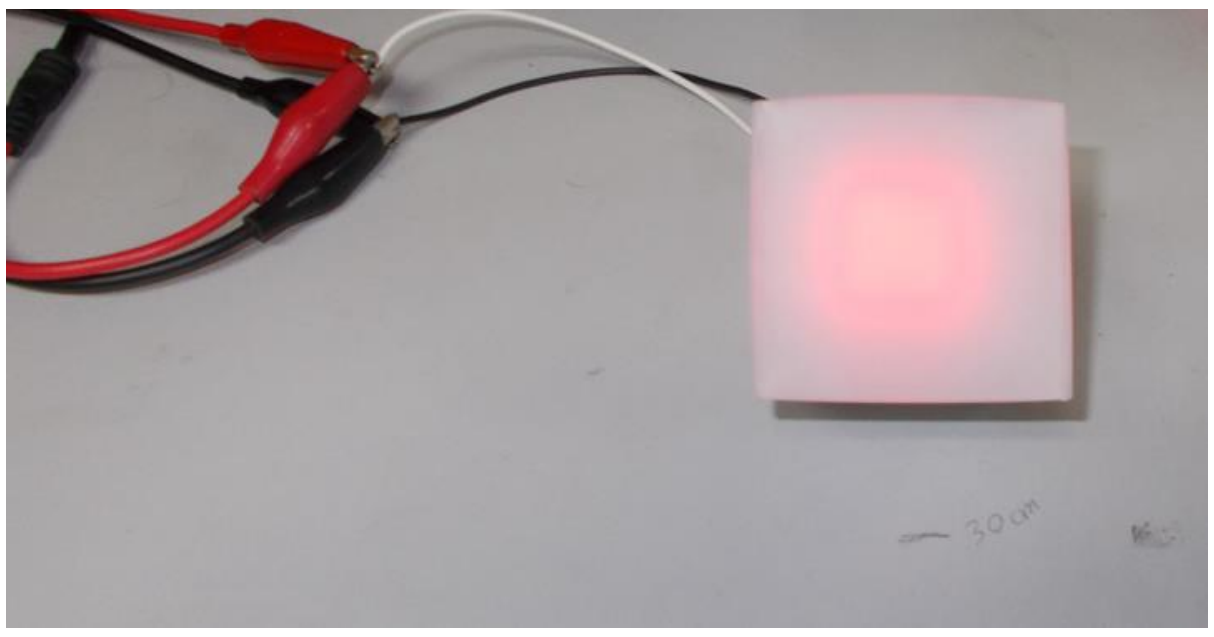

**Figure 36: Frosted Paper Box Covered LEDs**

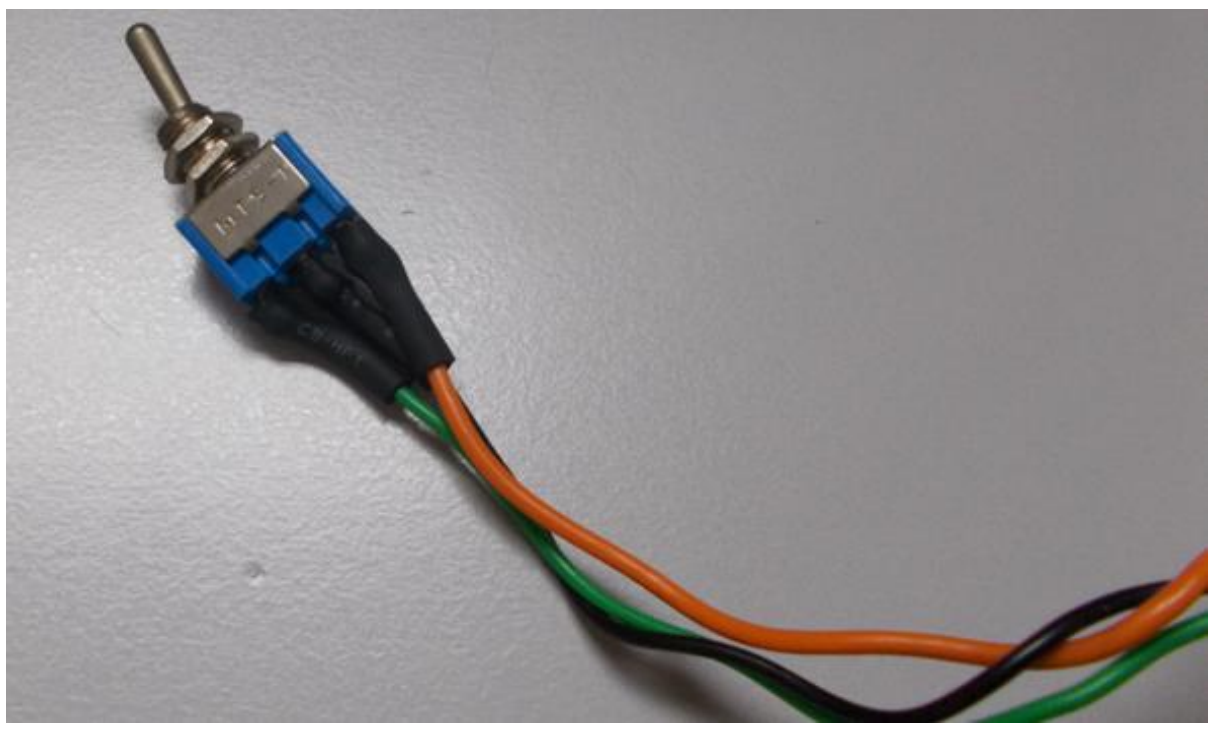

**Figure 37: Switch**

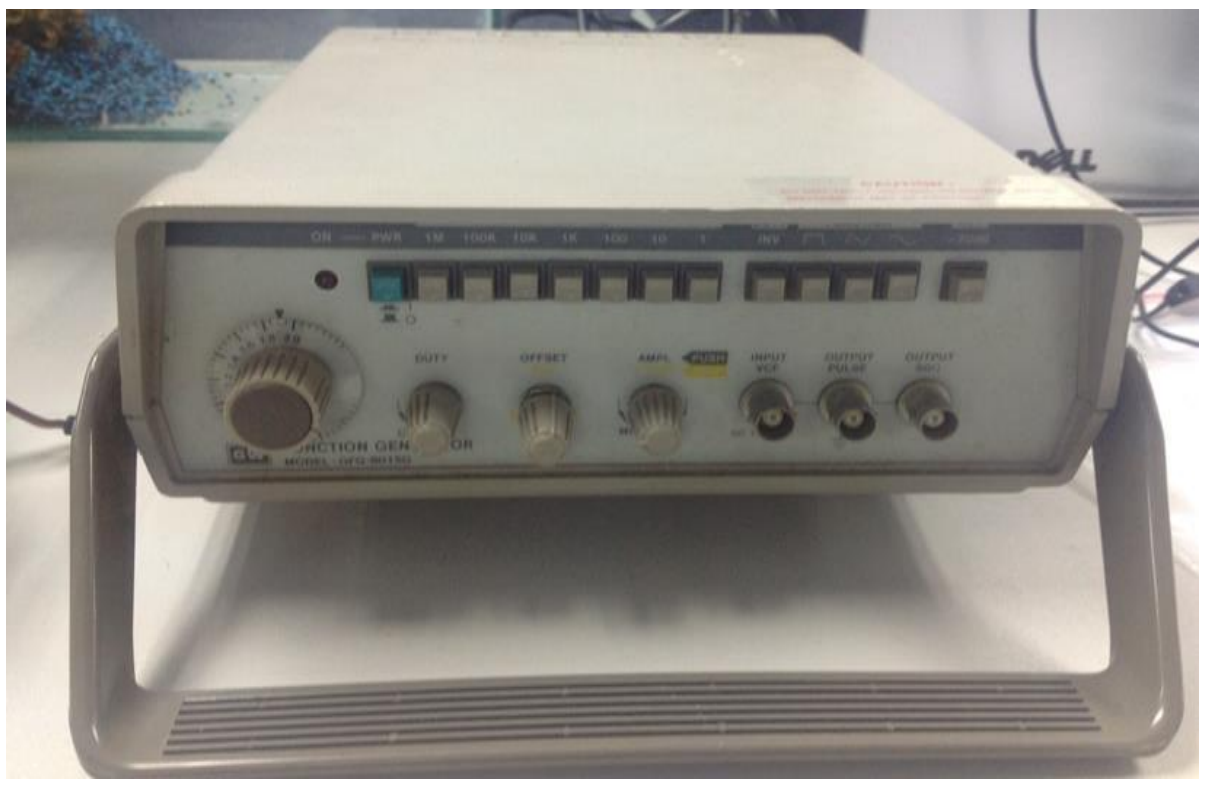

**Figure 38: Function Generator**

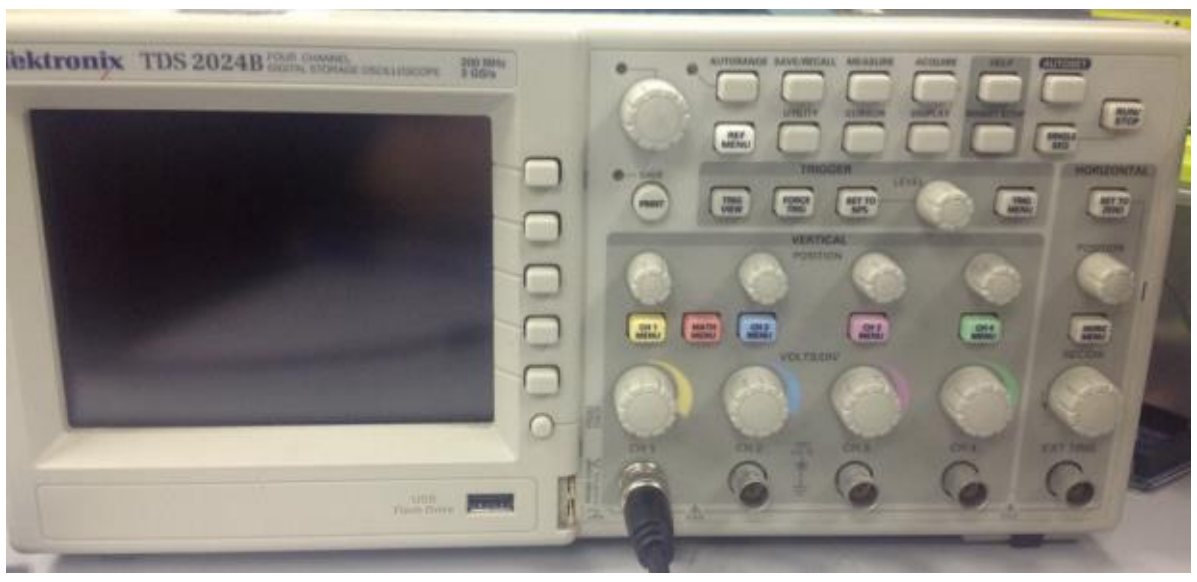

**Figure 39: Oscilloscope**

Meanwhile, a photodiode circuit is designed (shown in Figure 40) for measuring the output waveform shape of LEDs. Therefore it is guaranteed that the waveform shape of the flicking LEDs have the exact same waveform shape as the electrical signal generated by the function generator.

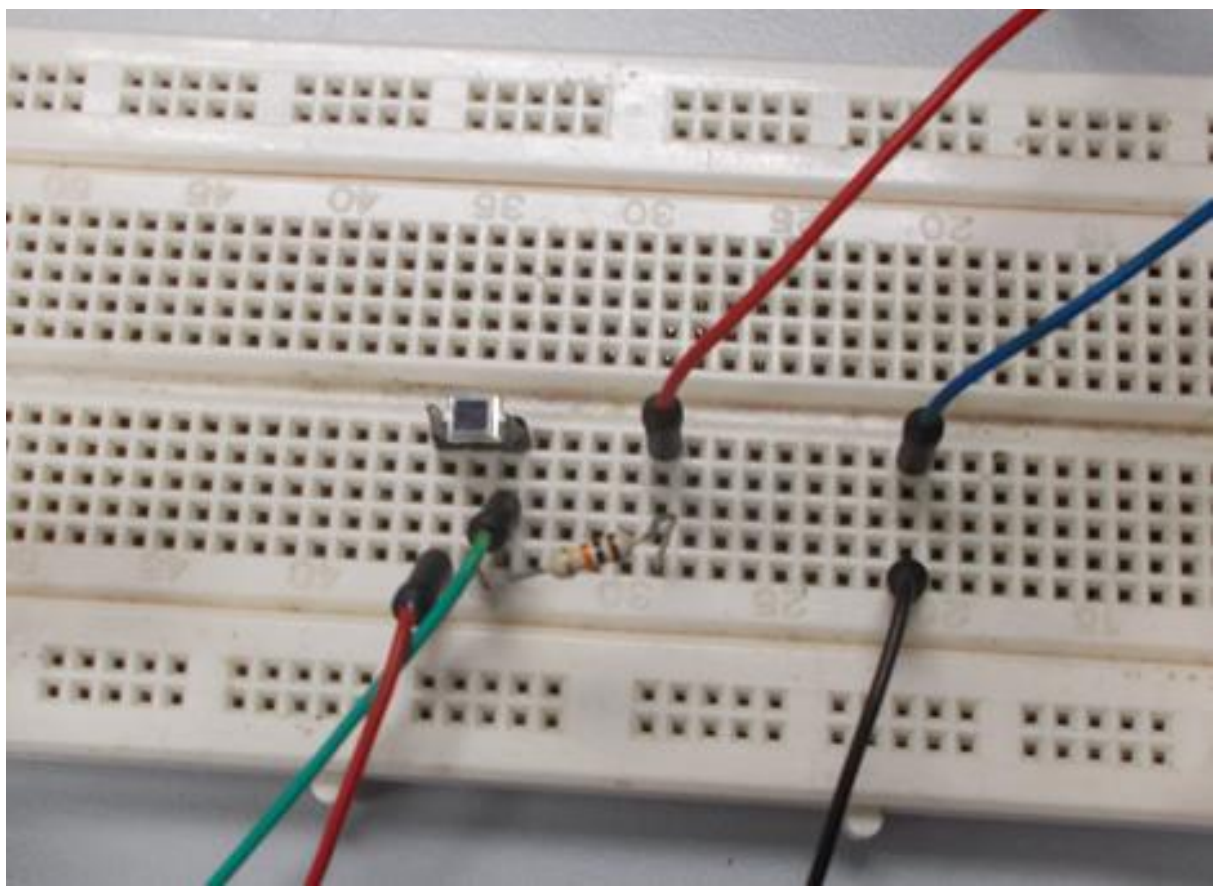

**Figure 40: Photodiode Circuit**

In order to avoid interference, the room is fully dark when the photodiode experiment starts. The photodiode was placed over the surface of the LED box in this test. A typical result is shown in Figure 41, Figure 42, Figure 43 and Figure 44, where yellow signal represents the signal from waveform generator and the blue signal shows the signal received from the photodiode.

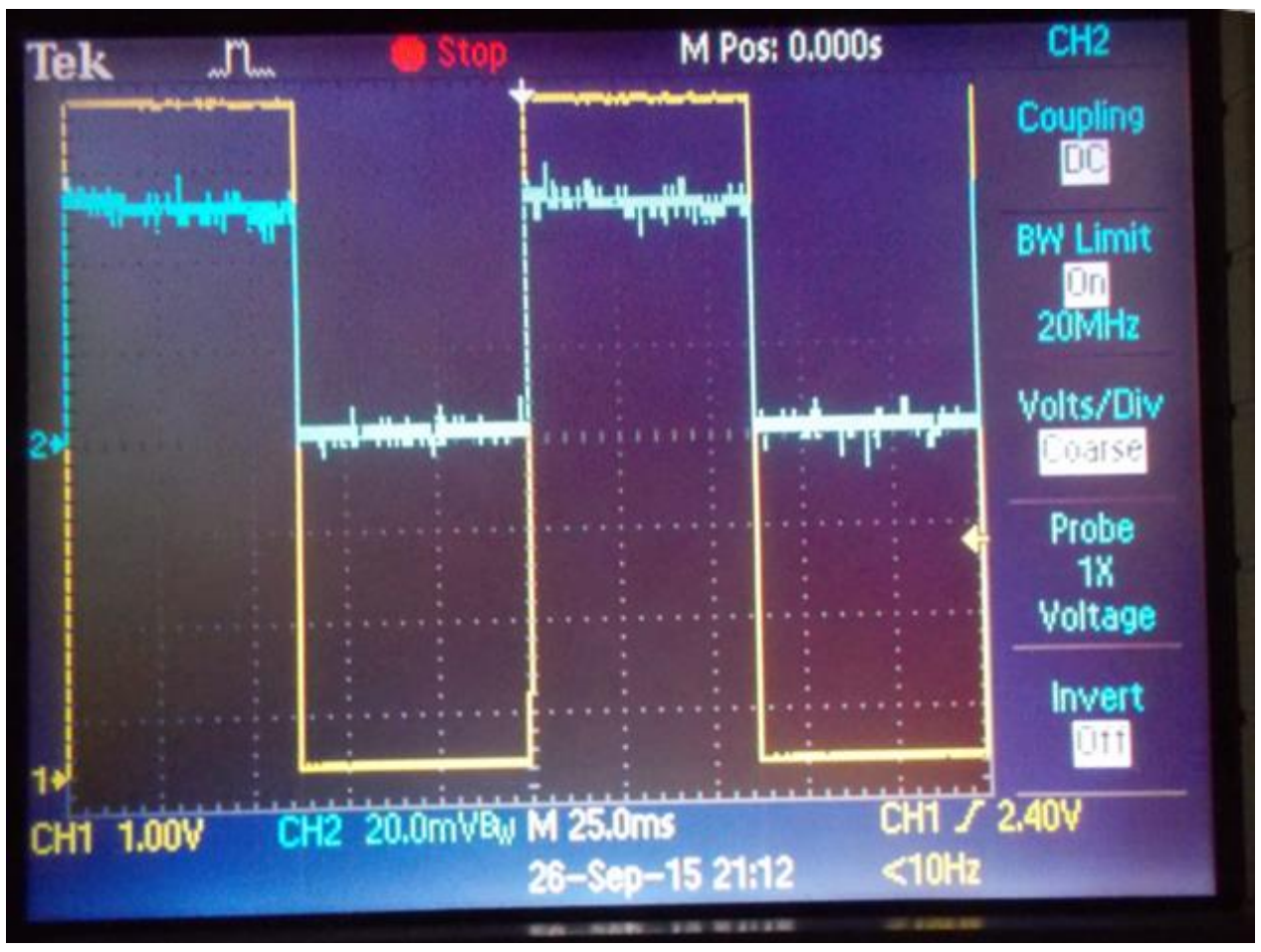

**Figure 41: Square Wave**

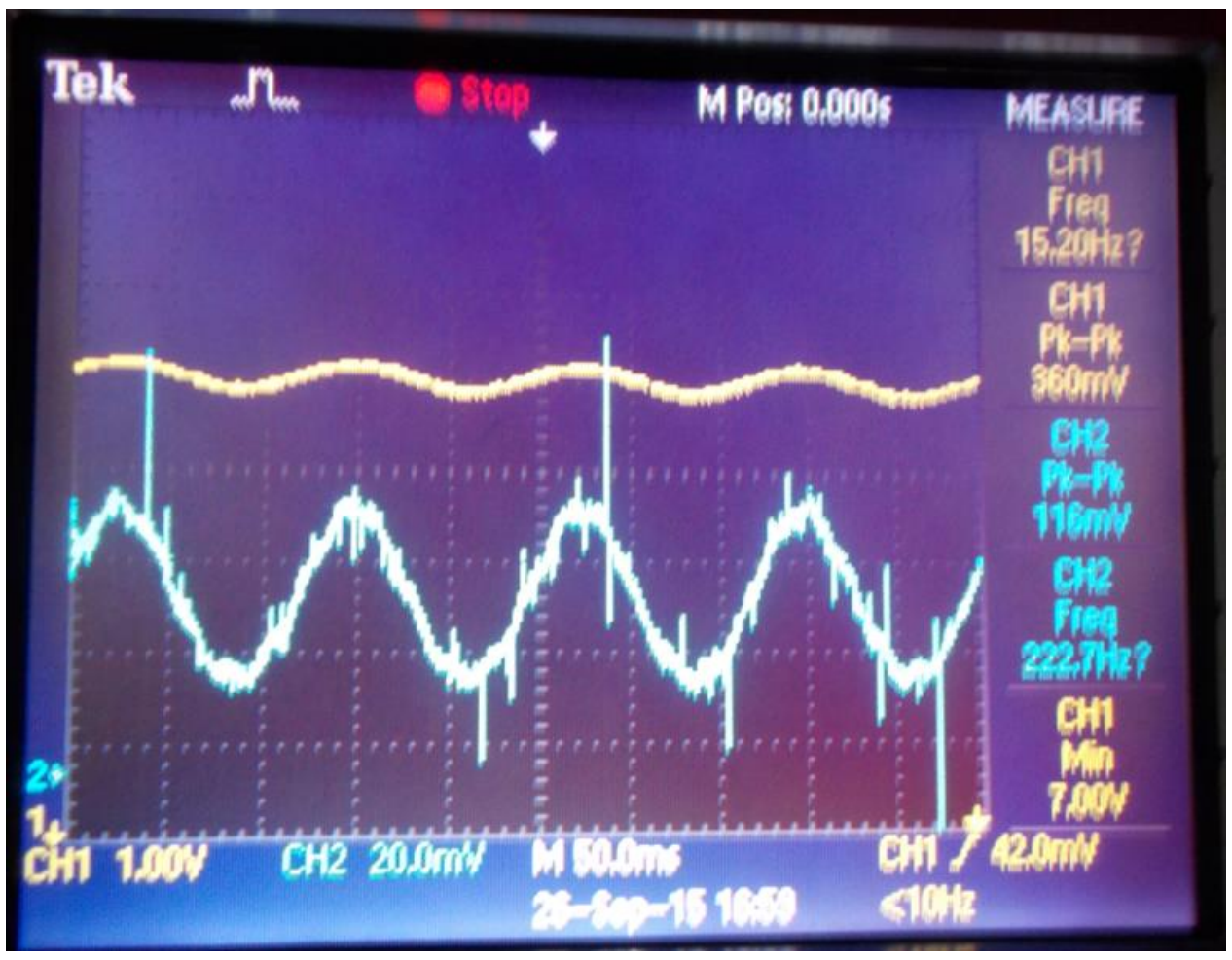

**Figure 42: Sine Wave**

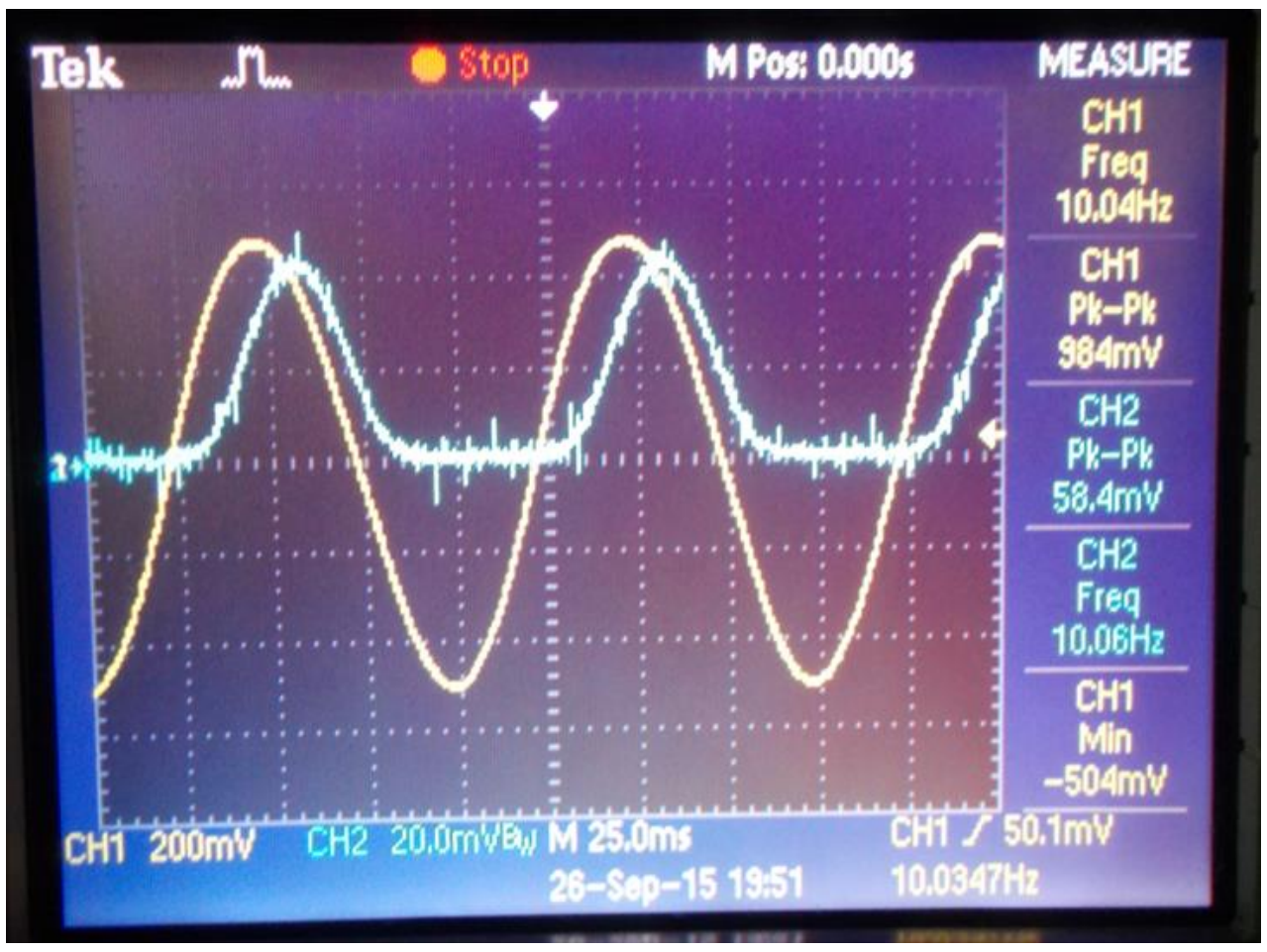

**Figure 43: Half Sine Wave**

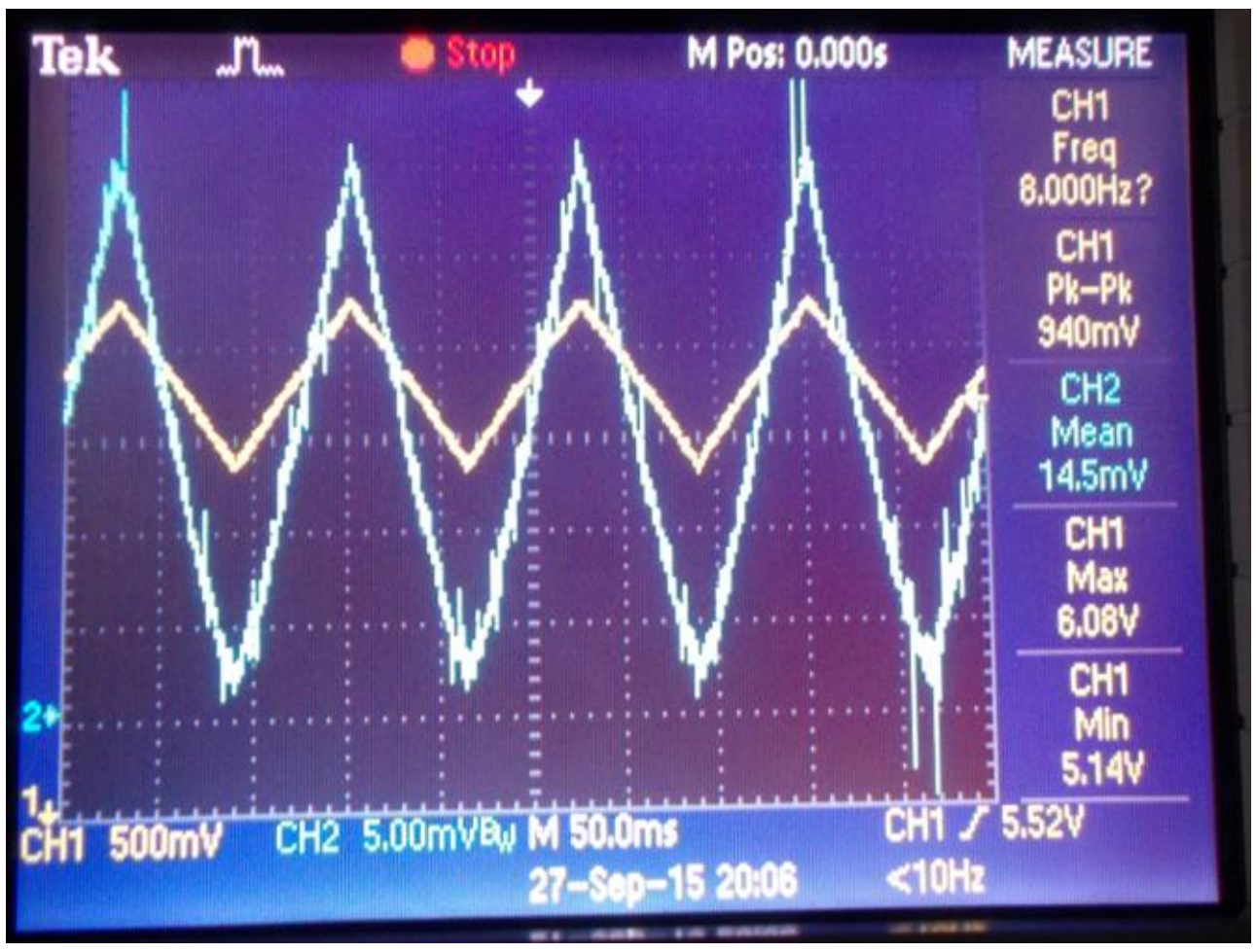

**Figure 44: Triangular Wave**

EEG signal acquisition and processing are performed in ASA-LAB™ software. EEG data are recorded with the sampling frequency of 256 Hz for 40 seconds in each trial. There are 10 seconds at the beginning and 10 seconds at the end for the subject to rest. 20 seconds in the middle is the testing period when the subject focuses on flickering LEDs. After getting the EEG data, they are edited in the ASA-LAb software. The useful 20 seconds EEG data are filtered by a bandpass filter with low cutoff frequency of 3 Hz and a high cutoff frequency of 60 Hz. FFT with the window size of 4 seconds is applied for the filtered EEG data. The FFT power spectrum is analyzed in the exported Excel files.

#### 4.1.2. Frequency Experiments

In order to verify the categories of SSVEP frequency result from WANG Yijun et al., red color LEDs with the frequency from 5 to 70 Hz are tested on one subject. The amplitudes from the FFT result corresponding to each testing frequency are gathered in Figure 45, Figure 46 and Figure 47 (these figures are presenting O1, Oz and O2 respectively). X-axis in these figures is stimulation frequency and Y-axis is amplitude response in frequency spectrum.

The results from electrodes O1 and O2 are very clear and they illustrate three different frequency regions, which are the low region from 6 to 12 Hz, the medium region from 13 to 22 Hz and the high region from 23 to 49 Hz. The result from electrode Oz is not similar to O1 and O2, therefore conclusion cannot be drawn. By comparing the results in Figure 45 to Regan’s results and the results from WANG Yijun et al.; it can be concluded that the results obtained in this experiment are closer to results reported by Regan.

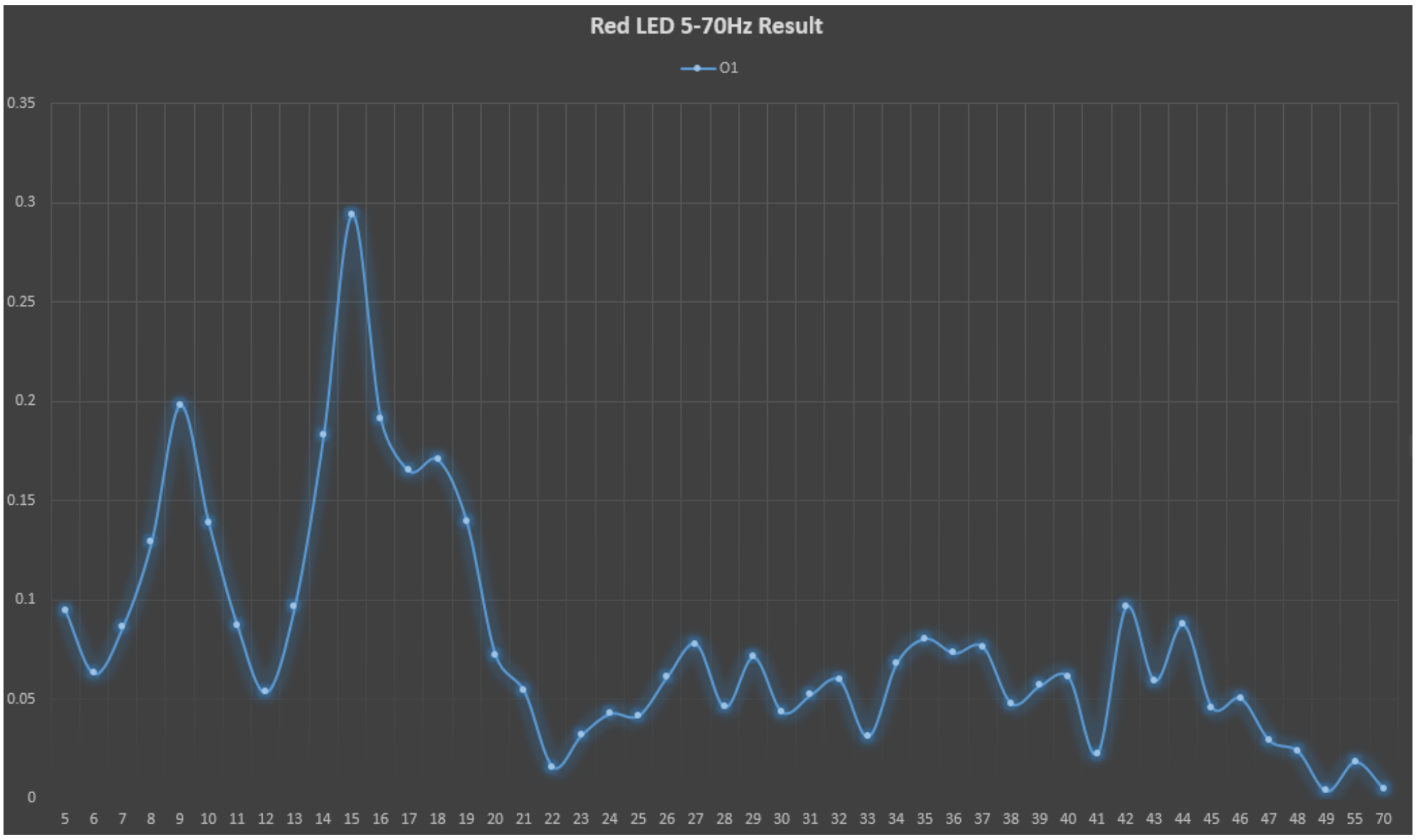

**Figure 45: The Result from Electrode O1**

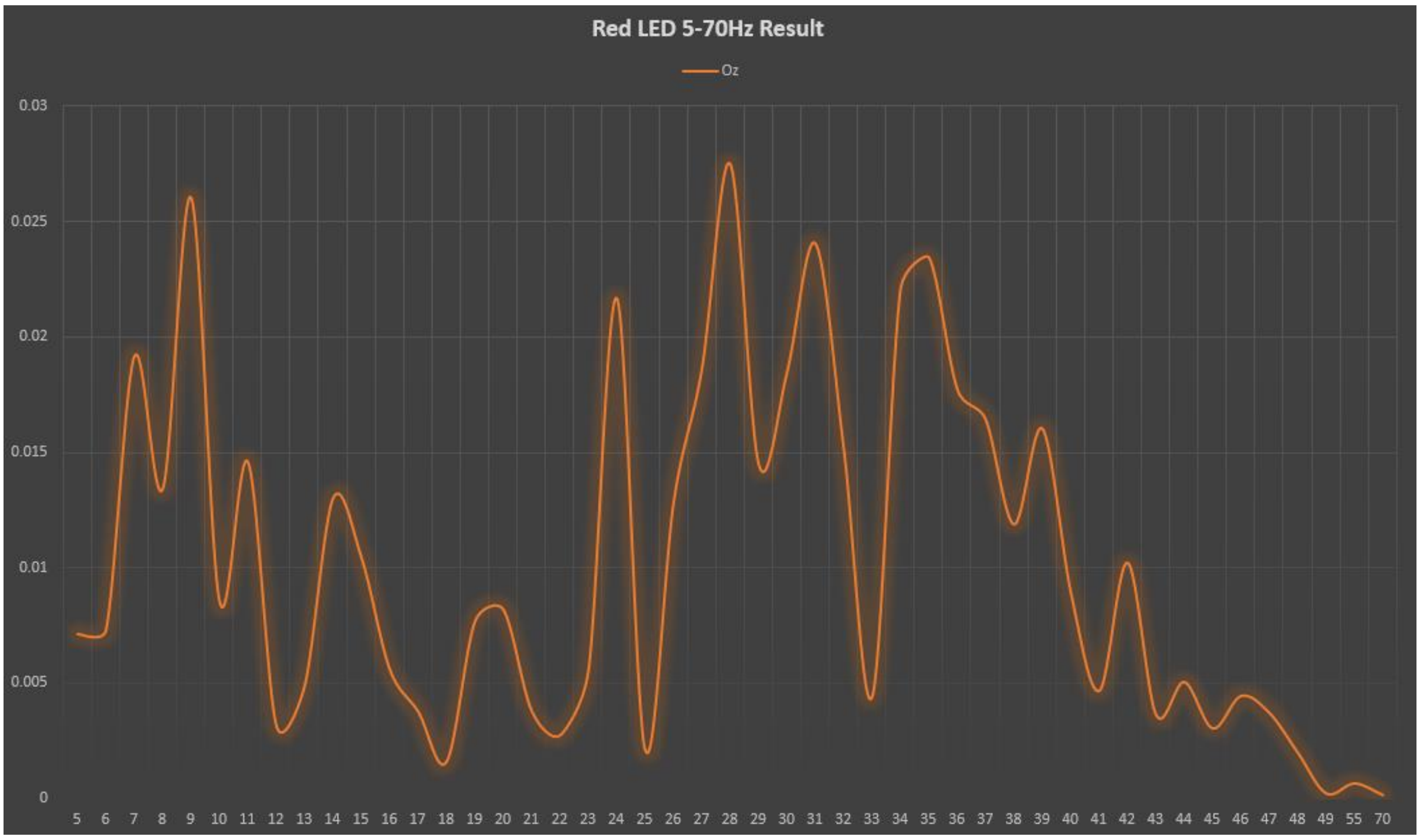

**Figure 46: The Result from Electrode Oz**

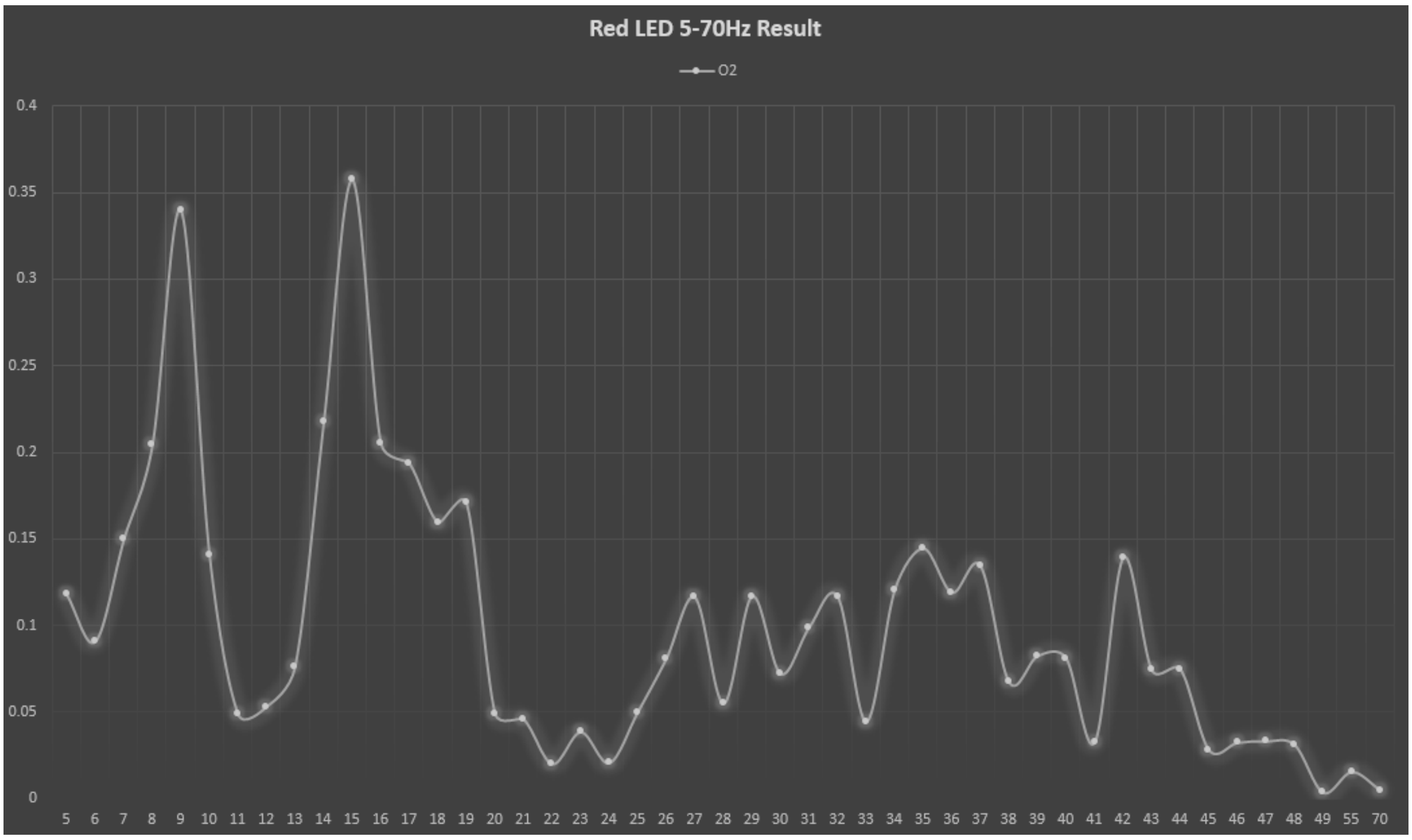

**Figure 47: The Result from Electrode O2**

### 4.1.3. Color Experiments

Red and white are the two different colors for studying the effects of colors in SSVEP. 4 subjects have been tested with these colors in frequency range of 8 to 12 Hz.

### 4.1.4. Waveforms Experiments

Waveform is one of the factors in SSVEP, which suffer from lacked of researches and studies. In the waveforms experiments, four volunteer subjects for the experiments have been tested with square wave, sine wave, half sine wave, triangular wave, ramp wave with 0% duty cycle and ramp wave with 100% duty cycle. Two different comparisons have been carried out to compare the response of each electrode to the same waveform and also to compare the same electrode's performance in different waveforms and frequencies.

Figure 48 shows that the same waveform performs differently on each electrode. The amplitudes on Oz are always lower than the other two. The most suitable frequency varies for different electrodes while the waveform remains the same. The performance of each of the waveform is totally different, which means waveforms with the same luminance will have different SSVEP performance. This result also implies that based on a specific waveform, the specific electrode can be used for better SSVEP result.

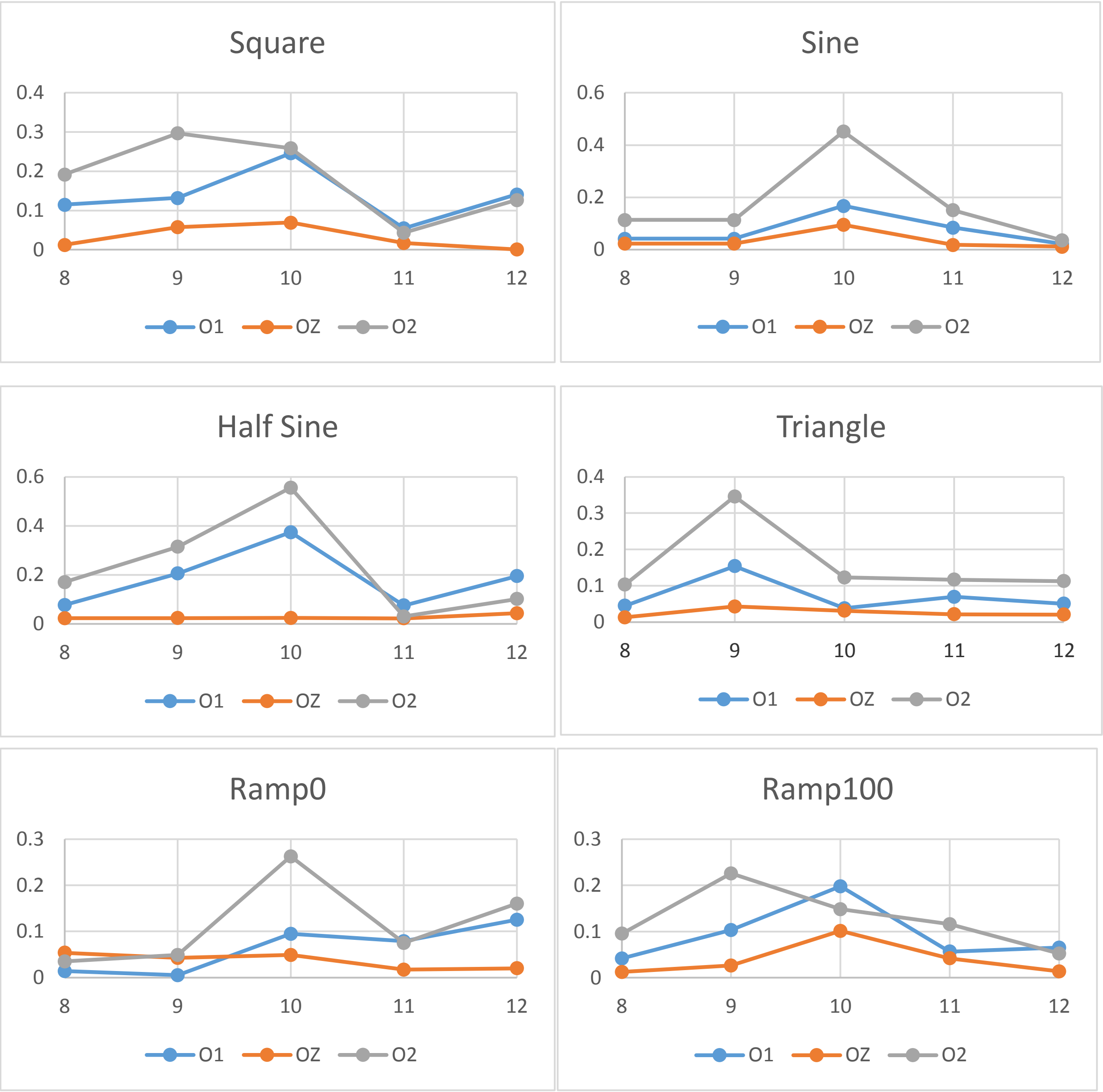

**Figure 48: The Performance of the Same Waveform on Different Electrodes for One of the Subjects**

Figure 49 shows the different waveforms performing differently in the same electrode. Most of the waveforms have the best performance frequency of 10 Hz. On the electrodes O1 and O2, half sine wave has the highest amplitude; while on the electrode Oz, ramp wave with 100% duty cycle has the highest amplitude. This implies that some electrodes work better with some waveform and this can be used to get better SSVEP responses.

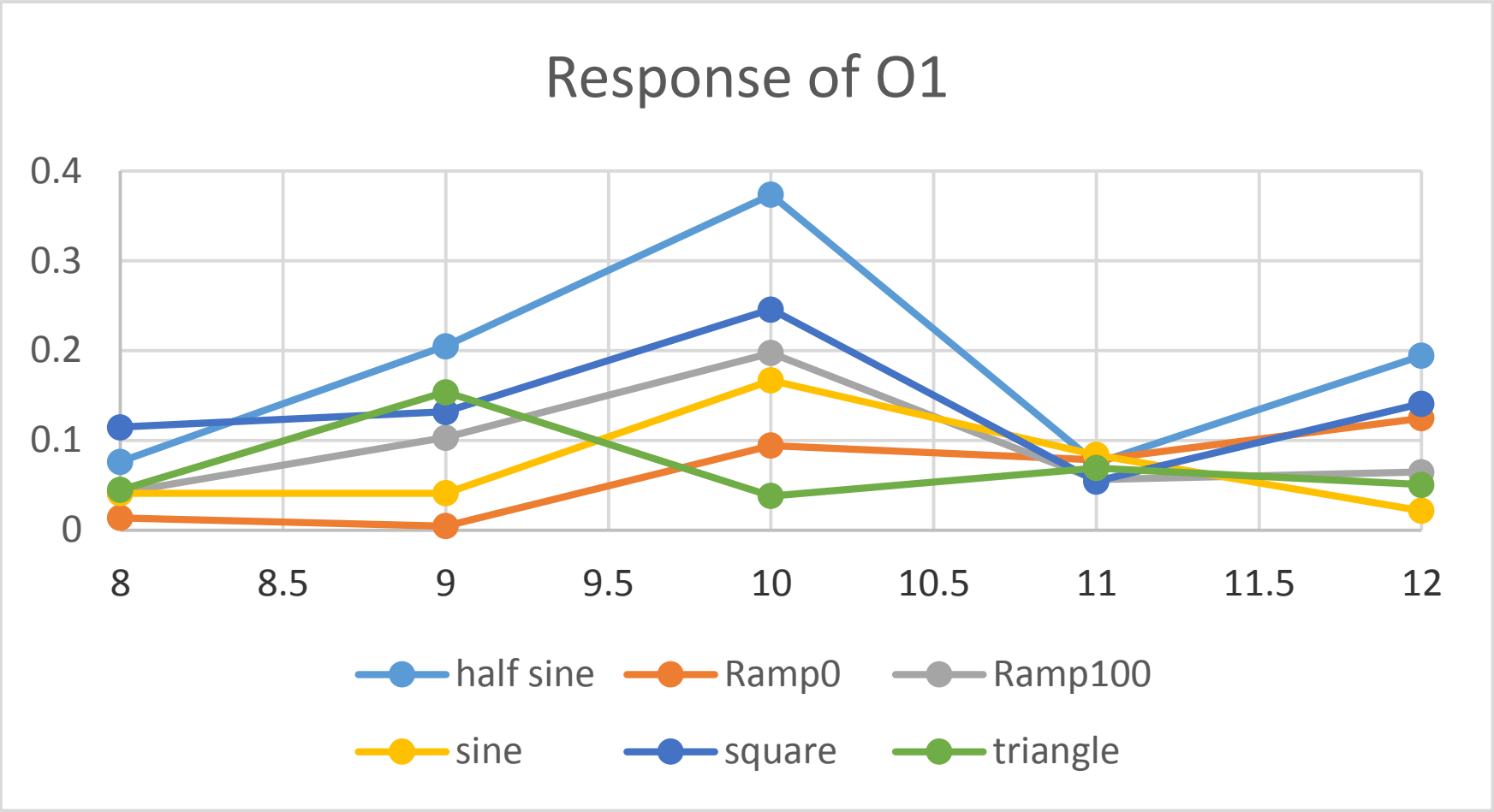

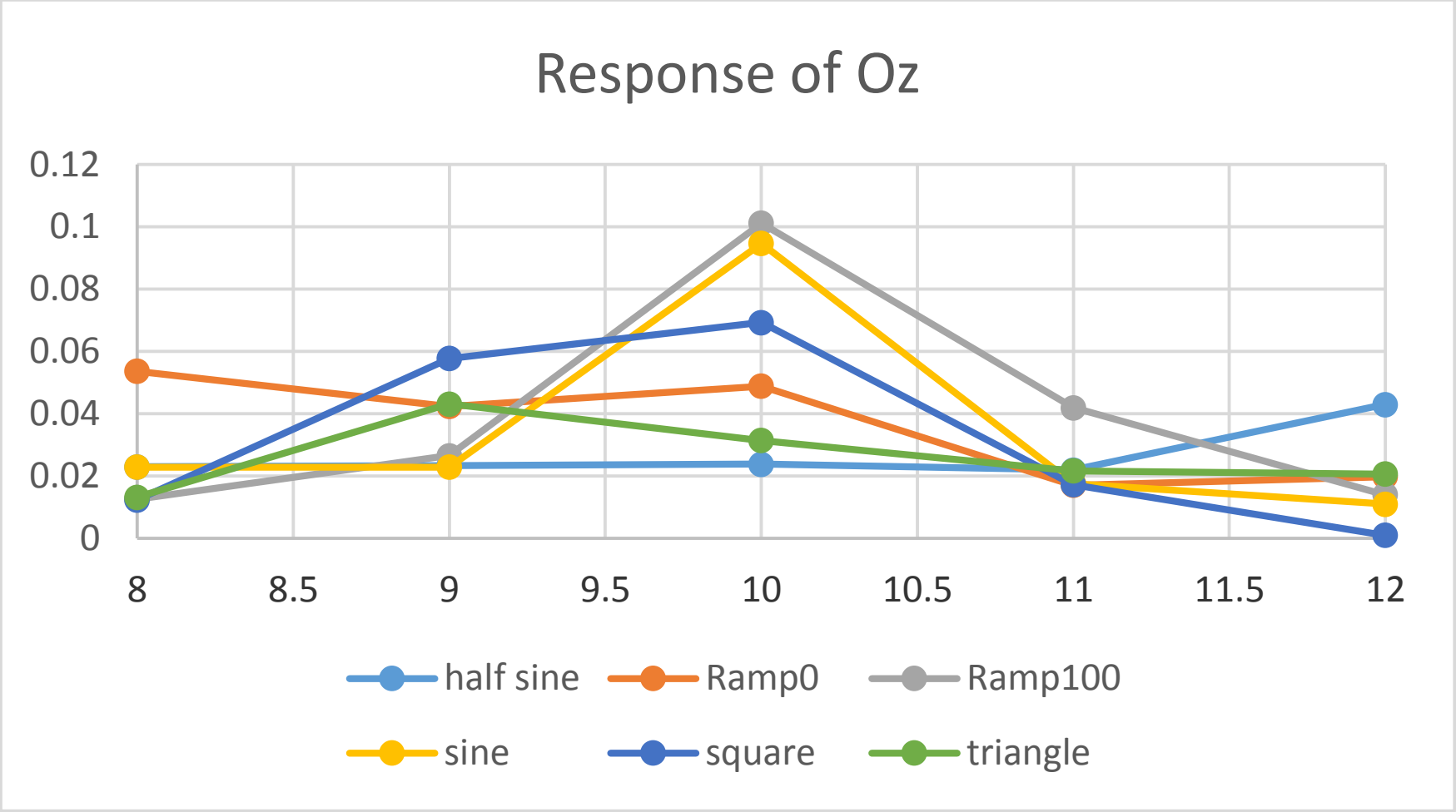

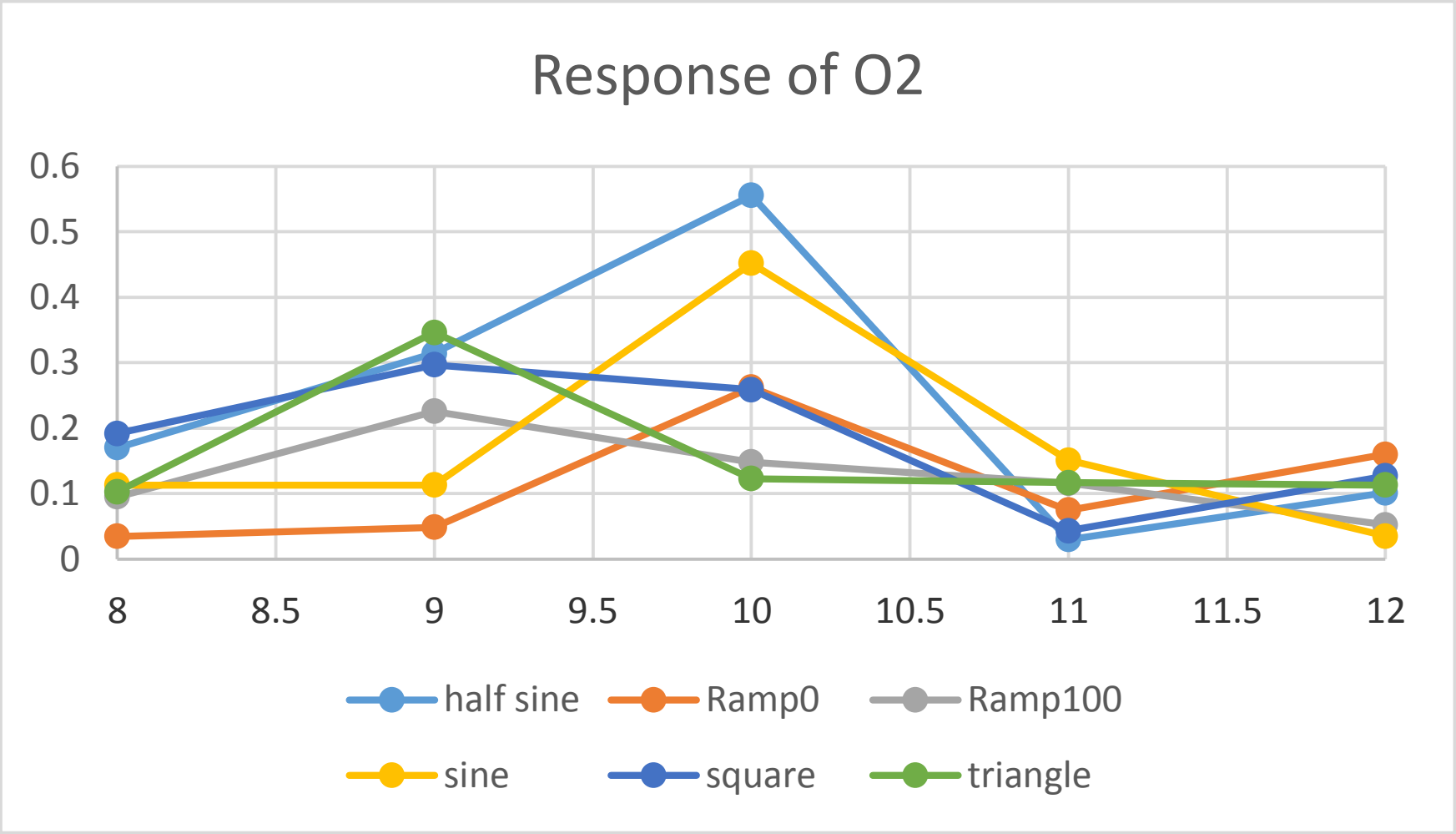

**Figure 49: The Performance of Different Waveforms on the Same Electrode for One of the Subjects**

### 4.1.5. Harmonics Experiments

Harmonics can be elicited in human brain when the fundamental frequency is stable.

Through analyzing the data from different subjects, it is found that all the waveforms used in the experiments can elicit different numbers of harmonics.

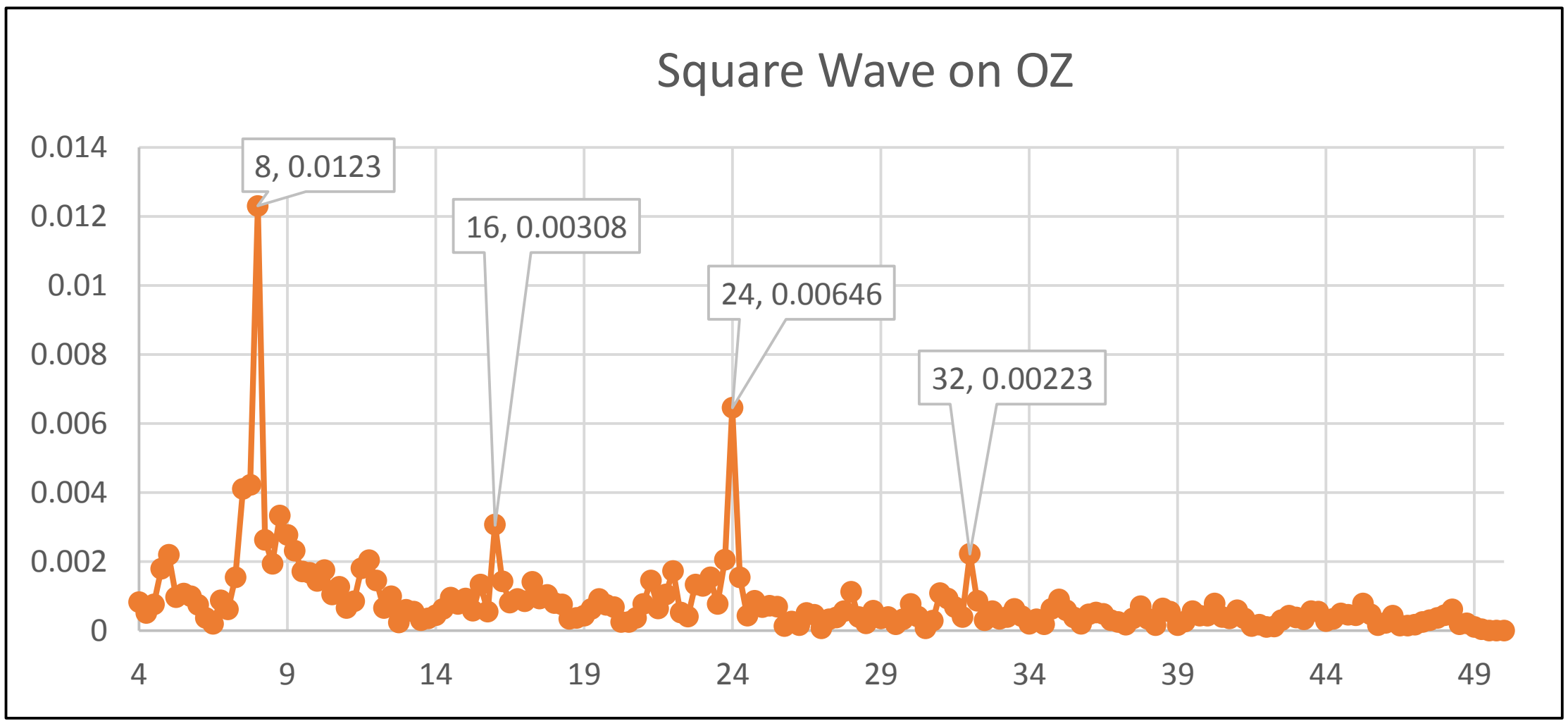

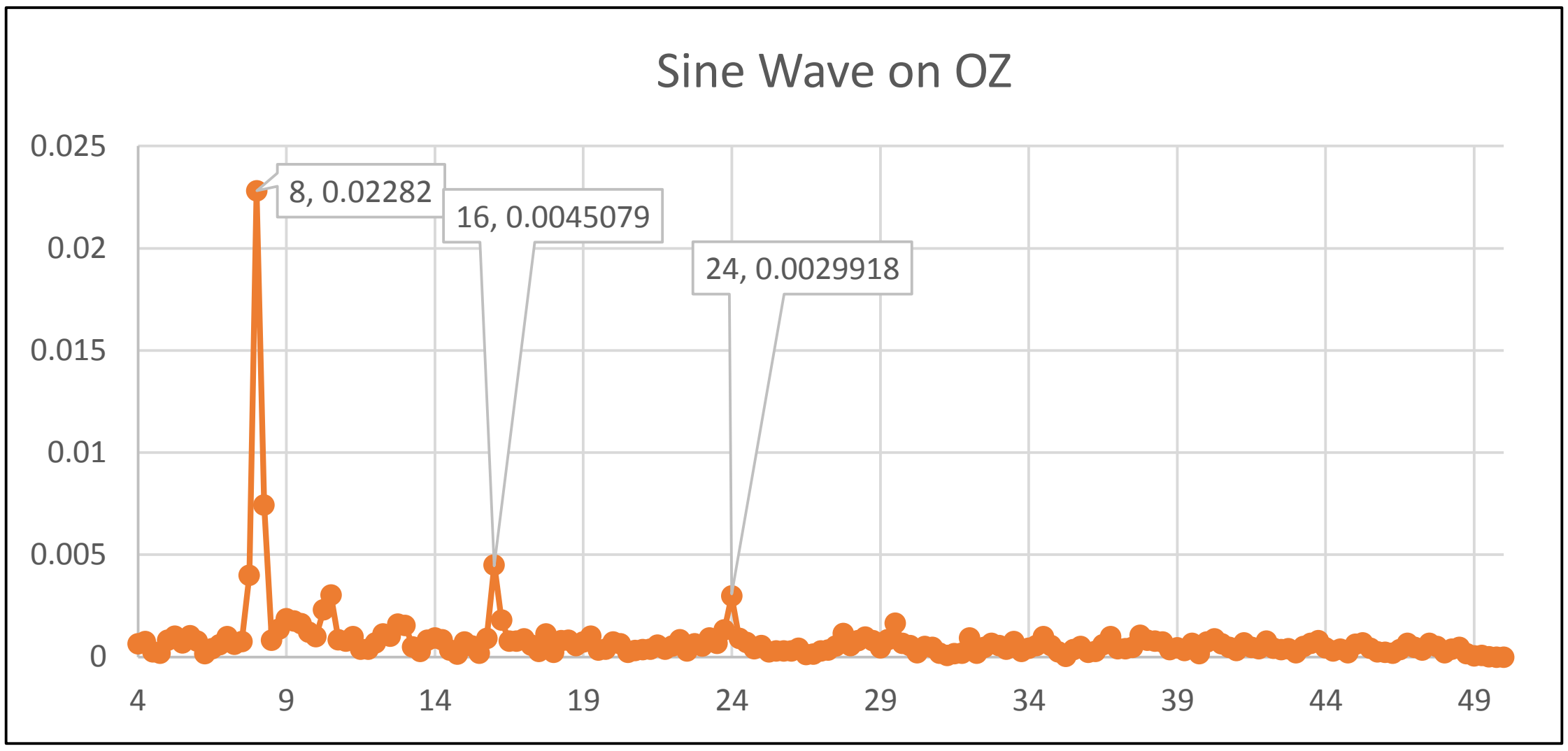

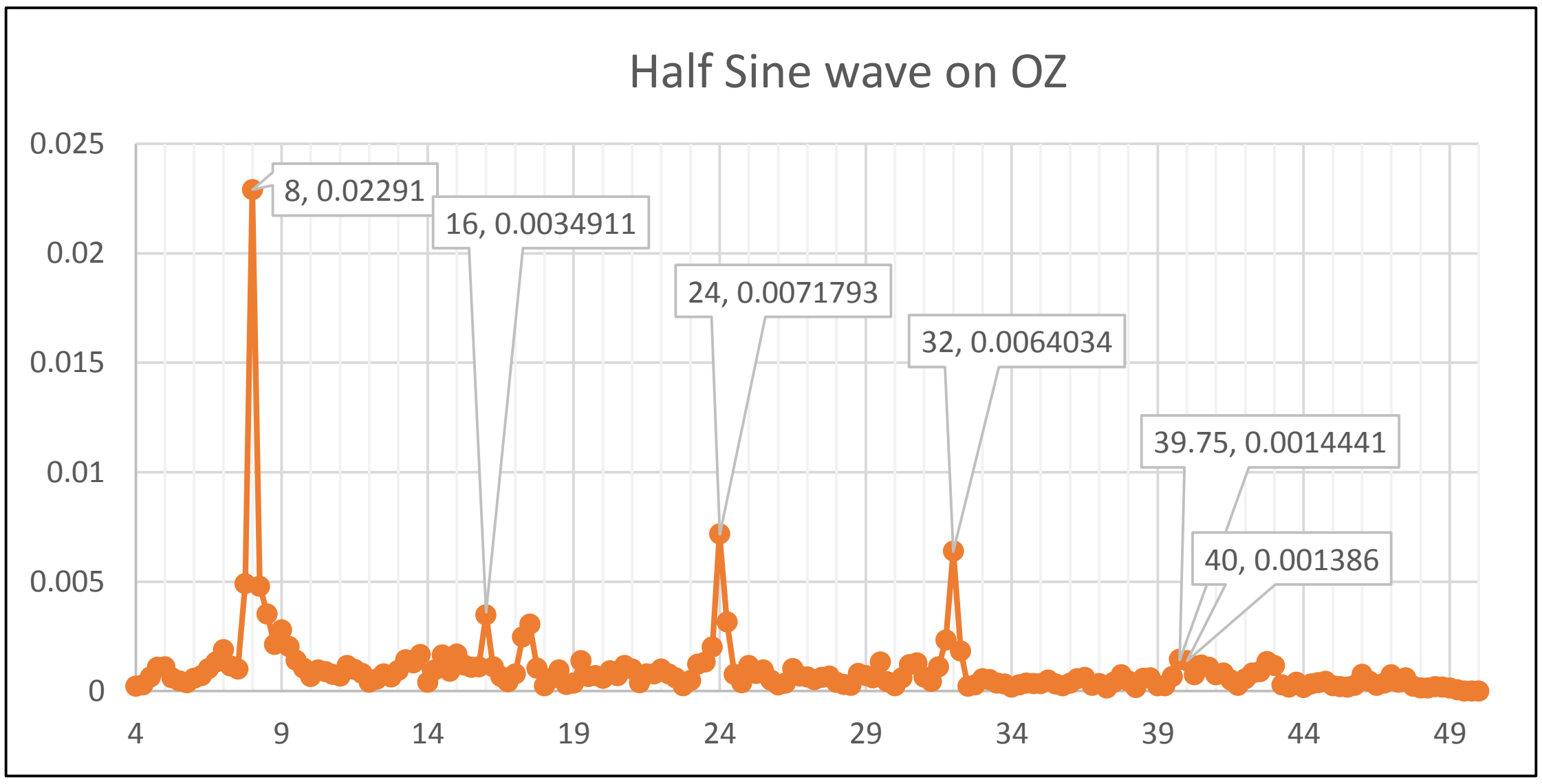

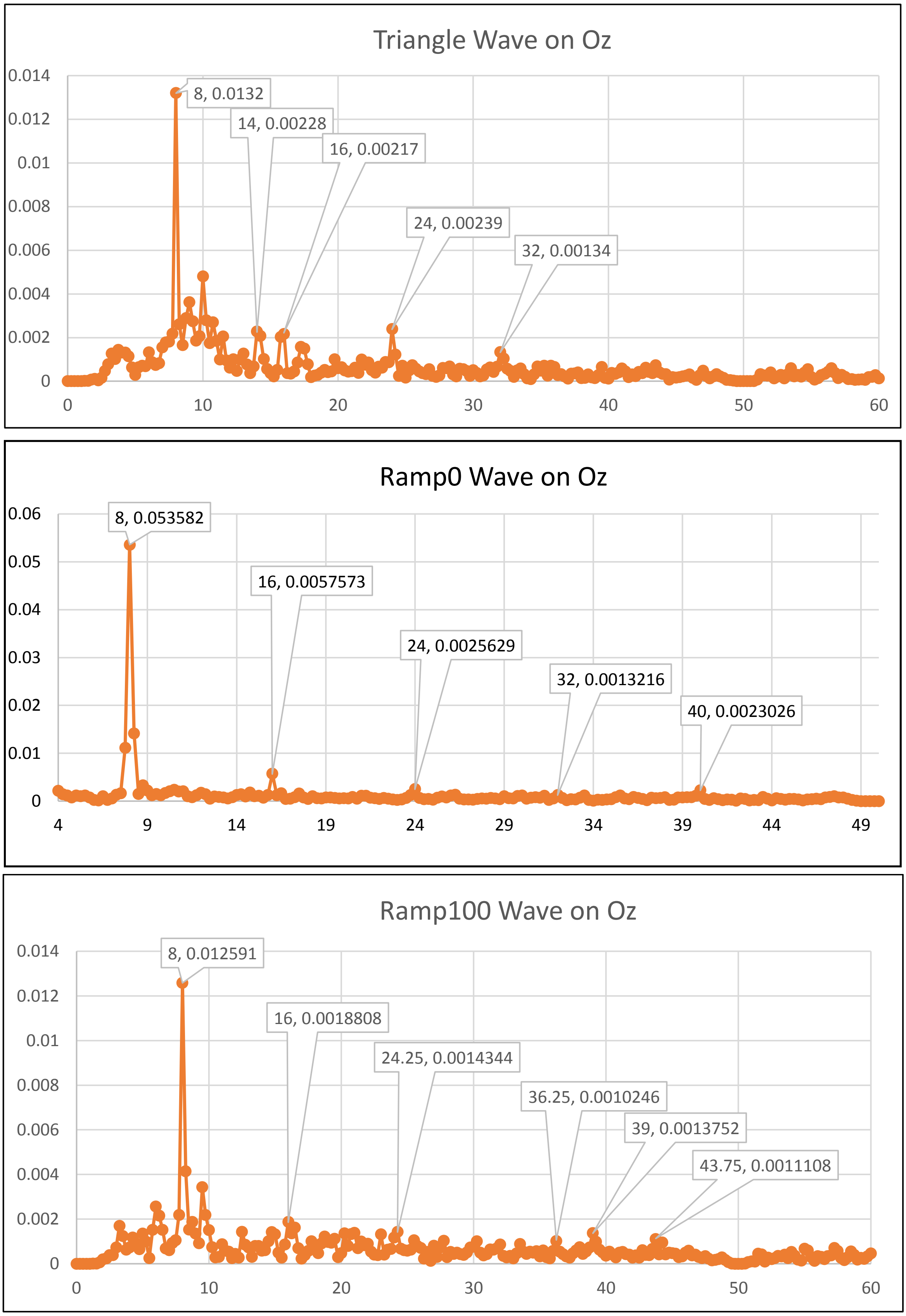

**Figure 50: Harmonics in Different Waveforms at 8Hz**

Figure 50 shows the fundamental frequency of 8 Hz and its harmonics corresponding to different waveforms.

## 4.2.SSVEP Visual Stimuli Panel

Based on the result of SSVEP visual stimuli experiments, it can be concluded that different factors influence the responses of SSVEP in different ways and these influences are not the same for different individuals, However some patterns are observed in the result but they are not helpful for the BCI system. Therefore, a SSVEP visual stimuli panel is made using multiple color LEDs so that different colors and waveforms can be displayed based on the subject's performance and preference with SSVEP visual stimuli.

### 4.2.1. Design of Adjustable Current Source

As shown in Figure 51, Figure 52, Figure 53 and Figure 54, four circuits have been designed to work as an adjustable current source. The experimental components for these circuits are microcontroller (Arduino Mega 2560), DAC TLC7226, op amp LM324, MOSFET IRF521, transistor 2N4401 and 50 Ω resistors.

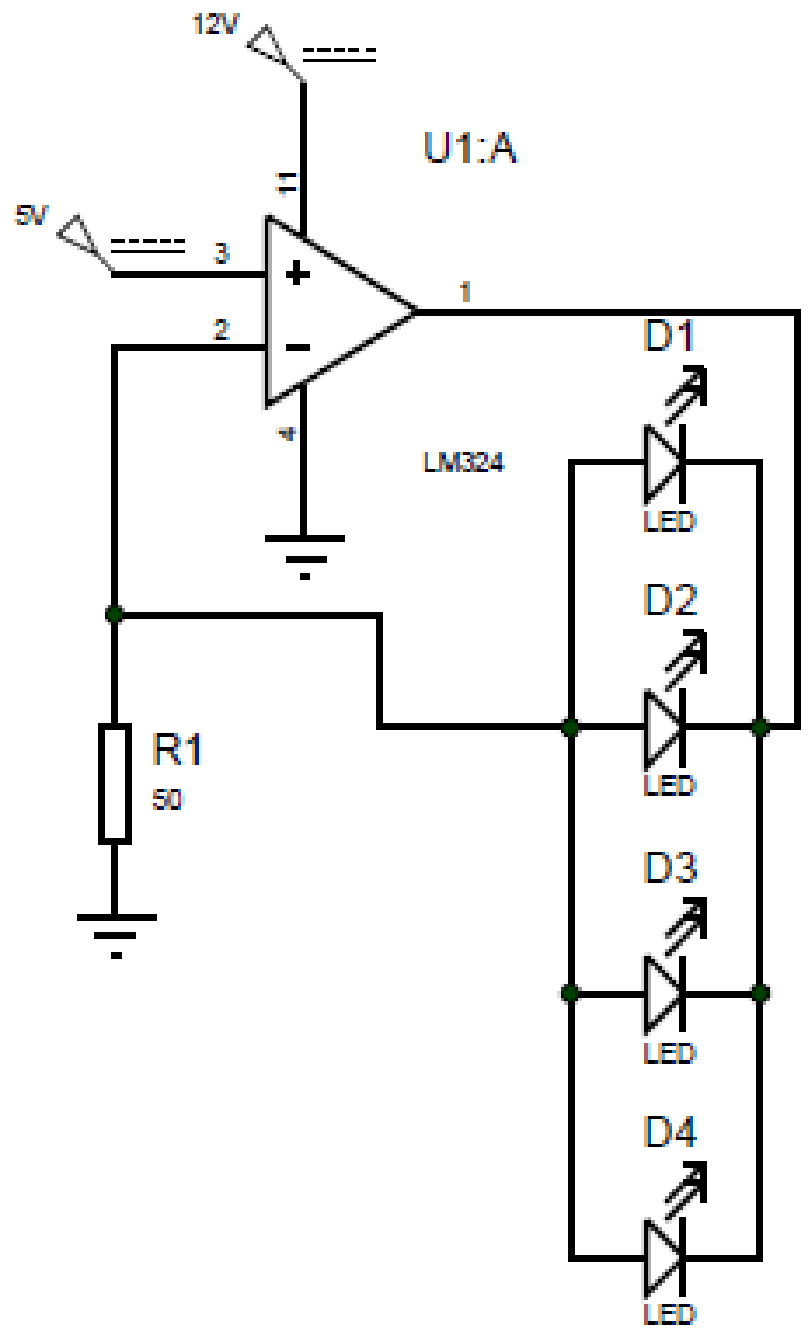

**Figure 51: First Design**

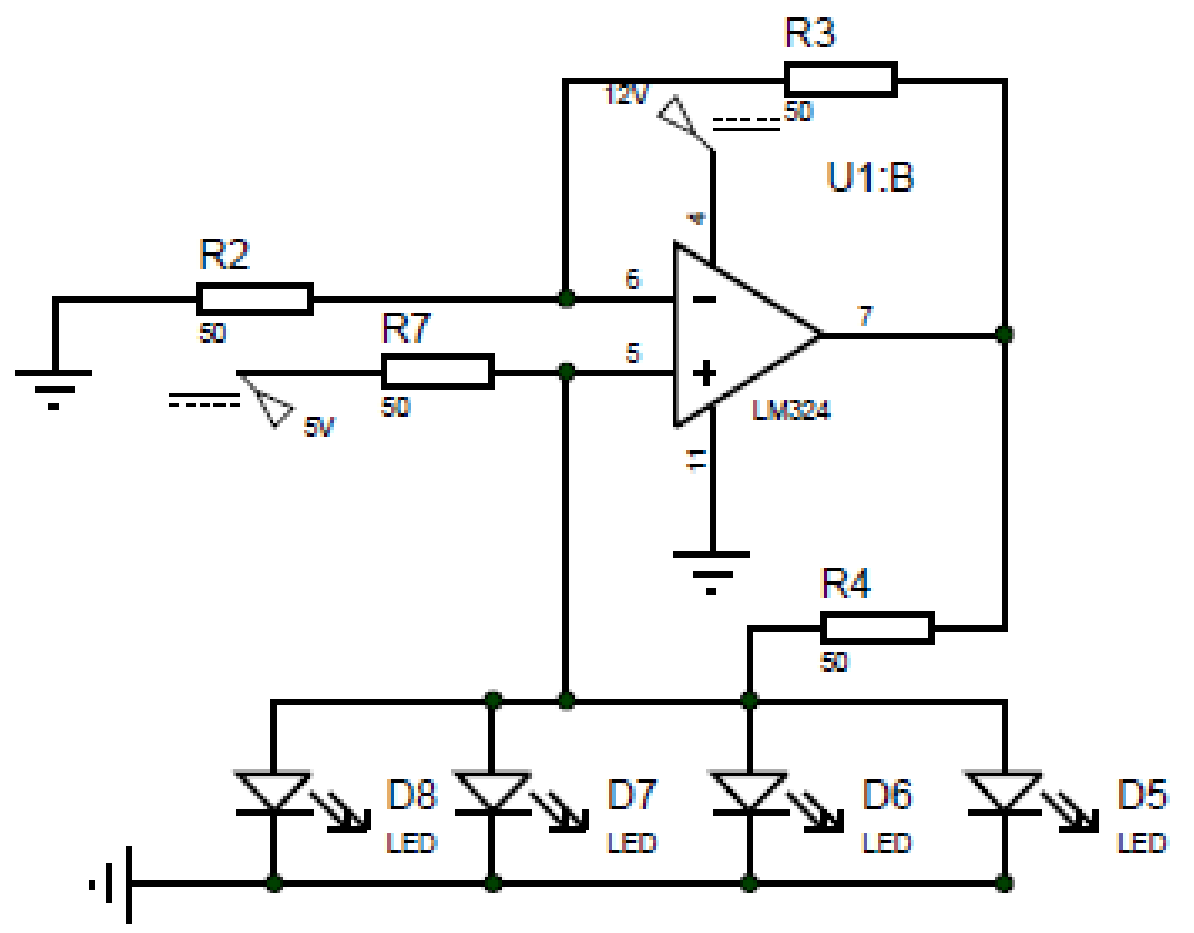

**Figure 52: Second Design**

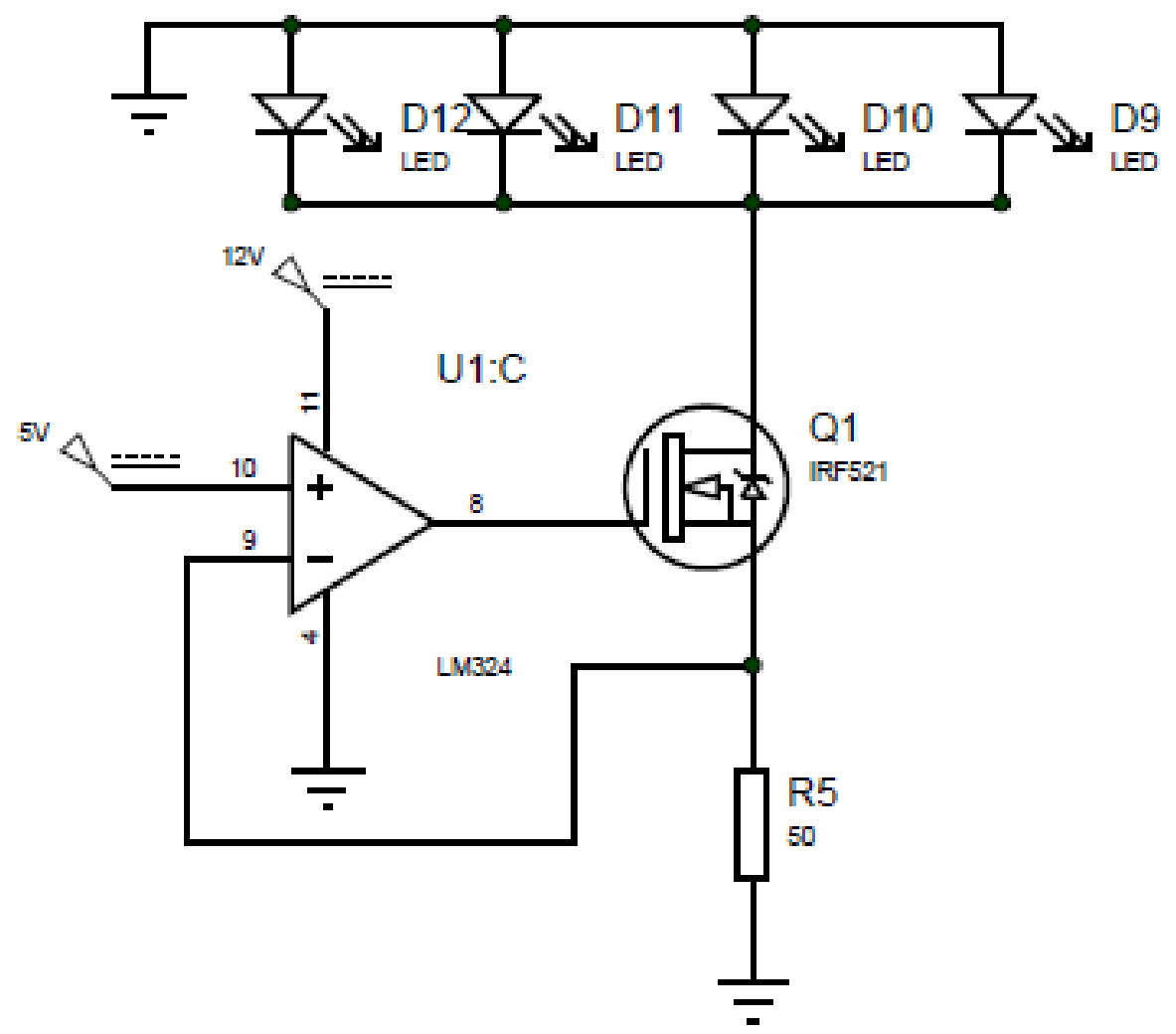

**Figure 53: Third Design**

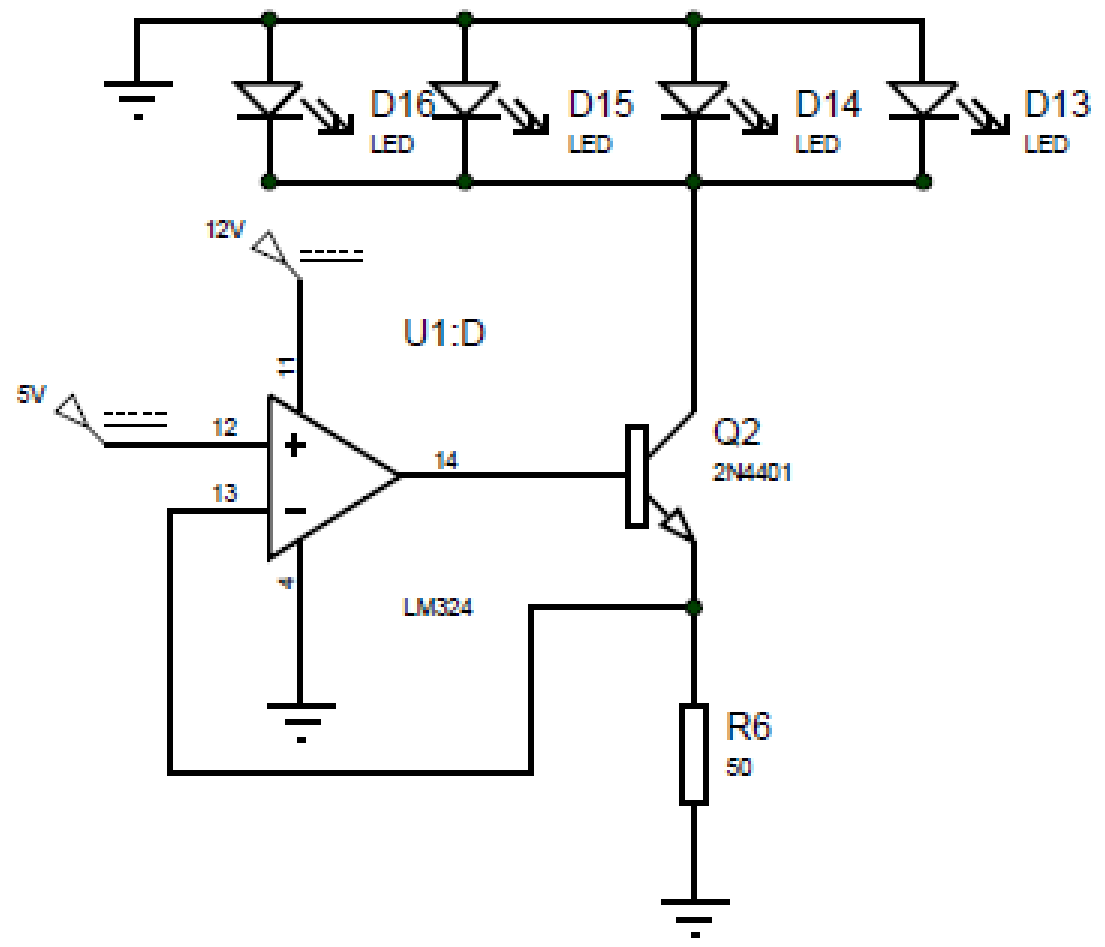

**Figure 54: Forth Design**

The problem with the second design is that the current in LEDs cannot go to zero, which means that the LEDs will never be fully off. The third and the forth designs are quite similar; the only difference is that the second design uses MOSFET, while the third design uses transistor. After testing the outputs for third and fourth design, it was noticed that these two designs do not have a stable output and it was observed that transistor circuit has more fluctuation than the MOSFET circuit. The main reason for the fluctuations is the nonlinearity of LEDs. Only the first design is capable of producing zero current when the input of the op amp is 0V. Moreover it is can produce stable current when the input voltage value is constant.

The calculation for the first design is shown as below:

$$P_D = (V_{cc} - V_{out})I_{out}$$

$$P_D = \left(V_{cc} - (V_{in} + V_f)\right)I_{out}$$

$$\frac{dP_D}{dV_{in}} = \frac{d(\frac{V_{cc}V_{in}}{R})}{dV_{in}} - \frac{d(\frac{{V_{in}}^2}{R})}{dV_{in}} - \frac{d(\frac{V_f V_{in}}{R})}{dV_{in}}$$

$$\frac{dP_D}{dV_{in}} = \frac{V_{cc}}{R} - \frac{2V_{in}}{R} - \frac{V_f}{R}$$

In order to find the maximum $V_{in}$, let $\frac{dP_D}{dV_{in}} = 0$, $\frac{V_{cc}}{R} - \frac{2V_{in}}{R} - \frac{V_f}{R} = 0$.

So, $V_{cc} - 2V_{in} - V_f = 0$, among them, $V_{cc} = 12V$, $V_{fmin} = 2.1$V. Thus, $V_{in} = 4.95V$.

Therefore, the maximum value of $P_D$ is $P_{Dmax} = \left(12 - (4.95 + 2.1)\right) \times \frac{5}{47} \approx 0.53W$.

In the calculation, $P_D$ is the power dissipation in the op amp, $V_{cc}$ is the power supply, $V_{in}$ and $V_{out}$ are the input and output of the op amp respectively, $I_{out}$ is output current, $V_f$ is the forward voltage of LEDs, $R$ is the power resister. According to the op amp's data sheet, thermal resistance junction-to-air $R_{\Theta JA}$ is $137^oC/W$, and junction-to-case $R_{\Theta JC}$ is $23^oC/W$. Thus, the junction temperature is $0.53 \times 137 + 23 = 95.61^oC$. The highest junction temperature must not exceed +$150^oC$. Since $95.61 < 150$, this op amp is suitable for this application.

### 4.2.2. PCB Design

Circuit experiments on bread board are very necessary. By doing the circuit experiments, the real-time circuit can be guaranteed to work. However, LED, DAC and voltage regulator are SMD components which means that their pins cannot fit on bread board directly. The way to solve this problem is to make them on small piece of PCBs with headers. Figure 55 shows the PCB design in 2D and 3D using Altium Designer. Altium Designer by default shows the top layer in red and bottom layer in blue in 2D view. The top layer in 3D vie is shown on the left while the bottom layer in 3D view is shown on the right. There are four LEDs on each PCB. The PCB is designed in double layers. Same color channels are connected paralleled with the same ground. So there are six pins in the headers of the PCB, including three color channels with their ground.

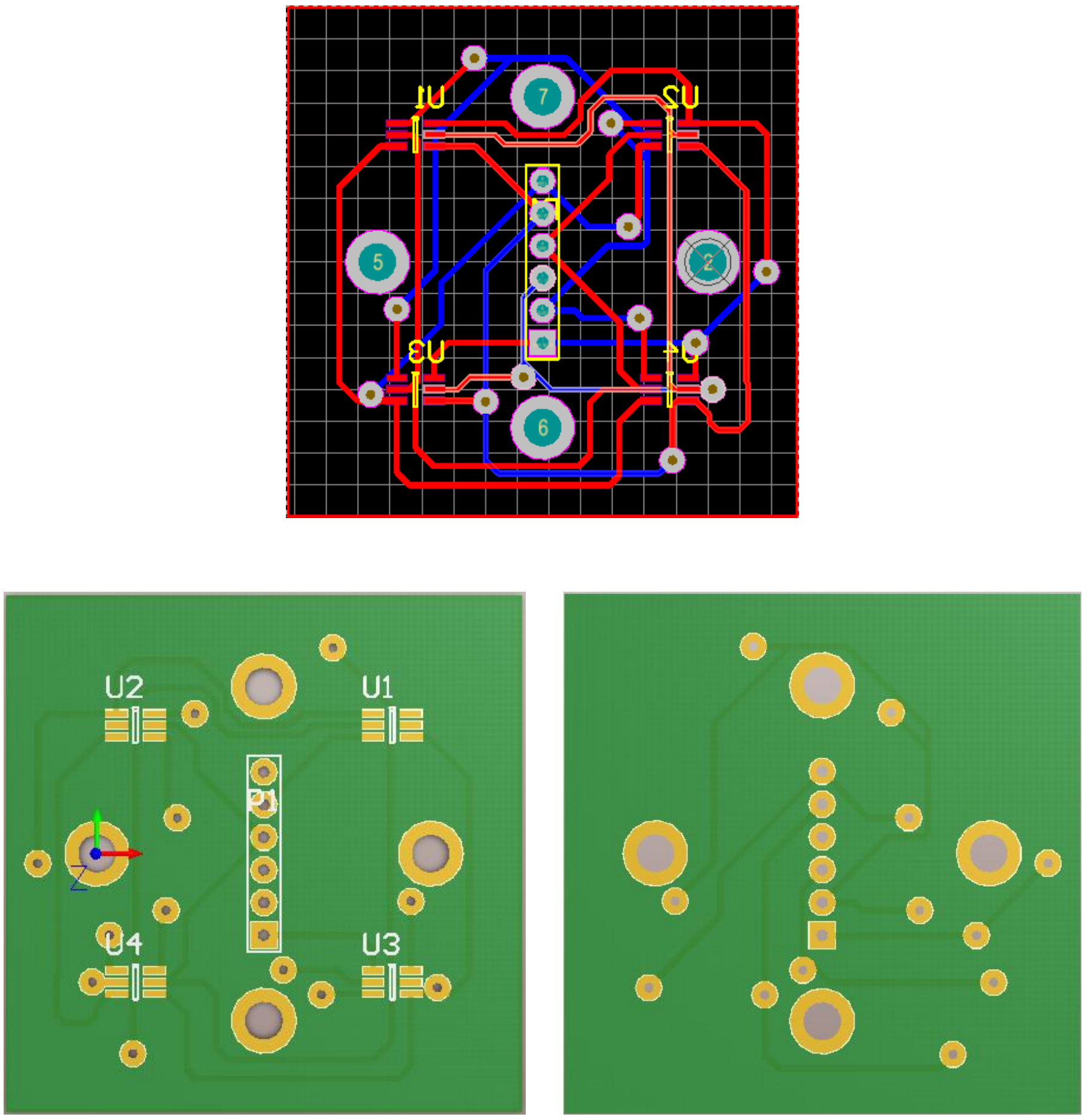

**Figure 55: LED Unit PCB Design**

The DAC PCB is only designed on top layer as shown in Figure 56. For best performance, the DAC power supply should be bypassed with at least a 1μF capacitor. Therefore, three 1μF capacitors are being used on this PCB.

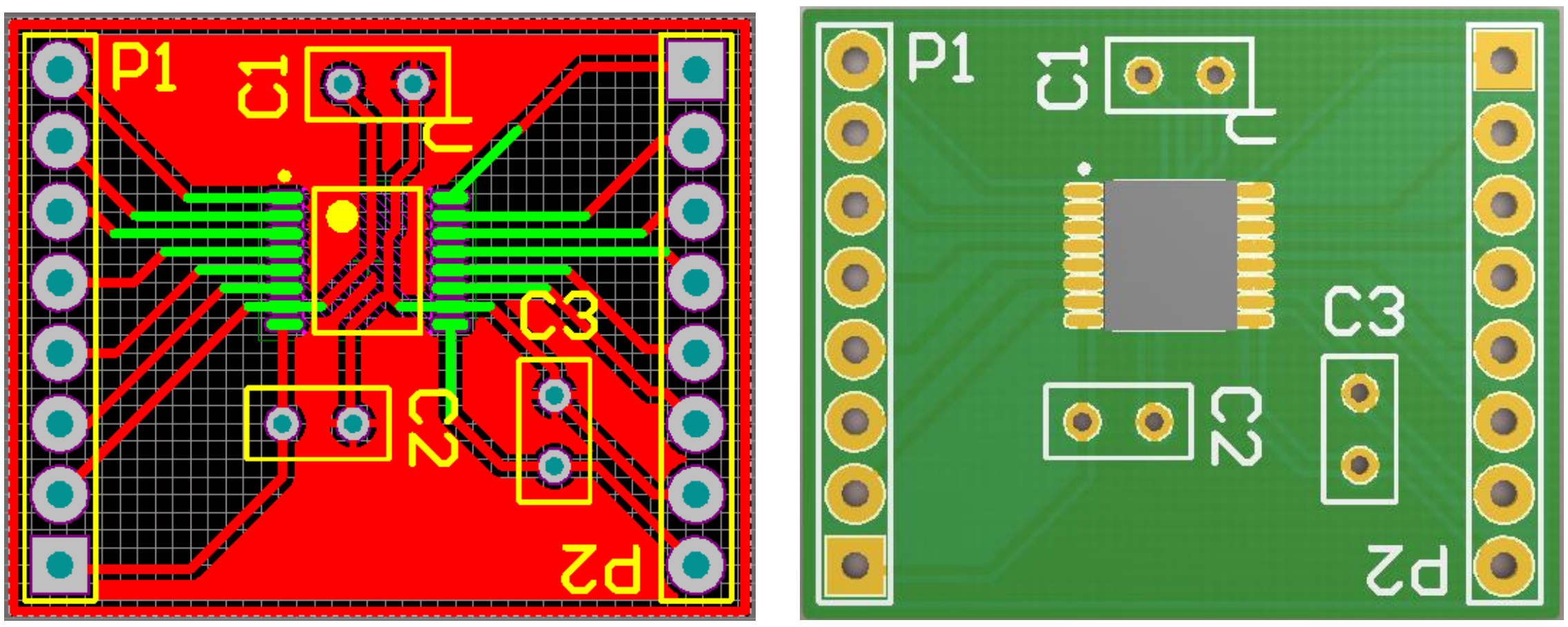

**Figure 56: DAC PCB Design**

The voltage regulator PCB is designed in bottom layer, and it should be soldered in bottom layer as shown Figure 57. The polygon is used for sharing the common ground. Two 10μF capacitors are used for filtering the high frequency components so that the input voltage and output voltage can be very clean DC values. The schematic of voltage regulator PCB is shown in Figure 58. In this project, $V_{in} = 12V$ and $V_{out} = 5V$. According to datasheet, the resister value needs to be calculated using the equation: $V_o = V_{REF}(1 + \frac{R2}{R1})$, where $V_{REF} = 1.25V$. The output here is 5 volts so the ratio of $R2$ and $R1$ is 3. Therefore, the values of $R2$ and $R1$ are given as 360 Ω and 120 Ω respectively.

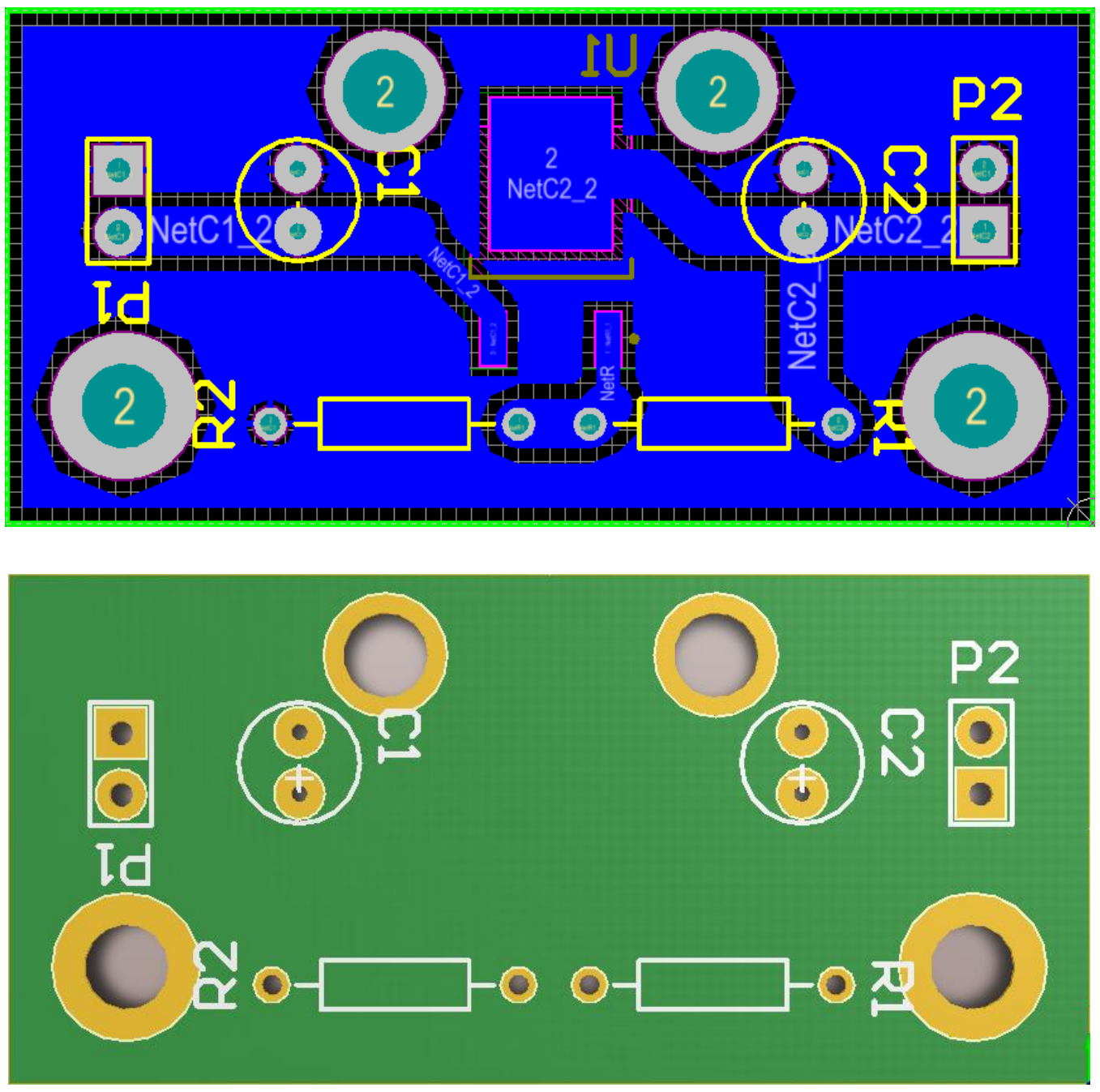

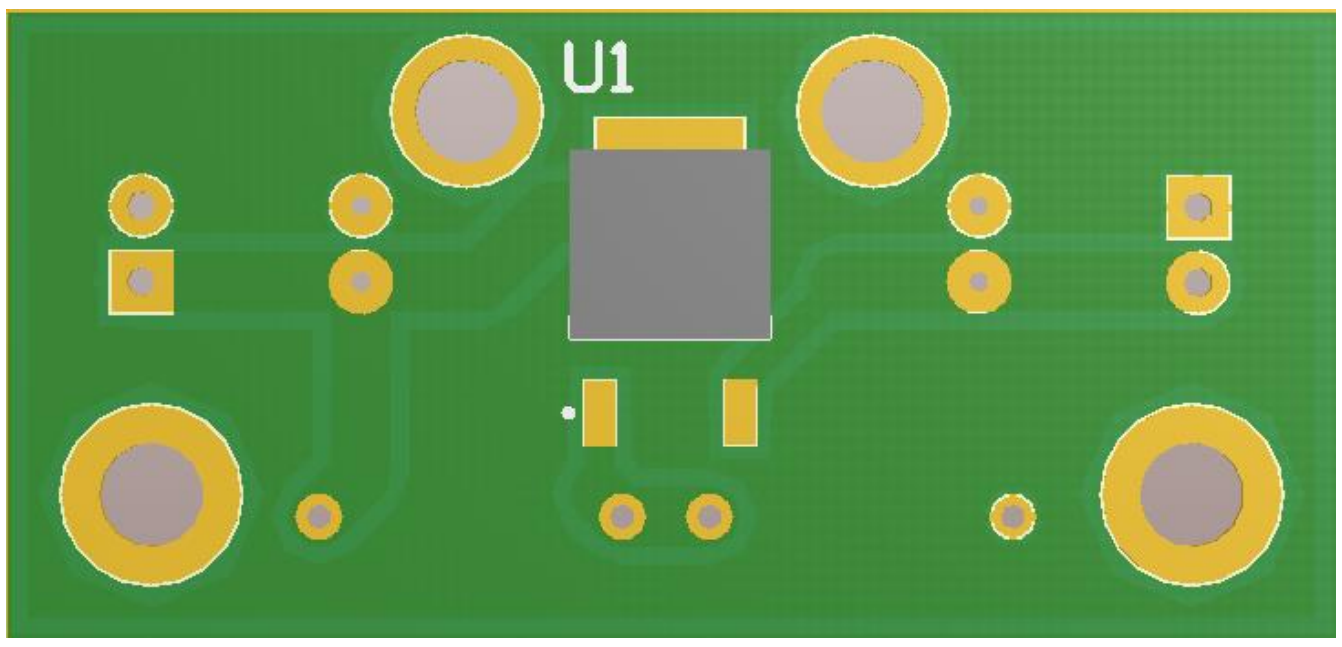

**Figure 57: Voltage Regulator PCB Design**

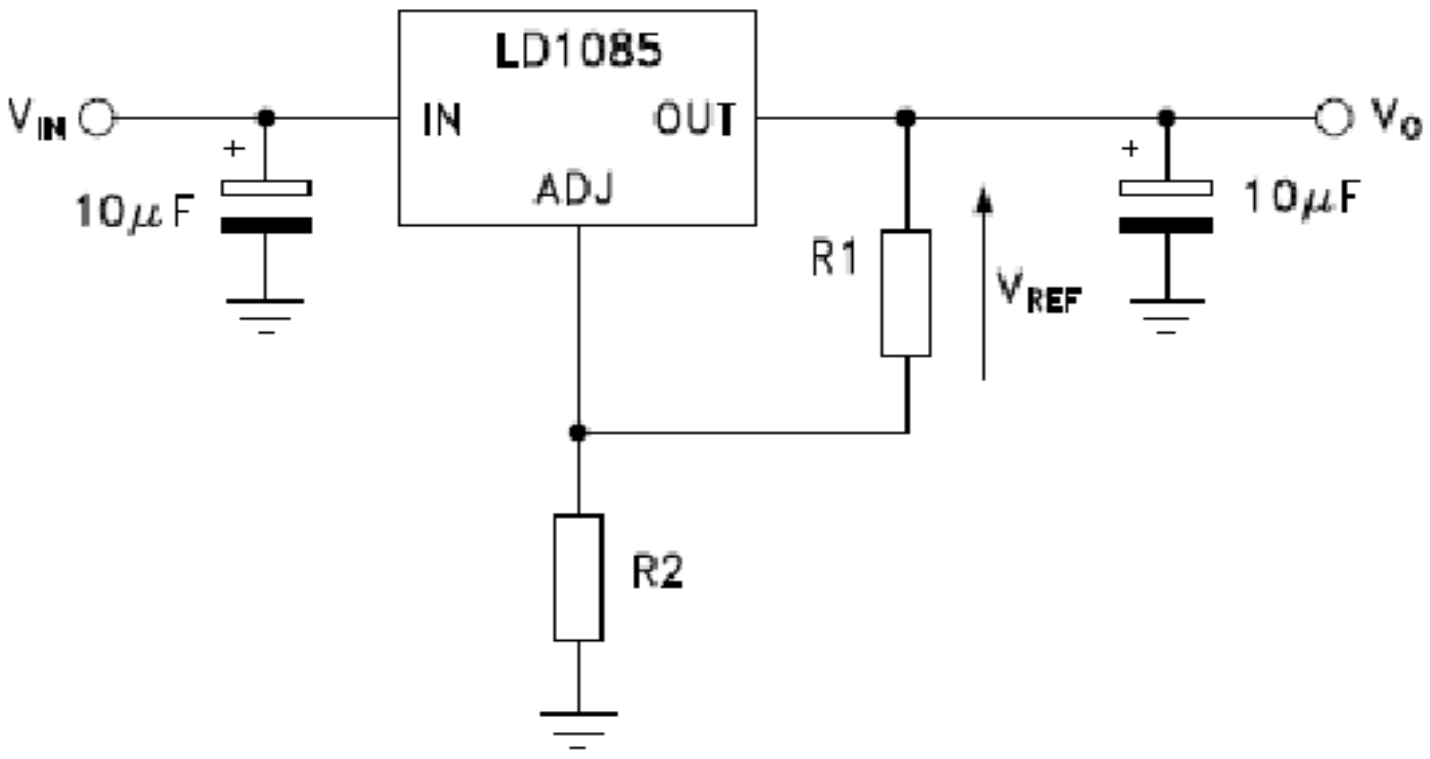

**Figure 58: Schematic of Voltage Regulator PCB Design**

### 4.2.3. Soldering the PCBs

After designing and printing the PCBs, the PCBs are ready to be soldered. Soldering SMD components is a very challenging work which needs a lot of concentration and focus. In this project, six PCBs of LED unit, four PCBs of DAC and one PCB of voltage regulator need to be soldered. The hardest part of this work is soldering the PCBs of LED unit. The reason is that each PCB has four LEDs with small J leads. A special procedure is followed to solder the LEDs. In short, place the flux on the pads, tack one line of pins and solder the rest of the pins. When there is too much flux core solder making two or more pins stick together, a desoldering wick need to be used for removing the extra flux core solder. However, it is very easy to burn the LEDs or parts of PCB board when the soldering iron's tip heat the desoldering wick for more than 5 seconds or when the same spot is repeatedly heated. Generally, only one or two channels are damaged when the LED is being soldered. Figure 59 shows the PCB of LED unit after soldering.

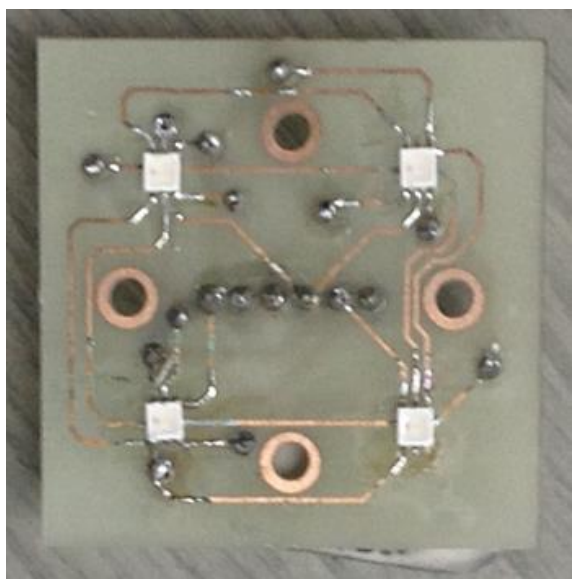

**Figure 59: Soldered PCB of LED Unit**

The step of soldering the DAC on PCB is almost same as soldering the LED on PCB. But soldering the DAC on PCB is a lot easier. Because the type of DAC is Thin-shrink small outline package (TSSOP) and size of it is bigger than LED. Figure 60 shows the PCB of DAC after soldering.

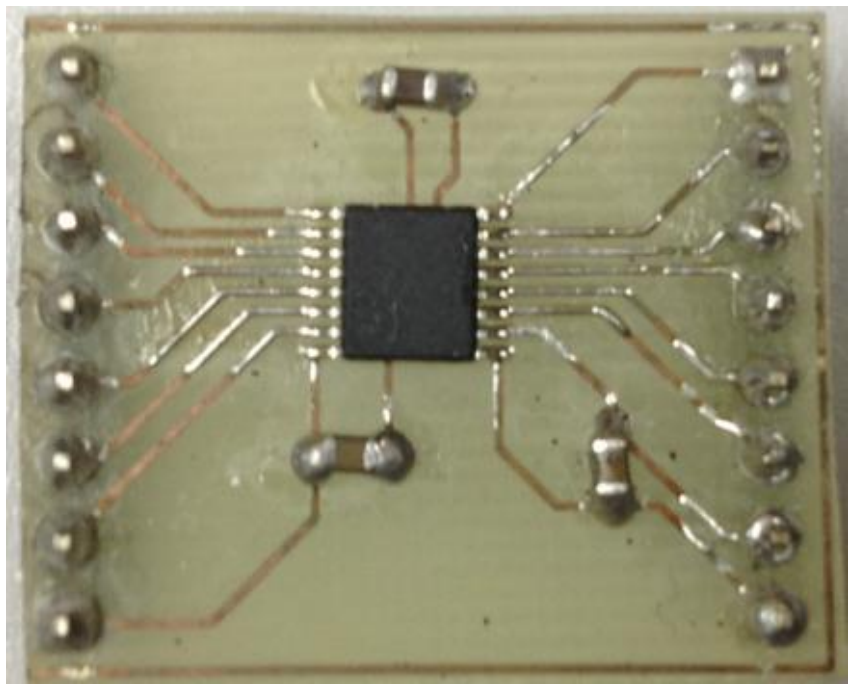

**Figure 60: Soldered PCB of DAC**

It is not difficult to solder the voltage regular, the only part that need to extra attention is the amount of flux core solder used in the output pin because excessive amount can create a short circuit. The voltage regular is shown in Figure 61.

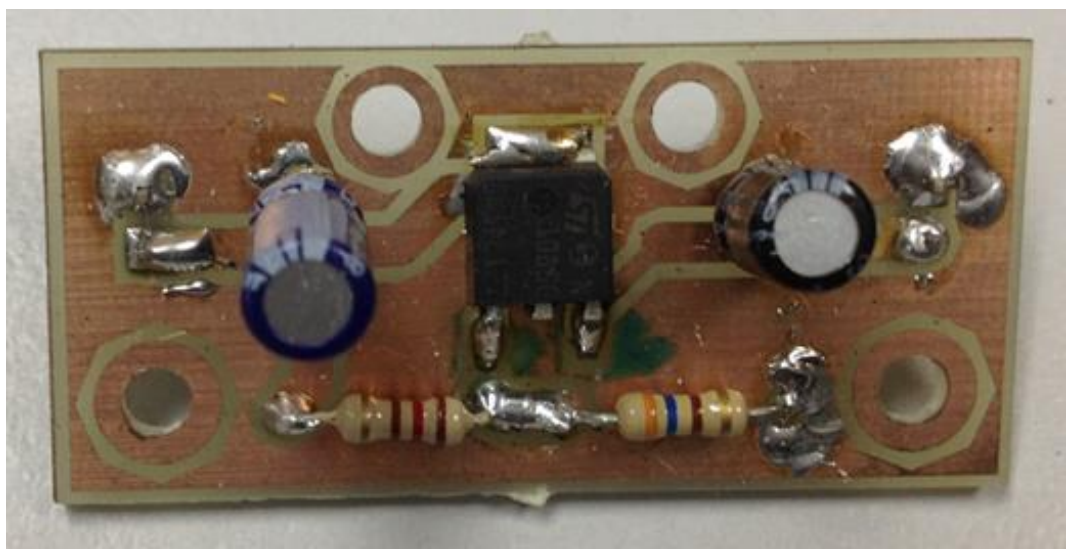

**Figure 61: Soldered PCB of Voltage Regulator**

### 4.2.4. Frequency Generation Circuit on Bread Broad

The components and the microcontroller are connected on the bread broad based on the frequency generation stimulation schematic. The Figure 62 shows the circuits of frequency generation (left) and LCD management (right) on the bread board.

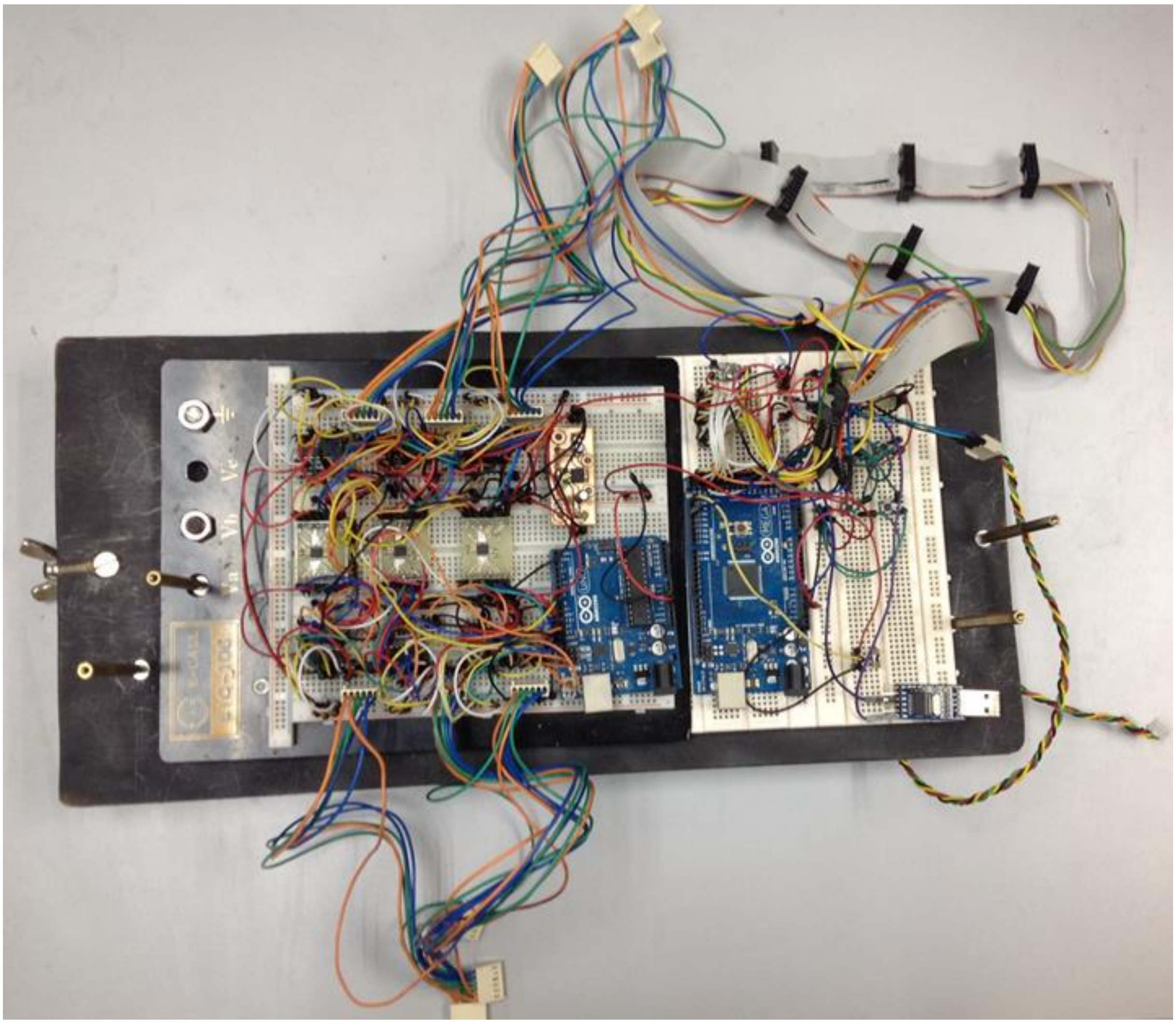

**Figure 62: Circuits Designed on Bread Board**

After uploading the codes, the microcontroller can successfully display six different frequencies in six LED units and the text information on LCDs.

### 4.2.5. Outlook of the Panel Design

The panel design is based on the second idea proposed which has been referred in 3.4.1. Figure 63 shows the outlook of panel's front and back.

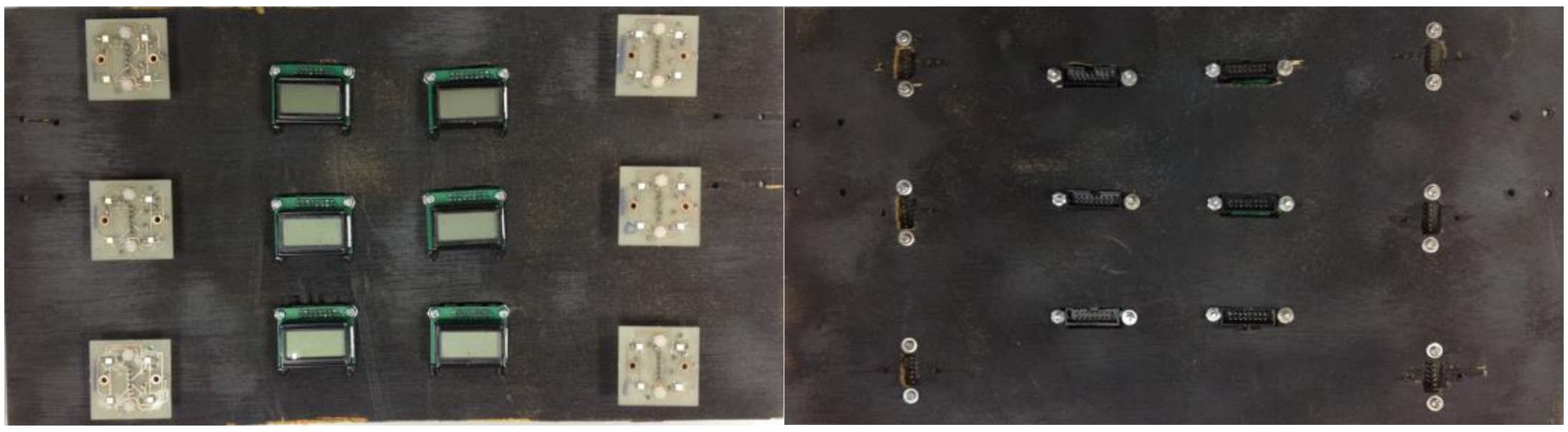

**Figure 63: The Outlook of Panel's Front (Left) and Back (Right)**

Connecters are used to connect the LED units and LCDs to the circuit on bread board. The full panel outlook is shown in Figure 64. To smooth the light intensity, six frosted paper boxes are made to cover LED units.

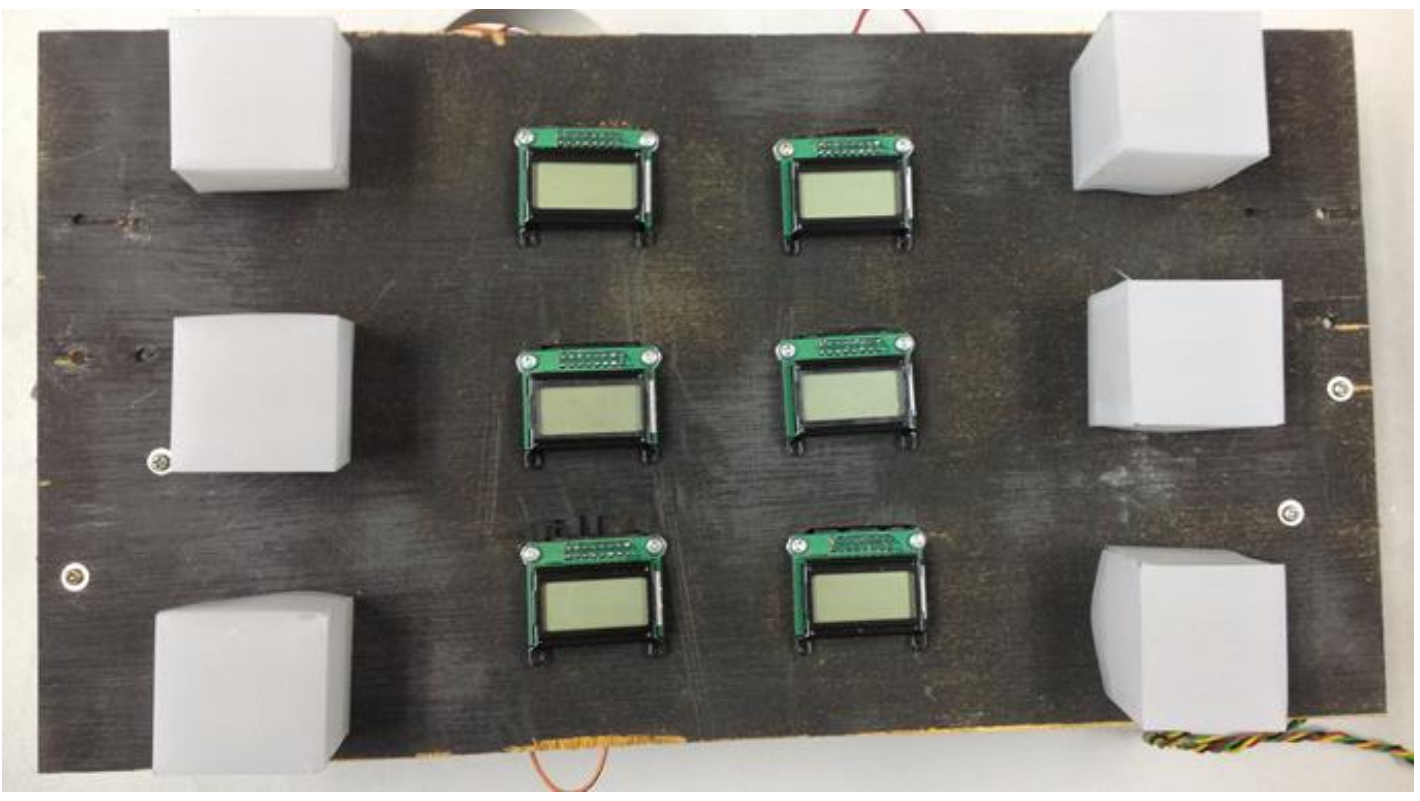

**Figure 64: SSVEP Visual Stimuli Plane Outlook**

In Figure 65, the whole panel (on the top) is banded with the circuits (on the bottom).

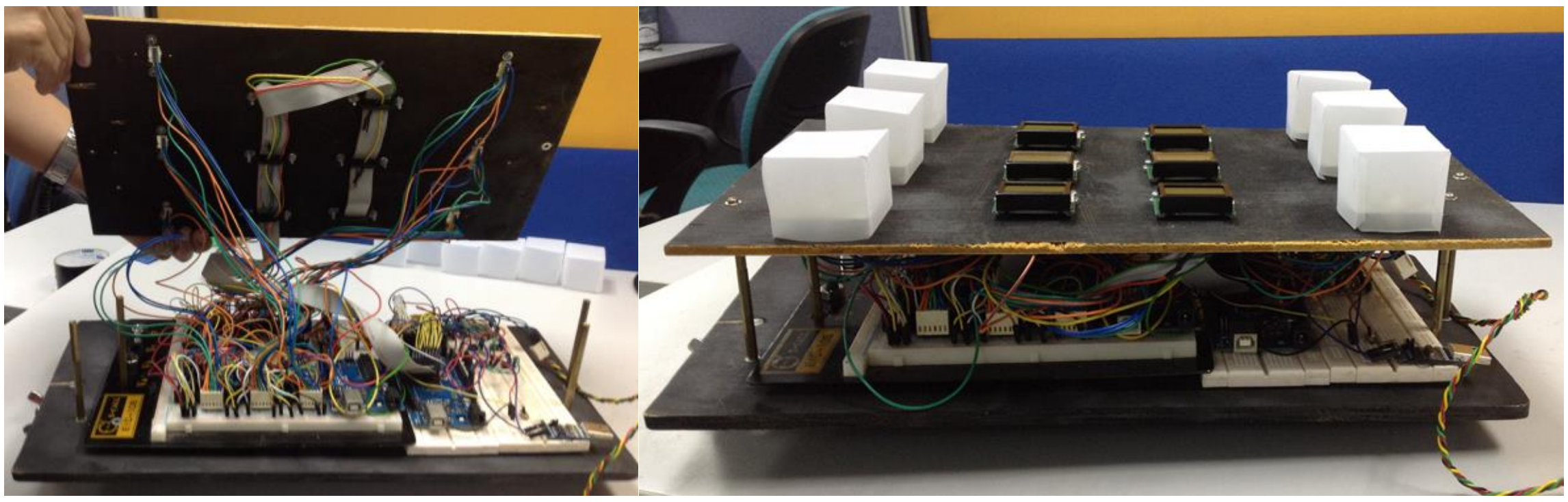

**Figure 65: The Whole Panel Outlook**

## 4.3. Real-time performance

### 4.3.1. Description of Subjects

In order to test the project in all aspects, the volunteer subjects with different ages and races are selected. Meanwhile, all subjects are healthy mentally and physically. No previous records of photosensitive epileptic seizure exist for all the subjects. The table below shows subjects' basic information.

**Table 4: Basic Information of Subjects**

| Subject | Name | Age | Race | Eye vision | Previous Experience Related to EEG |
|---|---|---|---|---|---|
| S1 | Alireza Safdari | 23 | Iranian | Good | Well-trained |
| S2 | Clement Kwan | 24 | Malaysian Chinese | Moderate myopia | Trained many times |
| S3 | Chen Jun Hui | 23 | Malaysian Chinese | Low to moderate myopia | No |
| S4 | Haroon Wardak | 21 | Afghans | Good | No |
| S5 | Hidayat Hambali | 27 | Malaysian Malay | Good | No |

### 4.3.2. Experiment Setup

The experiment is carried out inside the CRAE lab with smooth fluorescent light and quite environment. Both calibration and real-time experiments are carried out on a smart wheelchair. An overview of the wheelchair with the subjects is shown in Figure 66.

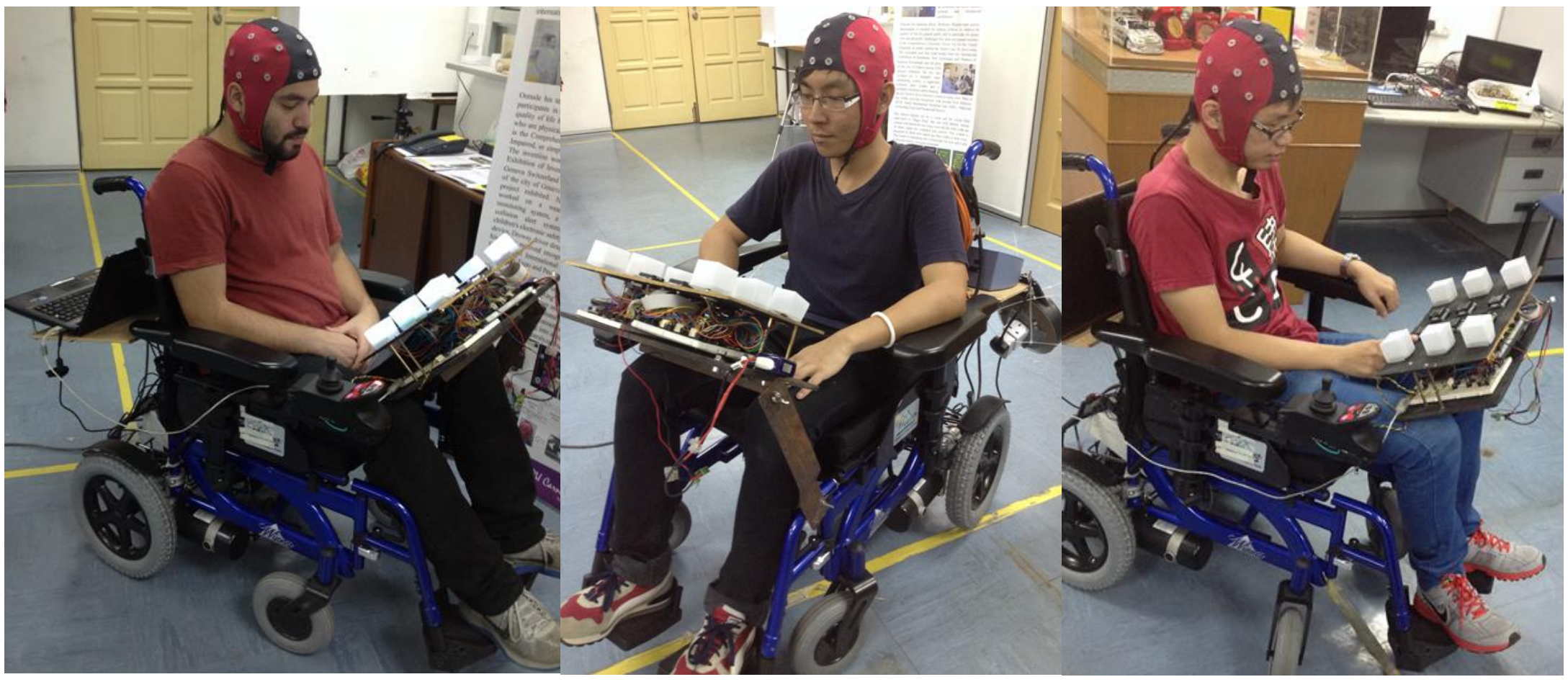

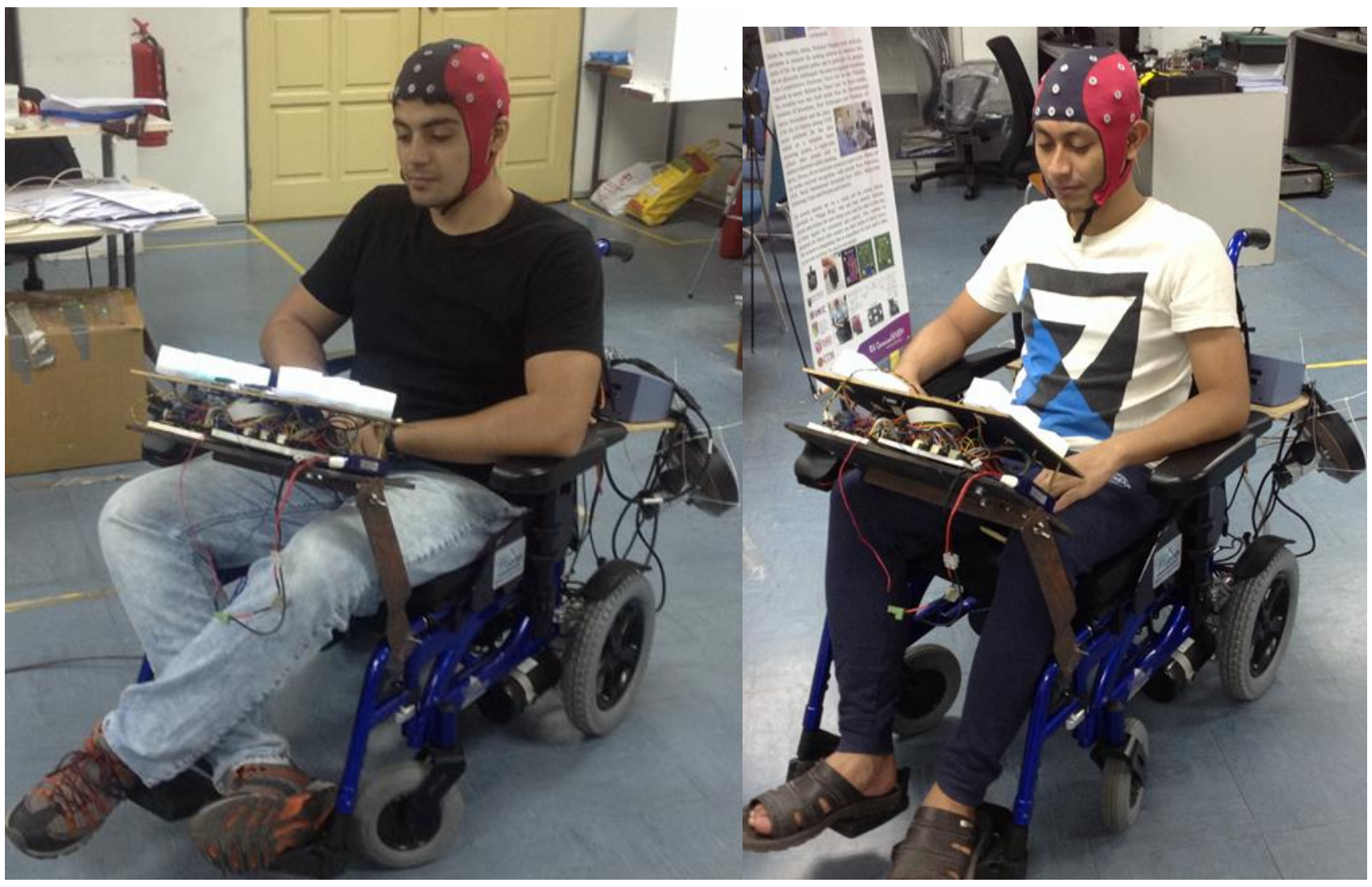

**Figure 66: An Overview of Subjects Sitting on the Wheelchair**

First, the EEG cap is worn on the subject. In order to make sure that the Cz electrode is always at the center of the participating subject, a measuring tape is used to measure the distances between inion and nasion, left and right ear lopes. Then, the conductive gels are poured inside the 4 used electrodes which are O1, Oz, O2 and the ground electrode. A wooden stick is used to reduce the impedances of all 3 electrodes below 5kΩ. The wooden stick is firmly placed inside the electrode hole to remove any barriers from the skin's surface. Figure 67 is a sample of electrodes' impendence tested on a subject using ASA LAB (after pouring gels in the electrodes).

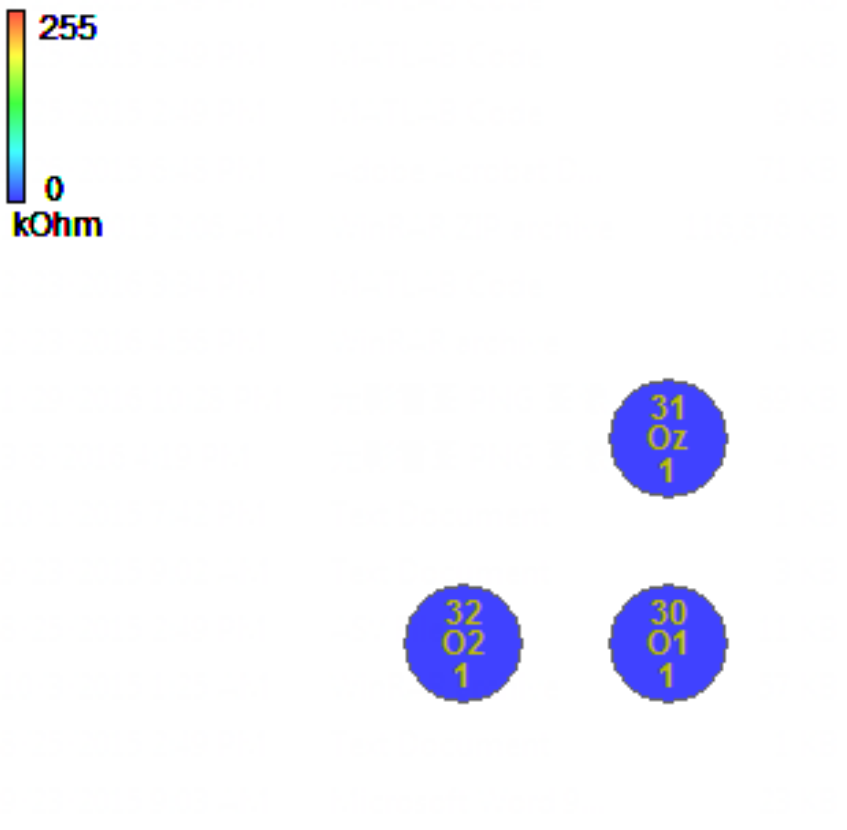

**Figure 67: Electrodes Impendences**

### 4.3.3. Calibration

In order to find the best thresholds for each subject, the calibration needs to be done separately from real-time experiments. An overview of calibration is explained below.

- First of all, EEG signals should be recorded by ASA LAB with 256Hz sampling frequency. The EEG signals with 10seconds rest followed by 20 seconds focus on visual stimuli are recorded.
- Then, ASA LAB is used to get the EEG signals' (during 20 seconds focusing time) power density with a 4 seconds window and a 3 to 60Hz bandpass filter. Although, in real time 2 seconds window and 3 to 50Hz bandpass filter are implemented. By enlarging window size and increasing the cutoff frequency range, threshold levels are more accurately calculated.
- Third, the FFT result can be exported into Excel file and each FFT point can be seen in the file.
- Finally, import the file for signal processing in Matlab, and the threshold levels are shown in the command windows.

Six threshold levels of different flickering frequencies can be obtained by repeating the steps mentioned above six times. The subject is supposed to look at specific frequency, so that the EEG signals with associated flickering frequency can be analyzed. Table 5 shows threshold levels of different flickering frequencies for participating subjects.

**Table 5: Threshold Levels from Calibration Result**

| Frequency / Subject | 7Hz | 8Hz | 9Hz | 10Hz | 11Hz | 12Hz |
|---|---|---|---|---|---|---|
| S1 | 0.41 | 0.62 | 0.48 | 0.61 | 0.53 | 0.53 |
| S2 | 0.19 | 0.10 | 0.14 | 0.33 | 0.40 | 0.27 |
| S3 | No recording | | | | | |
| S4 | 0.41 | 0.62 | 0.57 | 0.56 | 0.60 | 0.6 |
| S5 | No recording | | | | | |

The trend of the threshold level goes up with respect of frequency from 7 to 12Hz. It is due to the revolution of threshold level in different frequencies. The higher the frequency is, the lower the revolution is. Moreover, the error may occur more frequently. So it is better to use low frequencies with high accuracy.

The process of calibration is very complex. Subject 1, 2 and 4 have gone through the calibration while subject 3 and 5 have not done it. The reason of doing this is that S3 and S5 can compare their results with the other three in real time, so that the calibration may be cancelled in real-time operation in the future.

#### 4.3.4. Threshold Level Adjutancy

Although the thresholds of each frequency have already obtained in calibration session, the performance of each subject in real-time experiments decides the final threshold level.

Therefore, fine tuning the thresholds is very necessary. Here is the table of real-time threshold levels. The four used frequencies are recorded in all subjects while S2 has tested 10Hz and 12Hz too.

**Table 6: Threshold Levels from Real-time Experiment Result**

| Frequency / Subject | 7Hz | 8Hz | 9Hz | 10Hz | 11Hz | 12Hz |
|---|---|---|---|---|---|---|
| S1 | 0.24 | 0.24 | 0.24 | No recording | 0.35 | No recording |
| S2 | 0.16 | 0.16 | 0.16 | 0.30 | 0.34 | 0.30 |
| S3 | 0.16 | 0.16 | 0.16 | No recording | 0.2 | No recording |
| S4 | 0.25 | 0.25 | 0.26 | No recording | 0.27 | No recording |
| S5 | 0.26 | 0.26 | 0.27 | No recording | 0.30 | No recording |

The adjustment of threshold levels is based on the subject's performance in real time. When the commands have been given to the subject, the continuous movement of the wheelchair to the expected direction is the judgment of the final threshold level. In order to make the wheelchair move continuously, the threshold levels should be reduced. Additionally, the threshold levels of all frequencies for S3 and S5 are set to 0.25 initially.

### 4.3.5. Performance Evaluation Result

One important evaluation criterion in real-time experiment is to measure the delay time, and it is also one of the objectives in this project. So two trials in real-time experiments are recorded with the commands below:

- Trial 1: forward – left – right – reverse
- Trial 2: left – forward – reverse – right

In the real-time experiment, commands are given to the subject every time. The corresponding movement of each command lasts 5 seconds and then the subject stops controlling the wheelchair. The interval from the stop command given by subject to the wheelchair stopping moving is the time delay measured in this project. Table 7 shows the delay time measurement for each subject.

**Table 7: The Delay Time Measured for Each Subject**

| Commands | Subjects | Delay(s) | | |
|---|---|---|---|---|
| | | Trial 1 | Trial 2 | Average |
| Stop from forward | S1 | 1.95 | 2.10 | 2.03 |
| | S2 | Fail | | |
| | S3 | Fail | | |
| | S4 | 1.85 | 2.35 | 2.1 |
| | S5 | 2.05 | 1.78 | 1.92 |
| Stop from left | S1 | 2.02 | 2.43 | 2.23 |
| | S2 | Fail | | |
| | S3 | Fail | | |
| | S4 | 2.45 | 2.10 | 2.28 |
| | S5 | 1.79 | 1.89 | 1.84 |
| Stop from right | S1 | 2.38 | 2.06 | 2.22 |
| | S2 | Fail | | |
| | S3 | Fail | | |
| | S4 | 2.17 | 1.99 | 2.08 |
| | S5 | 2.44 | 2.32 | 2.38 |
| Stop from reverse | S1 | 2.33 | 2.15 | 2.24 |
| | S2 | Fail | | |
| | S3 | Fail | | |
| | S4 | 1.93 | 1.84 | 1.89 |
| | S5 | 1.67 | 1.85 | 1.76 |

The accuracy of the interval time is affected by the artificial behavior in the process of the time keeping. Meanwhile, the result can also be affected by the speed of the wheelchair.

During the experiments, no matter with glasses or without glasses, S2 and S3 fail in moving the wheelchair continuously. One possible reason is their vision in which both of them are weak. However, they may be good at other waveforms. For example, based on the previous research results (sampling frequency 256Hz, 4 seconds window, bandpass filter with the high cutoff frequency 60Hz and low cutoff frequency 3Hz), subject S2 has the best harmonics reaction to the triangular wave. Table 8 shows S2's threshold levels to the bright white LED while different waveforms and frequencies are used in visual stimuli.

**Table 8: Flickering Bright White LED Threshold Level**

| Waveform \ Frequency | 7Hz | 8Hz | 9Hz | 10Hz | 11Hz | 12Hz |
|---|---|---|---|---|---|---|
| Square wave | 0.19 | 0.10 | 0.14 | 0.33 | 0.40 | 0.27 |
| Sine wave | 0.26 | 0.33 | 0.29 | 0.39 | 0.40 | 0.13 |
| Half sine wave | 0.22 | 0.38 | 0.24 | 0.22 | 0.33 | 0.20 |
| Triangular wave | 0.30 | 0.34 | 0.38 | 0.33 | 0.47 | 0.33 |
| Ramp with 0 duty cycle | 0.15 | 0.14 | 0.38 | 0.39 | 0.40 | 0.47 |
| Ramp with 100 duty cycle | 0.30 | 0.19 | 0.19 | 0.33 | 0.2 | 0.40 |

Table 8 shows S2 has the best response to the triangular wave. Whereas, S2 has the best response to ramp with 100% duty cycle, when red LED is used, which is shown in Table 9. Therefore, it proves that the color has effect on subject's performance.

**Table 9: Flickering Red LED Threshold Level**

| Waveform \ Frequency | 7Hz | 8Hz | 9Hz | 10Hz | 11Hz | 12Hz |
|---|---|---|---|---|---|---|
| Square wave | 0.24 | 0.11 | 0.20 | 0.40 | 0.50 | 0.33 |
| Sine wave | 0.11 | 0.29 | 0.19 | 0.33 | 0.33 | 0.27 |
| Half sine wave | 0.26 | 0.43 | 0.14 | 0.33 | 0.47 | 0.27 |
| Triangular wave | 0.19 | 0.29 | 0.24 | 0.28 | 0.47 | 0.40 |
| Ramp with 0% duty cycle | 0.11 | 0.29 | 0.19 | 0.28 | 0.53 | 0.40 |
| Ramp with 100% duty cycle | 0.48 | 0.43 | 0.48 | 0.45 | 0.53 | 0.67 |

Another three subjects, S1, S4, S5 can successfully move the wheelchair which demonstrates that this project can be implemented on people without the limitation of races and ages. Besides that, the average delay time is around 2 seconds. As referred previously, this is mainly because of the window size. In order to verify this, S5 has tested the wheelchair with 1 second window (while the other configurations are remaining the same as before) and the results show the significant reduction in time delay. This result is shown in Table 10.

**Table 10: One Second Window Experiment on S5**

| Commands | Delay(s) | | |
|---|---|---|---|
| | Trial 1 | Trial 2 | Average |
| Stop from forward | 0.89 | 1.20 | 1.04 |
| Stop from left | 1.03 | 0.97 | 1.00 |
| Stop from right | 1.21 | 1.33 | 1.27 |
| Stop from reverse | 0.76 | 0.84 | 0.80 |

As the result shown in Table 10, the average delay time is approximately 1 second, which can verify that the window size of FFT is the main reason for the time delay in BCI system. And it also means this project can apply 1 second window in real time.

Moreover considering the successful SSVEP performances and the corresponding threshold levels, it can be concluded that the real-time threshold levels for S1, S4 and S5 are very similar. Therefore, setting all the threshold levels to 0.25 and fine tuning them can be implemented to avoid complex calibration.

#### 4.3.6. Summary

This chapter is about the experiments' results and discussions of the SSVEP-based control system. Three sections are included.

To achieve the first objective, investigation of the factors affecting the performance of SSVEP is carried out. These experiments are about LED flickering frequency, color, waveform and harmonic. The results can show these factors have different effects on the performance of SSVEP.

Second section relates to the design of SSVEP visual stimuli panel. First, four different adjustable current sources designs are introduced. And the suitable one can satisfy the requirement of the op amp power dissipation. Meanwhile, the design and soldering of the PCBs are also presented in this section. Finally, the outlook of the panel is shown.

The third section shows the real-time performance of the SSVEP in controlling the wheelchair. Five subjects with different races and ages are involved. The configurations of signal processing are 256Hz sampling frequency, bandpass filter with 3 to 50Hz cutoff frequencies and 2 seconds FFT window. The results show all the subjects who have good eye visions can operate the wheelchair successfully with approximately 2 seconds time delay. Besides that, the window size is the main element of the time delay. This is verified by applying a 1 second window in the real-time BCI controlled wheelchair system on one of subjects where the time delay is 1 second. On the other hand, complex calibration can be omitted by setting all the threshold values to 0.25 initially and fine tuning them in real-time later.

# 5. Future Work and Conclusion

## 5.1.Future Work

There are still a lot of works to be done for improving and completing this project. The flowing areas can open new doors for the future of this project.

- Testing the SSVEP-based BCI controlled wheelchair with 1 second FFT windows on more subjects to make sure the 1 second FFT can work properly.
- Instead of omitting the calibration, trying to calibrate the threshold level automatically in the signal processing.
- Testing the project on other Chinese who have good vision to check whether the quality of the vision has significant effect on SSVEP response.
- Testing the subjects S2 and S3 with different waveforms in real-time wheelchair controlled system and check whether the waveform of the flickering frequency affects the response of SSVEP as referred in 4.3.5.
- Adding the light, TV and air conditioner in the system and programming the information transfer SSVEP visual stimuli panel as referred in 3.4.1.

## 5.2.Conclusion

This project is aimed at improving the delay time of controlling the wheelchair and establishing a totally new SSVEP visual stimuli panel as visual stimulator. And these are the main objectives. In order to achieve the objectives, the theories related to background of BCI system have been learnt and explained in the chapter of literature review.

With respect of the knowledge of SSVEP and BCI, a SSVEP-based BCI wheelchair controlled system and the methods to realize it have been ascertained. SSVEP can be easily elicited in the vision cortical with the same frequency as the subject looking at. Based on this feature, the panel with six LED units combining six LCD units has been designed for visual stimulator. It mainly used two microcontrollers to control the flickers of the LED units and text information on LCDs. Meanwhile, a new method of classification utilizing the harmonics has been proposed and applied in the signal processing in Matlab.

The performance of SSVEP can be affected by factors like flickering frequency, color and waveforms. In the real-time experiments, the subjects with good eye vision can successfully control the movement of the wheelchair. And eventually, the time delay can be even reduced to approximately 1 second. In conclusion, the proposed system can be easily and successfully operated by disable people with very short time delay.

# Appendix 1: Matlab Program for Real-time EEG Acquisition, Signal Processing, and Command Generation

```
% Intro message
fprintf('Brain Controlled Wheelchair Project created by Zhou Ce\r');

% To create a serial port
UCOOB = serial('COM5', 'BaudRate', 112500, 'DataBits',8, 'Terminator', 'CR');

% Show serial port info to user
COM5_INFO =
get(UCOOB,{'BaudRate','DataBits','Parity','StopBits','Terminator'});
fopen(UCOOB);

disp('Press any key again to continue!');
pause;

% Clear input buffer
flushinput(UCOOB);

%Control commands
Forward_command = [255 170 1 1 254];
Reverse_command = [255 170 1 2 254];
Rot_Ccw_command = [255 170 1 3 254];
Rot_Cw_command = [255 170 1 4 254];
LED_on_command = [255 170 1 5 254];
LED_off_command = [255 170 1 6 254];
Stop_command = [255 170 1 0 254];

%Initializate the frequencies of commands
set_forward = 7;
set_reverse = 11;
set_Rot_Ccw = 9;
set_Rot_Cw = 8;
set_LED_on = 20;
set_LED_off = 12;

flickering_frequency = [set_forward set_reverse set_Rot_Ccw set_Rot_Cw
set_LED_on set_LED_off];
```

```
% Initializate the values of threshold
forward_threshold = 0.26;
reverse_threshold = 0.26;
Rot_Ccw_threshold = 0.25;
Rot_Cw_threshold = 0.22;

threshold_level = [forward_threshold reverse_threshold Rot_Ccw_threshold
Rot_Cw_threshold 1 1];


% Send an initial STOP command to wheelchair
fwrite(UCOOB, Stop_command, 'uchar', 'async');
fprintf('Stop\r');

% nMaxChannelCount  = 32;           % Show all 32 EEG channels
dWindow          = 2;          % Show n seconds of data

sHost           = 'tmsi';       % Change this to 'asa', 'eemagine', 'tmsi' or
                               % 'test' other work with other hosts
Fs = 256;                       % Sampling frequency of recorded EEG
L = Fs * dWindow;                % Total number of datapoints of raw data
NFFT = 2^nextpow2(L);            % Next power of 2 from length of raw_sti


Fp1 = 3;                        % Bandpass filter passband 1
Fp2 = 50;                       % Bandpass filter passband 2
order = 3;                      % Bandpass filter order, n = 2^order

% Creating a butterworth bandpass filter (recursive - IIR filter)
[b,a] = butter(order,[Fp1 Fp2]/(Fs/2),'bandpass');

oDevice = device(sHost);           % Connect to server
oDevice = connect(oDevice);        % Connect to EEG amplifier

%% set up search space for each frequency
search_space(6:10) = 0;
fundamental_and_harmonics(6:10) = 0;

for i = 1:6
```

```
    for j = 1:10
        search_space(i,j) = flickering_frequency(:,i)*(j - 0.5);
        fundamental_and_harmonics(i,j) = flickering_frequency(:,i)*j;
    end
end
%% get the seach space point
search_space_point = round(search_space.* dWindow + 1);
%% intialize the point
point(1:6) = 0;
%% make a sound for starting reminder
Fs_sound = 5000;                % Sampling frequency
freq_sound = 700;               % 700 kHz sine wave
dur_sound = 1;                  % Duration of tone play (second)
x = sin(linspace(0, 2*pi*freq_sound*dur_sound, round(dur_sound*Fs_sound)));

sound(x, Fs_sound);             % Play the tone, program starts!
%% the start of the loop
try

    bActive = true;
    hf = figure(1);
    % using tic toc for arrange the speed of sending the command
    tic;
    while (bActive)
        if toc > 0.1
           toc
           tic

    %read EEG
    mEEG = getEEG(oDevice, dWindow);
    mEEGavg = ones(L,1) * mean(mEEG,1);
    mEEG = mEEG - mEEGavg;

    vTime = dWindow * [0:size(mEEG,1)-1] / L;
    f = Fs/2*linspace(0,1,NFFT/2+1);    % The frequency vector for plotting
frequency axis

%% Oz:
    f_max_Oz(6:10)=0;

    %select the specific channel for filter Oz
```

```
    filt_eeg_Oz = filtfilt(b,a,mEEG(:,31));
    Y_filt_Oz = fft(filt_eeg_Oz, NFFT)/L;
    Y_Oz = abs(Y_filt_Oz(1:NFFT/2+1)).*2;

    %Finding Max in Oz
    for i = 1:6
        for j = 1:9
            if search_space(i,j) <= 50
            [max_value, f_max_ind]=
max(Y_Oz(search_space_point(i,j):search_space_point(i,j+1)));
            f_max_Oz(i,j) = (f_max_ind + search_space_point(i,j) - 2)/NFFT*Fs;
            end
        end
    end

%% O1
    f_max_O1(6:10)=0;

    %select the specific channel for filter O1
    filt_eeg_O1 = filtfilt(b,a,mEEG(:,30));
    Y_filt_O1 = fft(filt_eeg_O1, NFFT)/L;
    Y_O1 = abs(Y_filt_O1(1:NFFT/2+1)).*2;

    %Finding Max in O1
    for i = 1:6
        for j = 1:9
            if search_space(i,j) <= 50
            [max_value, f_max_ind]=
max(Y_O1(round(search_space_point(i,j)):round(search_space_point(i,j+1))));
            f_max_O1(i,j) = (f_max_ind + search_space_point(i,j) - 2)/NFFT*Fs;
            end
        end
    end

 %% O2
    f_max_O2(6:10)=0;

    %select the specific channel for filter O2
    filt_eeg_O2 = filtfilt(b,a,mEEG(:,32));
    Y_filt_O2 = fft(filt_eeg_O2, NFFT)/L;
    Y_O2 = abs(Y_filt_O2(1:NFFT/2+1)).*2;
```

```
%Finding Max in O2
for i = 1:6
    for j = 1:9
        if search_space(i,j) <= 50
        [max_value, f_max_ind]=
max(Y_O2(round(search_space_point(i,j)):round(search_space_point(i,j+1))));
        f_max_O2(i,j) = (f_max_ind + search_space_point(i,j) - 2)/NFFT*Fs;
        end
    end
end

%% adding the corresponding frequency and harmonics
 point(1:6) = 0;

 for i = 1:6
     for j = 1:10
         if (f_max_Oz(i,j) == fundamental_and_harmonics(i,1)||...
             f_max_Oz(i,j) == fundamental_and_harmonics(i,2)||...
             f_max_Oz(i,j) == fundamental_and_harmonics(i,3)||...
             f_max_Oz(i,j) == fundamental_and_harmonics(i,4)||...
             f_max_Oz(i,j) == fundamental_and_harmonics(i,5)||...
             f_max_Oz(i,j) == fundamental_and_harmonics(i,6)||...
             f_max_Oz(i,j) == fundamental_and_harmonics(i,7)||...
             f_max_Oz(i,j) == fundamental_and_harmonics(i,8)||...
             f_max_Oz(i,j) == fundamental_and_harmonics(i,9)||...
             f_max_Oz(i,j) == fundamental_and_harmonics(i,10))
             point(i) = point(i) + 1;
         end
         if (f_max_O1(i,j) == fundamental_and_harmonics(i,1)||...
             f_max_O1(i,j) == fundamental_and_harmonics(i,2)||...
             f_max_O1(i,j) == fundamental_and_harmonics(i,3)||...
             f_max_O1(i,j) == fundamental_and_harmonics(i,4)||...
             f_max_O1(i,j) == fundamental_and_harmonics(i,5)||...
             f_max_O1(i,j) == fundamental_and_harmonics(i,6)||...
             f_max_O1(i,j) == fundamental_and_harmonics(i,7)||...
             f_max_O1(i,j) == fundamental_and_harmonics(i,8)||...
             f_max_O1(i,j) == fundamental_and_harmonics(i,9)||...
             f_max_O1(i,j) == fundamental_and_harmonics(i,10))
             point(i) = point(i) + 1;
         end
```

```
            if (f_max_O2(i,j) == fundamental_and_harmonics(i,1)||...
                f_max_O2(i,j) == fundamental_and_harmonics(i,2)||...
                f_max_O2(i,j) == fundamental_and_harmonics(i,3)||...
                f_max_O2(i,j) == fundamental_and_harmonics(i,4)||...
                f_max_O2(i,j) == fundamental_and_harmonics(i,5)||...
                f_max_O2(i,j) == fundamental_and_harmonics(i,6)||...
                f_max_O2(i,j) == fundamental_and_harmonics(i,7)||...
                f_max_O2(i,j) == fundamental_and_harmonics(i,8)||...
                f_max_O2(i,j) == fundamental_and_harmonics(i,9)||...
                f_max_O2(i,j) == fundamental_and_harmonics(i,10))
                point(i) = point(i) + 1;
            end
        end
    end
%% get the point
    for i = 1:6
        point(i) = point(i)/(3*floor(50/flickering_frequency(i)));
    end
%% send to serial port
       [value, index]= max(point);
       if point(index) > threshold_level(index)
           if index == 1
               fwrite(UCOOB, Forward_command, 'uchar', 'async');
               fprintf('Forward\r');
           elseif index == 2
               fwrite(UCOOB, Reverse_command, 'uchar', 'async');
               fprintf('Reverse\r');
           elseif index == 3
               fwrite(UCOOB, Rot_Ccw_command, 'uchar', 'async');
               fprintf('left\r');
           elseif index == 4
               fwrite(UCOOB, Rot_Cw_command, 'uchar', 'async');
               fprintf('right\r');
           elseif index == 5
               fwrite(UCOOB, LED_on_command, 'uchar', 'async');
               fprintf('LED_on\r');
           elseif index == 6
               fwrite(UCOOB, LED_off_command, 'uchar', 'async');
               fprintf('LED_off\r');
           end
       else
```

```
            fwrite(UCOOB, Stop_command, 'uchar', 'async');
            fprintf('Stop\r');
        end


        % force to update figure
        drawnow;

        % check if all figures still exists
        if (~ishandle(hf))
          bActive = false;

        end

        end
    end

catch ME
    stopasync(UCOOB);
    fclose(UCOOB);
    delete(UCOOB);
    clear UCOOB;
    oDevice = disconnect(oDevice);
    error(ME.message)

end

stopasync(UCOOB);
oDevice = disconnect(oDevice);
fclose(UCOOB);
delete(UCOOB);
clear UCOOB;
```

# Appendix 2: Daisy Chain of DAC Functions in Frequency Generation

```
void set_channel_A_update_all(uint8_t value_DAC1, uint8_t value_DAC2, uint8_t value_DAC3 )
{
   uint16_t command = 0xB000, command_DAC1, command_DAC2, command_DAC3;
   command_DAC1 = command | ( (uint16_t) value_DAC1 << 4);
   command_DAC2 = command | ( (uint16_t) value_DAC2 << 4);
   command_DAC3 = command | ( (uint16_t) value_DAC3 << 4);
   digitalWrite (SYNCpin, LOW);
   SPI.transfer16(command_DAC3);
   SPI.transfer16(command_DAC2);
   SPI.transfer16(command_DAC1);
   digitalWrite (SYNCpin, HIGH);
}

void set_channel_B(uint8_t value_DAC1, uint8_t value_DAC2, uint8_t value_DAC3)
{
   uint16_t command = 0x1000, command_DAC1, command_DAC2, command_DAC3;
   command_DAC1 = command | ( (uint16_t) value_DAC1 << 4);
   command_DAC2 = command | ( (uint16_t) value_DAC2 << 4);
   command_DAC3 = command | ( (uint16_t) value_DAC3 << 4);
   digitalWrite (SYNCpin, LOW);
   SPI.transfer16(command_DAC3);
   SPI.transfer16(command_DAC2);
   SPI.transfer16(command_DAC1);
   digitalWrite (SYNCpin, HIGH);
}

void set_channel_C(uint8_t value_DAC1, uint8_t value_DAC2, uint8_t value_DAC3)
{
   uint16_t command = 0x2000, command_DAC1, command_DAC2, command_DAC3;
   command_DAC1 = command | ( (uint16_t) value_DAC1 << 4);
   command_DAC2 = command | ( (uint16_t) value_DAC2 << 4);
   command_DAC3 = command | ( (uint16_t) value_DAC3 << 4);
   digitalWrite (SYNCpin, LOW);
   SPI.transfer16(command_DAC3);
   SPI.transfer16(command_DAC2);
```

```
    SPI.transfer16(command_DAC1);
    digitalWrite (SYNCpin, HIGH);
}

void set_channel_D(uint8_t value_DAC1, uint8_t value_DAC2, uint8_t value_DAC3)
{
    uint16_t command = 0x3000, command_DAC1, command_DAC2, command_DAC3;
    command_DAC1 = command | ( (uint16_t) value_DAC1 << 4);
    command_DAC2 = command | ( (uint16_t) value_DAC2 << 4);
    command_DAC3 = command | ( (uint16_t) value_DAC3 << 4);
    digitalWrite (SYNCpin, LOW);
    SPI.transfer16(command_DAC3);
    SPI.transfer16(command_DAC2);
    SPI.transfer16(command_DAC1);
    digitalWrite (SYNCpin, HIGH);
}

void set_channel_E(uint8_t value_DAC1, uint8_t value_DAC2, uint8_t value_DAC3)
{
    uint16_t command = 0x4000, command_DAC1, command_DAC2, command_DAC3;
    command_DAC1 = command | ( (uint16_t) value_DAC1 << 4);
    command_DAC2 = command | ( (uint16_t) value_DAC2 << 4);
    command_DAC3 = command | ( (uint16_t) value_DAC3 << 4);
    digitalWrite (SYNCpin, LOW);
    SPI.transfer16(command_DAC3);
    SPI.transfer16(command_DAC2);
    SPI.transfer16(command_DAC1);
    digitalWrite (SYNCpin, HIGH);
}

void set_channel_F(uint8_t value_DAC1, uint8_t value_DAC2, uint8_t value_DAC3)
{
    uint16_t command = 0x5000, command_DAC1, command_DAC2, command_DAC3;
    command_DAC1 = command | ( (uint16_t) value_DAC1 << 4);
    command_DAC2 = command | ( (uint16_t) value_DAC2 << 4);
    command_DAC3 = command | ( (uint16_t) value_DAC3 << 4);
    digitalWrite (SYNCpin, LOW);
    SPI.transfer16(command_DAC3);
    SPI.transfer16(command_DAC2);
    SPI.transfer16(command_DAC1);
    digitalWrite (SYNCpin, HIGH);
```

```
}

void set_channel_G(uint8_t value_DAC1, uint8_t value_DAC2, uint8_t value_DAC3)
{
   uint16_t command = 0x6000, command_DAC1, command_DAC2, command_DAC3;
   command_DAC1 = command | ( (uint16_t) value_DAC1 << 4);
   command_DAC2 = command | ( (uint16_t) value_DAC2 << 4);
   command_DAC3 = command | ( (uint16_t) value_DAC3 << 4);
   digitalWrite (SYNCpin, LOW);
   SPI.transfer16(command_DAC3);
   SPI.transfer16(command_DAC2);
   SPI.transfer16(command_DAC1);
   digitalWrite (SYNCpin, HIGH);
}

void set_channel_H(uint8_t value_DAC1, uint8_t value_DAC2, uint8_t value_DAC3)
{
   uint16_t command = 0x7000, command_DAC1, command_DAC2, command_DAC3;
   command_DAC1 = command | ( (uint16_t) value_DAC1 << 4);
   command_DAC2 = command | ( (uint16_t) value_DAC2 << 4);
   command_DAC3 = command | ( (uint16_t) value_DAC3 << 4);
   digitalWrite (SYNCpin, LOW);
   SPI.transfer16(command_DAC3);
   SPI.transfer16(command_DAC2);
   SPI.transfer16(command_DAC1);
   digitalWrite (SYNCpin, HIGH);
}
```

# Appendix 3: Main Function of Frequency Generation

```
// inslude the SPI library:
#include <SPI.h>

const int SYNCpin = 7;
const uint8_t number_of_frequency = 6;
double frequence[6] = {7,12,9,11,12,8};//array
uint16_t interrupt_counter[6] = {0};
const uint16_t f_interrupt = 2500;//T=0.4ms.
uint16_t max_counter_value[6] = {0};
uint16_t half_counter_value[6] = {0};
uint8_t DAC_next_value[6] = {0};
uint8_t interrupt_status = 0;
byte temp_state;
byte header[3]={0xFF,0xAB,0x01};

void setup() {

   pinMode(SYNCpin, OUTPUT);

   Serial.begin(250000);

   //initialize SPI:
  SPI.begin();
  SPI.beginTransaction( SPISettings( 40000000, MSBFIRST, SPI_MODE1 ) );
  for(uint8_t i=0; i<number_of_frequency; i++)
  {
    max_counter_value[i] = round((double)f_interrupt/frequence[i]);
    half_counter_value[i] = round((double)f_interrupt/frequence[i]/2);
    //Serial.println( max_counter_value[i]);
    //Serial.println( half_counter_value[i]);
    }
  noInterrupts();
  //Timer/Counter 1 initialization
  // Clock source: System Clock
  // Clock value: 16000.000 kHz
  // Mode: Fast PWM top=ICR1
  // OC1A output: Discon.
  // OC1B output: Discon.
```

```
    // OC1C output: Discon.
    // Noise Canceler: Off
    // Input Capture on Falling Edge
    // Timer1 Overflow Interrupt: On
    // Input Capture Interrupt: Off
    // Compare A Match Interrupt: Off
    // Compare B Match Interrupt: Off
    // Compare C Match Interrupt: Off
    TCCR1A = 0x02;
    TCCR1B = 0x19;
    TCNT1H = 0x00;
    TCNT1L = 0x00;
    ICR1H = 0x18;
    ICR1L = 0xFF;
    OCR1AH = 0x00;
    OCR1AL = 0x00;
    OCR1BH = 0x00;
    OCR1BL = 0x00;
//    OCR1CH = 0x00;
//    OCR1CL = 0x00;

    // Timer/Counter 1 Interrupt(s) initialization
    TIMSK1=0x01;
    interrupts(); // enable all interrupts
}

ISR(TIMER1_OVF_vect)          // interrupt service routine that wraps a user defined
function supplied by attachInterrupt
{
    interrupt_status = 1;
  }


void loop() {

    if( interrupt_status == 1 )


{ set_channel_A_update_all(DAC_next_value[0],DAC_next_value[1],DAC_next_value[2]);

          uint16_t remainder = 0;
```

```
 for(uint8_t i = 0; i<number_of_frequency; i++)
 {
   interrupt_counter[i]++;
   remainder = interrupt_counter[i] % max_counter_value[i];
   if (remainder == 0)
   {
     interrupt_counter[i] = 0;//0
     }
   if (remainder < half_counter_value[i])
   {
     DAC_next_value[i] = 15;
     }
   else
   {

     DAC_next_value[i] = 0;//0
     }
   }
   set_channel_B(DAC_next_value[0],DAC_next_value[1],DAC_next_value[2]);
   set_channel_C(DAC_next_value[0],DAC_next_value[1],DAC_next_value[2]);
   set_channel_E(DAC_next_value[3],DAC_next_value[4],DAC_next_value[5]);
   set_channel_F(DAC_next_value[3],DAC_next_value[4],DAC_next_value[5]);
   set_channel_G(DAC_next_value[3],DAC_next_value[4],DAC_next_value[5]);

   interrupt_status = 0;
}

// send data only when you receive data:
   if (Serial.available() > 0) {
     Serial.println ("data");
           // read the incoming byte:

           if(header[0] == Serial.read()){
             Serial.println ("1");
             if(header[1] == Serial.read()){
             Serial.println ("2");
              if(header[2] == Serial.read()){
              Serial.println ("3");
               temp_state = Serial.read();
               if (temp_state == 0x00) {
                 Serial.println ("4");
```

```
                    if (Serial.read() == 0xFE) {
                    Serial.println ("5");
                    //noInterrupts();
                      Timer_and_Interrupt_Off();
                      for(uint8_t i = 0; i<number_of_frequency; i++)
                      {
                        DAC_next_value[i] = 0;//0
                      }

set_channel_B(DAC_next_value[0],DAC_next_value[1],DAC_next_value[2]);

set_channel_C(DAC_next_value[0],DAC_next_value[1],DAC_next_value[2]);

set_channel_E(DAC_next_value[3],DAC_next_value[4],DAC_next_value[5]);

set_channel_F(DAC_next_value[3],DAC_next_value[4],DAC_next_value[5]);

set_channel_G(DAC_next_value[3],DAC_next_value[4],DAC_next_value[5]);

set_channel_A_update_all(DAC_next_value[0],DAC_next_value[1],DAC_next_value[2]);
                      interrupt_status = 0;

                    }
                    }

                  if (temp_state == 0x01) {
                    Serial.println ("6");
                    if (Serial.read() == 0xFE) {
                    Serial.println ("7");
                    //interrupts();
                    Timer_and_Interrupt_On();
                    }
                   }
                 }
                }
              }

        }
}
```

# Appendix 4: Manually Control of the Wheelchair Function

```
void Manually_Control() {

  if(digitalRead (Forward_pin) == LOW){
    set_channel_A(increase_mode);
    delay(1);
    set_channel_B(stable_mode);
    delay(1);
    }
  else if(digitalRead (Backward_pin) == LOW){
    set_channel_A(decrease_mode);
    delay(1);
    set_channel_B(stable_mode);
    delay(1);
    }
  else if(digitalRead (Left_pin) == LOW){
    set_channel_B(left_mode);
    delay(1);
    set_channel_A(stable_mode);
    delay(1);
    }
  else if(digitalRead (Right_pin) == LOW){
    set_channel_B(right_mode);
    delay(1);
    set_channel_A(stable_mode);
    delay(1);
    }
  else
    {
    set_channel_B(stable_mode);
    delay(1);
    set_channel_A(stable_mode);
    delay(1);
      }
  }
```

# Appendix 5: Automatic Control of the Wheelchair Function

```
void Automatic_Control()
{
   switch(EEG_command)
   {

      case 1:
         set_channel_A(increase_mode);
         delay(1);
         set_channel_B(stable_mode);
         delay(1);
         break;
      case 2:
         set_channel_A(decrease_mode);
         delay(1);
         set_channel_B(stable_mode);
         delay(1);
         break;
      case 3:
         set_channel_B(left_mode);
         delay(1);
         set_channel_A(stable_mode);
         delay(1);
         Serial.print("3");
         break;
      case 4:
         set_channel_B(right_mode);
         delay(1);
         set_channel_A(stable_mode);
         delay(1);
         break;
      case 0:
         set_channel_B(stable_mode);
         delay(1);
         set_channel_A(stable_mode);
         delay(1);
         break;
      }
   }
```

# Appendix 6: DAC Functions in Arduino Mega 2560

```
void set_dac_to_WTM()
{
   digitalWrite (SYNCpin, LOW);
   SPI.transfer16(0x9000);
   digitalWrite (SYNCpin, HIGH);
}

void set_channel_A(uint8_t value_DAC1)
{
   uint16_t command = 0x0000, command_DAC1;
   command_DAC1 = command | ( (uint16_t) value_DAC1 << 4);
   digitalWrite (SYNCpin, LOW);
   SPI.transfer16(command_DAC1);
   digitalWrite (SYNCpin, HIGH);
}

void set_channel_B(uint8_t value_DAC1)
{
   uint16_t command = 0x1000, command_DAC1;
   command_DAC1 = command | ( (uint16_t) value_DAC1 << 4);
   digitalWrite (SYNCpin, LOW);
   SPI.transfer16(command_DAC1);
   digitalWrite (SYNCpin, HIGH);
}

void set_channel_C(uint8_t value_DAC1)
{
   uint16_t command = 0x2000, command_DAC1;
   command_DAC1 = command | ( (uint16_t) value_DAC1 << 4);
   digitalWrite (SYNCpin, LOW);
   SPI.transfer16(command_DAC1);
   digitalWrite (SYNCpin, HIGH);
}

void set_channel_D(uint8_t value_DAC1)
{
   uint16_t command = 0x3000, command_DAC1;
   command_DAC1 = command | ( (uint16_t) value_DAC1 << 4);
```

```
   digitalWrite (SYNCpin, LOW);
   SPI.transfer16(command_DAC1);
   digitalWrite (SYNCpin, HIGH);
}

void set_channel_E(uint8_t value_DAC1)
{
   uint16_t command = 0x4000, command_DAC1;
   command_DAC1 = command | ( (uint16_t) value_DAC1 << 4);
   digitalWrite (SYNCpin, LOW);
   SPI.transfer16(command_DAC1);
   digitalWrite (SYNCpin, HIGH);
}

void set_channel_F(uint8_t value_DAC1)
{
   uint16_t command = 0x5000, command_DAC1;
   command_DAC1 = command | ( (uint16_t) value_DAC1 << 4);
   digitalWrite (SYNCpin, LOW);
   SPI.transfer16(command_DAC1);
   digitalWrite (SYNCpin, HIGH);
}

void set_channel_G(uint8_t value_DAC1)
{
   uint16_t command = 0x6000, command_DAC1;
   command_DAC1 = command | ( (uint16_t) value_DAC1 << 4);
   digitalWrite (SYNCpin, LOW);
   SPI.transfer16(command_DAC1);
   digitalWrite (SYNCpin, HIGH);
}

void set_channel_H(uint8_t value_DAC1)
{
   uint16_t command = 0x7000, command_DAC1;
   command_DAC1 = command | ( (uint16_t) value_DAC1 << 4);
   digitalWrite (SYNCpin, LOW);
   SPI.transfer16(command_DAC1);
   digitalWrite (SYNCpin, HIGH);
}
```

# Appendix 7: Decrypting Communication Function

```
void Decrypting_Communication()
{
   byte temp_state;

      if(header[0] == Serial1.read()){
         delayMicroseconds(100);//100us=12/115200
        if(header[1] == Serial1.read()){
           delayMicroseconds(100);//100us
          if(header[2] == Serial1.read()){
             delayMicroseconds(100);//100us
             temp_state = Serial1.read();
             delayMicroseconds(100);//100us
           if (Serial1.read() == 0xFE) {
              switch (temp_state)
               {
                  case 0x01:
                     EEG_command = 1;
                     break;
                  case 0x02:
                     EEG_command = 2;
                     break;
                  case 0x03:
                     EEG_command = 3;
                     break;
                  case 0x04:
                     EEG_command = 4;
                     break;
                  case 0x05:
                     EEG_command = 5;
                     break;
                  case 0x06:
                     EEG_command = 6;
                     break;
                  case 0x00:
                     EEG_command = 0;
                     break;
               }
            }
```

```
                }
            }
        }
    }
```

# Appendix 8: LCD Information Transfer Function

```
  void LCD_information()
{
if (Serial1.available() > 0)
            {
              Decrypting_Communication();
              switch (EEG_command)
              {
                 case 1:
                    if (selection == 1)
                    {
                    }
                    else
                    {
                       lcd_1.setCursor(0,1);
                       lcd_1.write("selected");
                       lcd_2.clear();
                       lcd_3.clear();
                       lcd_4.clear();
                       lcd_4.print("right");
                       lcd_3.print("left");
                       lcd_2.print("backward");
                       selection = 1;
                    }
                 break;

                 case 2:
                    if (selection == 2)
                    {
                    }
                    else
                    {
                    lcd_2.setCursor(0,1);
                    lcd_2.write("selected");
                    lcd_1.clear();
                    lcd_3.clear();
                    lcd_4.clear();
                    lcd_4.print("right");
```

```
    lcd_3.print("left");
    lcd_1.print("forward");
    selection = 2;
    }
break;

case 3:
    if (selection == 3)
    {
    }
    else
    {
    lcd_3.setCursor(0,1);
    lcd_3.write("selected");
    lcd_2.clear();
    lcd_1.clear();
    lcd_4.clear();
    lcd_4.print("right");
    lcd_1.print("forward");
    lcd_2.print("backward");
    selection = 3;
    }
break;

case 4:
    if (selection == 4)
    {
    }
    else
    {
    lcd_4.setCursor(0,1);
    lcd_4.write("selected");
    lcd_2.clear();
    lcd_3.clear();
    lcd_1.clear();
    lcd_1.print("forward");
    lcd_3.print("left");
    lcd_2.print("backward");
    selection = 4;
    }
break;
```

```
        case 0:
          if (selection == 0)
          {
          }
          else if (selection == 1)
          {
            lcd_1.clear();
            lcd_1.print("forward");
            selection = 0;
              }
          else if (selection == 2)
          {
            lcd_2.clear();
            lcd_2.print("backward");
            selection = 0;
              }
          else if (selection == 3)
          {
            lcd_3.clear();
            lcd_3.print("left");
            selection = 0;
              }
          else if (selection == 4)
          {
            lcd_4.clear();
            lcd_4.print("right");
            selection = 0;
              }
         break;
      }

//          case 5:
//            break;
//          case 6:
//            break;

    }
}
```

# Appendix 9: Main Codes in Arduino Mega 2560

```
// include the library code:
#include <LiquidCrystal.h>
// inslude the SPI library:
#include <SPI.h>

// initialize the library with the numbers of the interface pins
LiquidCrystal lcd_1(22, 23, 24, 25, 26, 27, 28, 29, 30, 31, 32);
LiquidCrystal lcd_2(22, 23, 37, 25, 26, 27, 28, 29, 30, 31, 32);
LiquidCrystal lcd_3(22, 23, 34, 25, 26, 27, 28, 29, 30, 31, 32);
LiquidCrystal lcd_4(22, 23, 35, 25, 26, 27, 28, 29, 30, 31, 32);
LiquidCrystal lcd_5(22, 23, 33, 25, 26, 27, 28, 29, 30, 31, 32);
LiquidCrystal lcd_6(22, 23, 36, 25, 26, 27, 28, 29, 30, 31, 32);

byte header[3]={0xFF,0xAA,0x01};
byte EEG_command = 0;
const int mode_select = 21, SYNCpin = 47, Forward_pin = 40, Backward_pin = 41, Left_pin
= 42, Right_pin = 43;
const int stable_mode = 128, increase_mode = 175, decrease_mode = 80, right_mode = 185,
left_mode = 70;
int mode_state, previous_state = 2;

int selection = 0;

void setup() {
   Serial.begin(112500);
   Serial1.begin(112500);
   //initialize SPI:
   SPI.begin();
   SPI.beginTransaction( SPISettings( 30000000, MSBFIRST, SPI_MODE1 ) );
   pinMode(mode_select,INPUT);
   pinMode(SYNCpin, OUTPUT);
   pinMode(Forward_pin, INPUT_PULLUP);
   pinMode(Backward_pin, INPUT_PULLUP);
   pinMode(Left_pin, INPUT_PULLUP);
   pinMode(Right_pin, INPUT_PULLUP);

   // set up the LCD's number of columns and rows:
   lcd_1.begin(8, 2);
```

```
  lcd_2.begin(8, 2);
  lcd_3.begin(8, 2);
  lcd_4.begin(8, 2);
  lcd_5.begin(8, 2);
  lcd_6.begin(8, 2);

  set_dac_to_WTM();
  delay(1);
  set_channel_B(stable_mode);
  delay(1);
  set_channel_A(stable_mode);
  delay(1);
  interrupts();

}

void loop() {
  if (digitalRead(mode_select) == HIGH)
  {
      if (Serial1.available() > 0)
      {
      mode_state = 0;
      if (mode_state == previous_state)
        {
        LCD_information();
        Automatic_Control();
        }
      else{
        lcd_1.clear();
        lcd_2.clear();
        lcd_3.clear();
        lcd_4.clear();
        lcd_5.clear();
        lcd_6.clear();
        previous_state = 0;
        lcd_1.print("forward");
        lcd_2.print("backward");
        lcd_3.print("left");
        lcd_4.print("right");
        lcd_5.print("");
        lcd_6.print("");
```

```
        // send data only when you receive data:
            LCD_information();
            Automatic_Control();
                  }
          EEG_command = 0;
          }
    }

    else
    {
       mode_state = 1;
       if (mode_state == previous_state)
          {
          Manually_Control();
          }
       else
       {
          lcd_1.clear();
          lcd_2.clear();
          lcd_3.clear();
          lcd_4.clear();
          lcd_5.clear();
          lcd_6.clear();
          previous_state = 1;
          lcd_1.print("manual");
          lcd_2.print("manual");
          lcd_3.print("manual");
          lcd_4.print("manual");
          lcd_5.print("manual");
          lcd_6.print("manual");
          Manually_Control();
       }
    }
}
```